\documentclass[%
reprint,nofootinbib,
amsmath,amssymb,
aps,prd]{revtex4-2}
\usepackage[latin9]{inputenc}

\usepackage{bm}
\usepackage{graphicx}
\usepackage{hyperref}

\usepackage{mathtools}

\usepackage{yhmath}

\begin{document}
\title{Gravitational waves from corrugated bubble walls}

\author{Ariel Mégevand}
\email{megevand@mdp.edu.ar}

\author{Federico Agustín Membiela}
\email{membiela@mdp.edu.ar}

\affiliation{Instituto de Investigaciones Físicas de Mar del Plata (IFIMAR) CONICET,
	UNMDP, Funes 3350, 7600 Mar del Plata, Argentina}

\begin{abstract}
We study the production of gravitational waves during a first-order cosmological phase transition,
taking into account a hydrodynamic instability that occurs when the velocity of a bubble wall is below a critical threshold.
This instability causes exponential growth of wall deformations on a length scale much smaller than the bubble radius.
Additionally, fluid perturbations grow within a thin region next to the bubble wall.
We assume that the perturbations stop growing once they become nonlinear.
While the gravitational wave spectrum from bubble collisions peaks at a frequency related to the bubble separation, the instability causes an additional peak at a higher frequency, related to the corrugation scale.
Wall corrugation also affects the gravitational wave spectrum from bubble collisions, causing it to decrease faster beyond the new scale.
We discuss the possibility of observing these effects with future gravitational wave observatories.
\end{abstract}
\maketitle

\section{Introduction}

First-order phase transitions (FOPTs) in the early universe involve
the nucleation and expansion of bubbles within a relativistic fluid
\cite{quiros07field,abf+24cosmological}. 
The motion of these bubbles and the surrounding fluid can generate a stochastic gravitational wave (GW)
background 
\cite{hogan86gravitational,tw90relic,ktw92gravitational,ktw92gravitationala,kt93gravitational,kkt94gravitational}
that may be detectable by future GW observatories \cite{maggiore00gravitational}.
Specifically, the space-based  interferometer LISA \cite{a+17laser} could
detect GWs from a FOPT at the electroweak scale ($\sim 100\,\mathrm{GeV}$),
and pulsar timing array experiments \cite{jhm+15gravitational}
may detect GWs generated at the QCD scale ($\sim 100\,\mathrm{MeV}$).
In the Standard Model of particle physics, the QCD and electroweak
phase transitions are smooth crossovers \cite{klr+96is,klr+97nonperturbative,aef+06order}.
However, many extensions of the Standard Model allow for FOPTs (see, e.g.,
\cite{amn+01supersymmetric,ekn+08some,hk08production,lms12gravitational,dhn14cosmological,kkm15gravitational,hkn+16detectable,lm16gravitational,gq16strong,tyy17gravitational,dhk+17second,csw17gravitational,agk17gravitational,hs18qcd,mnq18cosmological,cpq+25primordial}).

Three main sources of GWs from a phase transition have been investigated
in the literature, namely, bubble collisions, sound waves, and turbulence.
The bubble collision mechanism \cite{tw90relic,ktw92gravitational,ktw92gravitationala,kt93gravitational,hk08gravitational}
considers the motion of uncollided thin bubble walls. 
This approach is known as the envelope approximation \cite{kt93gravitational}.
This mechanism dominates GW production in FOPTs with significant supercooling \cite{lm16gravitational,mr17bubble,eln+19gravitational,eln19maximal,elv20updated}.
In this case, most of the energy released during the phase transition
goes to the bubble walls, which reach ultra-relativistic velocities
\cite{ekn+10energy,megevand13friction,lm16hydrodynamics}. In FOPTs
with moderate supercooling, the bubble walls are slower and a significant
fraction of the released energy goes to bulk fluid motions \cite{dk84supercooling,eik+92nucleation,kl96bubble,megevand08gravitational,ms10velocity,lm15hydrodynamics,mr18bubble}.
Although the released energy is smaller than in extremely supercooled FOPTs,
the resulting sound waves keep producing GWs after the phase transition
is completed \cite{hhr+15numerical,hhr+14gravitational,hhr+17shape,hindmarsh18sound,hh19gravitational}.
Furthermore, if the fluid becomes turbulent, the eddies provide
an additional source of GWs \cite{kkt94gravitational,dgn02relic,kmk02gravitational,cd06gravitational,cds09stochastic,kks10gravitational,gkk07spectrum,rmb+20numerical}.

The details of the generated GW spectrum depend on the specific mechanism.
However, under the assumption of homogeneous nucleation of spherical bubbles that
grow at a constant wall velocity $v_{w}$, the relevant scale that
determines the wavelength of GWs is the duration of the phase transition,
usually denoted by $\beta^{-1}$.
More precisely, the generated GW spectrum has
a peak at a frequency $\omega\sim\beta$. 

On the other hand, it is well known that the steady motion of a bubble wall that
propagates as a deflagration is unstable \cite{LandauLifshitz87Fluid}.
In the context of cosmological FOPTs, this instability was studied
in Refs.~\cite{fa90hadron,link92deflagration,kf92instability,hkl+93hydrodynamic,fa03hydrodynamic,mm14stability,mms15lower}.
Below a critical wall velocity, small perturbations in the wall and fluid grow exponentially. 
This occurs for perturbations on
wavelengths larger than a certain critical value $\lambda_c$ which is
several orders of magnitude smaller than the bubble radius $R\sim v_{w}\beta^{-1}$.
In Ref.~\cite{mm14stability}, we noted that this hydrodynamic
instability could create a peak in the GW spectrum at $\omega\gg\beta$
(see \cite{mm21gravitational} for simple modeling of this phenomenon).

The first rigorous stability analysis of a cosmological phase transition front
that correctly accounted for the forces acting on the bubble wall 
was conducted in  Ref.~\cite{hkl+93hydrodynamic}.
A more detailed study was carried out in Ref.~\cite{mm14stability}.
Unfortunately, these studies only considered the linear stability of planar walls.
The spherical case and the nonlinear regime remain to be studied.
A simulation in 2+1 dimensions was performed in Ref.~\cite{fa03hydrodynamic}.
No unstable behavior was observed, in tension with the linear stability
analysis. 
However, this type of computation is constrained by computing
resources which limit the modeling of the system and the use of realistic
size and time scales.
The same kind of instability is known to develop in a combustion front for spherical bubbles
(see, e.g., Ref.~\cite{bl00dynamics}).
The dynamics of the instability beyond the linear
regime is far from understood. 
For a phase transition front, a dendritic growth model was proposed
in Ref.~\cite{fa90hadron}, while a transition from deflagration
to detonation was assumed in Ref.~\cite{kf92instability}. In a
combustion front, Rayleigh-Taylor or Darrieus-Landau instabilities
are known to cause velocity amplification and fractal front structure
\cite{bl00dynamics}.

In this work, we study the spectrum of GWs produced during a FOPT
in the presence of this instability. 
In Ref.~\cite{mm26gravitational}, we model bubble nucleation as occurring simultaneously
during a phase transition at the electroweak scale.
Here, we consider various temperature scales and use a more realistic exponential nucleation model.
Additionally, we provide more detail on some of the derivations and approximations used in Ref.~\cite{mm26gravitational}.
Furthermore, we analyze the stochastic background of GWs that would be observed today.

We will use the results of the linear stability analysis of planar walls.
This approximation is valid because the length scale of wall deformations is much smaller than the bubble radius
and the unstable fluid perturbations are confined to a thin layer adjacent to the bubble walls.
With the-thin wall approximation, 
our calculation resembles the envelope approximation for the bubble-collision mechanism.
Several phenomena that may occur in the nonlinear regime, such as cusp formation, fragmentation, area increase, and velocity amplification, may affect the GW spectrum. 
However, we will assume that perturbations stop growing once they become nonlinear.
We will consider other effects elsewhere.
The fluid perturbations may cause changes in the turbulence and sound wave spectra.
We will consider the GWs generated by these mechanisms elsewhere.

The paper is organized as follows. In the next section, we briefly
review the dynamics of a FOPT, the hydrodynamics of a moving bubble
wall, and the instability of deflagrations. In Sec.~\ref{sec:stochastic},
we perform a statistical analysis of the interface deformations, adapt
the planar-wall results to spherical bubbles, and model the nonlinear regime.
In Sec.~\ref{sec:gravrad}, we consider deformed bubbles
during a phase transition as a stochastic source of GWs. We assume
small deformations and use a perturbative calculation. In Secs.~\ref{sec:bbcolldef}
and \ref{sec:Pi11}, we calculate contributions up to second order.
In Sec.~\ref{sec:spectrum}, we calculate the GW spectrum generated
during the phase transition and discuss the GW background that would be observed
today. 
In Sec.~\ref{sec:conclu} we summarize our conclusions. In
App.~\ref{app:corrfunc}, we provide details on the stochastic equations and correlation functions.
In App.~\ref{app:Approximations}, we present analytic approximations and provide fit functions. 

\section{Deflagrations in phase transitions}

\label{sec:instadefla}

\subsection{Phase transition dynamics}

The key element in describing a phase transition is the
free energy or finite-temperature effective potential $V(\phi,T)$,
where $T$ is the temperature and $\phi$ is a scalar field that
acts as an order parameter. 
The effective potential can be obtained through standard field-theoretical calculations from a specific model. 
The state of the system is determined by the free energy minimum $\phi_{m}$.
The equation of state is obtained by expressing the
pressure $p$ as a function of $T$ through the relation $p=-V(\phi_{m}(T),T)\equiv-V(T)$.
In the case of a first-order phase transition, there is a range of
temperatures in which $V$ has two minima $\phi_{\pm}(T)$,
and therefore there are two coexisting phases. 
The smaller of the corresponding values $V_{\pm}(T)$ determines the stable phase.
We use the index $+$ to denote the phase that is stable at higher temperatures
and the index $-$ to denote the phase that is stable at lower temperatures.
At the critical temperature $T_{c}$, we have $V_{+}=V_{-}$.

The equation of state (EOS) of each phase is often modeled using the
bag EOS,
\begin{equation}
p_{+}=\frac{\pi^{2}}{90}g_{*}T_{+}^{4}-\frac{L}{4},\quad p_{-}=\left(\frac{\pi^{2}}{90}g_{*}-\frac{L}{4T_{c}^{4}}\right)T^{4},\label{eq:bag}
\end{equation}
where $g_{*}$ is the number of relativistic degrees of freedom of
the plasma and $L$ is the latent heat. 
In this model, $L$ is related to the false
vacuum energy density $\rho_{\mathrm{vac}}$ by $L=4\rho_{\mathrm{vac}}$.
We define the quantity
\begin{equation}
\alpha_{T}=\frac{\rho_{\mathrm{vac}}}{\rho_{\mathrm{rad}}}=\frac{L}{3w_{+}},\label{eq:alphaT}
\end{equation}
where $\rho_{\mathrm{rad}}=\frac{\pi^{2}}{30}g_{*}T^{4}$ is the radiation
energy density in the high-temperature phase, and $w_{+}=4\rho_{\mathrm{rad}}/3$
is the enthalpy.
The parameters of the bag EOS can be related to the effective potential for a realistic model.

Below the critical temperature, the metastable phase $+$ becomes
supercooled and bubbles of the $-$ phase begin to nucleate at a rate $\Gamma(T)$
which can be obtained from the effective potential \cite{coleman77fate,cc77fate,linde81fate,linde83decay}.
The temperature decrease is governed by the adiabatic expansion of the universe.
The function $\Gamma(T(t))$ is usually modeled as an exponential,
\begin{equation}
\Gamma=\Gamma_{*}e^{\beta(t-t_{*})}.\label{eq:nuclexp}
\end{equation}
In general, this is a good approximation near a conveniently chosen
time $t_{*}$. The quantity $\beta$ depends on the specific model.
In general, $\beta^{-1}$ is a fraction of the Hubble time, and we
typically have $\beta\sim10-1000H$. The Hubble rate is given by
\begin{equation}
H=\sqrt{\frac{8\pi}{3}G\rho_{\mathrm{tot}}}\simeq\sqrt{\frac{8\pi^{3}g_{*}}{90}(1+\alpha_{T})},\label{eq:Friedmann}
\end{equation}
where $\rho_{\mathrm{tot}}$ is the total energy density, $G=1/M_{P}^{2}$
is Newton's constant, and $M_{P}\simeq1.22\times10^{19}\mathrm{GeV}$
is the Planck mass.

A nucleated bubble is a spherically symmetric configuration in which the field takes
the value $\phi_{-}$ at the center and the value $\phi_{+}$ at infinity.
During the phase transition, the region in the $-$ phase (the bubble interior) grows to a size of the order of $\beta^{-1}$.
The region where the field varies between $\phi_+$ and $\phi_-$ (the bubble wall) has a width of the order of $T^{-1}$.
This width is several orders of magnitude smaller than the final bubble size.
Therefore, the bubble wall can be approximated as a thin interface.
Even if the wall deforms during the phase transition,
the thin-wall approximation is valid as long as the wall width is much smaller than the curvature radius of the deformations
\cite{mm23thin,mm24thinb,mm25thin}.

The bubble expansion is driven by the pressure difference $p_{-}-p_{+}$
and is obstructed by the back-reaction of the plasma \cite{gkk+84deflagrations,kurkisuonio85deflagration,kk86bubble,kl95supersonic}.
The interaction of the wall with plasma particles at the microscopic level
can be effectively described with a friction force \cite{lmt92bubble,turok92electroweak,dlh+92towards,khlebnikov92fluctuation,arnold93one,mp95bubble,mp95how,bbm00can,bm17electroweak}.
Thus, the dynamics of the field-fluid system can be modeled using the equations
\cite{ikk+94growth}
\begin{align}
\partial_{\mu}\left(-T\frac{\partial V}{\partial T}u^{\mu}u^{\nu}+g^{\mu\nu}V\right)+\partial_{\mu}\partial^{\mu}\phi\partial^{\nu}\phi & =0,\label{fluideq}\\
\partial_{\mu}\partial^{\mu}\phi+\frac{\partial V}{\partial\phi}+\tilde{\eta}u^{\mu}\partial_{\mu}\phi & =0,\label{fieldeq}
\end{align}
where $\ensuremath{u^{\mu}=(\gamma,\gamma\mathbf{v})}$ is the fluid four-velocity.
For a constant $\phi$, Eq.~(\ref{fluideq}) yields conservation of the fluid's energy-momentum tensor,
\begin{equation}
\partial_{\mu}\left(wu^{\mu}u^{\nu}-pg^{\mu\nu}\right)=0,\label{eq:fluideqsinphi}
\end{equation}
where $w$ is the enthalpy.
The last term in Eq.~(\ref{fluideq}) represents the energy transfer between the fluid and the field. 
The field equation, Eq.~(\ref{fieldeq}), includes a phenomenological damping
term. A suitable function $\tilde{\eta}=f(\phi,u^{\mu})$ can be selected
to model the actual damping obtained from microphysics calculations
\cite{megevand13friction}. For subsonic wall velocities, a constant
coefficient $\tilde{\eta}$ is appropriate. 

In the thin-wall approximation,
we can assume a field profile of the form $\phi=\phi_{0}(n)$, where
$n$ is the distance to the hypersurface describing the worldvolume
of the wall. 
In a coordinate system that includes $n$ as a coordinate,
Eq.~(\ref{fieldeq}) becomes
\begin{equation}
\phi_{0}''(n)-\mathcal{K}(n)\phi_{0}'(n)=V'(\phi_{0}).\label{fieldeqGNC}
\end{equation}
The quantity $\mathcal{K}$ contains contributions from the first and third
terms in Eq.~(\ref{fieldeq}).
The former are related to the curvature of the hypersurface
\cite{mm23thin,mm24thinb}. 
A good estimate of $\phi_0$ can be obtained by
neglecting the term with a single
derivative, $\phi_{0}'(n)$, in Eq.~(\ref{fieldeqGNC}), which corresponds
to the lowest-order approximation in the wall width \cite{mm23thin,mm24thinb}.
We thus obtain the same equation as for the profile of a planar wall
at rest,
\begin{equation}
\phi_{0}''(n)=V'(\phi_{0}).\label{fieldeqthinwall}
\end{equation}
On the other hand, if we multiply Eq.~(\ref{fieldeqGNC}) by $\phi_{0}'(n)$
and integrate, the first term vanishes. Since the quantity $\mathcal{K}$
is approximately constant within the range in which $\phi_{0}'$ does
not vanish, we obtain the equation
\begin{equation}
-\sigma_{0}\mathcal{K}_{n=0}= F_{\mathrm{dr}},\label{eq:walleqgral}
\end{equation}
where 
$\sigma_{0}=\int[\phi_{0}'(n)]^{2}dn$ and $F_{\mathrm{dr}} = \int V'(\phi_{0})\phi_{0}^{\prime}dn$.
The wall dynamics is governed by Eq.~(\ref{eq:walleqgral}). 
The quantity $\sigma_0$ is the  surface tension.
The quantity $\mathcal{K}_{n=0}$
depends on the acceleration and the spatial curvature of the surface.
The right-hand side gives the driving force $F_{\mathrm{dr}}$ that
makes the wall move towards the metastable phase. 
For a vacuum phase transition, $V$ does not depend on $T$,
and we obtain $F_{\mathrm{dr}}=V_{+}-V_{-}$. 

\subsection{Stationary deflagration fronts}

\label{subsec:stationary}

Due to the friction with the plasma, the bubble walls generally reach a terminal velocity.
In this regime, the acceleration term in the quantity $\mathcal{K}|_{n=0}$ vanishes.
For spherical wall of radius $R\sim \beta^{-1}\gg T^{-1}$,
we can also ignore the spatial curvature, and we have $\mathcal{K}|_{n=0}=\tilde{\eta}\gamma v$,
where
$v$ is the velocity of the fluid relative to the wall and $\gamma$
is the Lorentz factor $\gamma=1/\sqrt{1-v^{2}}$.
Thus, Eq.~(\ref{eq:walleqgral}) becomes
\begin{equation}
	F_{\mathrm{fr}} = F_{\mathrm{dr}}.\label{eq:eqvw}
\end{equation}
where 
$F_{\mathrm{fr}}
=\tilde{\eta}\sigma_{0}\gamma v$ is the friction force.
We estimate
the value of $\gamma v$ within the wall by taking the average of
the values in front of and behind it. 
Therefore, we have $F_{\mathrm{fr}}=\eta\overline{\gamma v}$,
with $\eta=\tilde{\eta}\sigma_{0}$ and
$\overline{\gamma v}=\frac{1}{2}(\gamma_{+}v_{+}+\gamma_{-}v_{-})$.
With the same kind of approximations, the driving force is given by
(see, e.g., \cite{mm14stability}) 
\begin{equation}
F_{\mathrm{dr}}=p_{-}-p_{+}+\langle dp/dT^{2}\rangle(T_{+}^{2}-T_{-}^{2}).\label{eq:Fdr}
\end{equation}
Besides the pressure difference between the two phases, there is a
term arising from the temperature variation inside the wall. 

In the steady state,
we can assume that the solutions to the fluid equations are spherically symmetric.
The equations for the radial fluid profiles in each phase can be derived from Eq.~(\ref{eq:fluideqsinphi}).
The boundary conditions are those of an unperturbed fluid far away
from the bubble and of a fluid at rest at the bubble center. 
In particular, far in front of the wall, the temperature is the nucleation temperature $T_{N}$.
The matching conditions at the thin interface
can be obtained by integrating Eq.~(\ref{fluideq}). 
In the wall reference frame, we obtain
\begin{align}
w_{-}v_{-}\gamma_{-}^{2} & =w_{+}v_{+}\gamma_{+}^{2},\label{eq:matching1}\\
w_{-}v_{-}^{2}\gamma_{-}^{2}+p_{-} & =w_{+}v_{+}^{2}\gamma_{+}^{2}+p_{+}.\label{eq:matching2}
\end{align}

There are different kinds of solutions to Eqs.~(\ref{eq:matching1})-(\ref{eq:matching2}),
which are classified as weak detonations, Jouguet detonations, Jouguet
deflagrations, and weak deflagrations. We are interested in the latter,
for which we have $T_{+}>T_{-}$ and $v_{+}<v_{-}$. 
Considering the fluid profiles and boundary conditions, the weak deflagration
also fulfills $T_{+}>T_{N}$ and $v_{-}=v_{w}<c_{s}$, where $c_{s}$
is the speed of sound ($c_{s}=1/\sqrt{3}$ for the bag EOS). 
This means that the wall is subsonic, with a shock wave preceding it that reheats the plasma and sets the fluid in motion. Meanwhile, the fluid behind the wall remains at rest.

Solving Eqs.~(\ref{eq:eqvw})-(\ref{eq:matching2}) yields the wall velocity as a
function of the phase transition parameters. 
In this work, we treat the stationary wall velocity as a free parameter. However, it is
worth noting its parametric dependence. For small $v_{w}$ and $\alpha_{T}$,
we have
\begin{equation}
v_{w}\simeq F_{\mathrm{dr}}/\eta,\quad F_{\mathrm{dr}}\simeq L(1-T_{+}/T_{c})\equiv Lv_{c}^{2},\label{eq:FdrNR}
\end{equation}
where we have defined the quantity 
\begin{equation}
v_{c}=\sqrt{(T_{c}-T_{+})/T_{c}},\label{eq:vc}
\end{equation}
which will play the role of a critical velocity for the stability
of the deflagration. 
The plasma is reheated in front of the wall.
For a planar wall, we have
$T_{+}\simeq T_{N}+\sqrt{3}T_{c}\alpha_{T}v_{w}$.
For a spherical bubble, the relationship between $T_+$ and $T_N$ is similar, although the temperature profile is different
\cite{lm11spherical}.
Therefore, we have
\begin{equation}
v_{w}=Lv_{c}^{2}/\eta,\quad v_{c}^{2}=\frac{1-T_{N}/T_{c}}{1+\sqrt{3}\alpha_{T}L/\eta}.
\label{eq:vwvc}
\end{equation}
The quantities $L$ and $\eta$, as well as the degree of supercooling
$T_{N}/T_{c}$, can be calculated for each specific model. Since these
quantities are not directly related, we consider $v_{w}$
and $v_{c}$ as independent parameters. 

The initial bubble radius is of the order of $T^{-1}$, which is several orders of magnitude smaller than the final bubble radius.
The terminal velocity $v_w$ is reached in a time that is also of this order and can be neglected \cite{heckler95effects}.
Therefore, a bubble nucleated
at time $t_{N}$ has a radius $R=v_{w}(t-t_{N})$ at time $t$.
The volume fraction occupied by the metastable phase is given by (see, e.g., Ref.~\cite{gw81cosmological}) 
\begin{equation}
	f_{+}=e^{-I},\quad I=\int_{-\infty}^{t}\negthickspace dt_{N}\Gamma(t_{N})\frac{4\pi}{3}\left[v_{w}(t-t_{N})\right]^{3}.\label{defI}
\end{equation}
For the exponential nucleation rate, we have
\begin{equation}
	I(t)=8\pi v_{w}^{3}\beta^{-4}\Gamma_{*}e^{\beta(t-t_{*})}.\label{eq:Iexp}
\end{equation}
The phase transition ends when $f_+ = 0$, which formally takes infinite time.
However, $f_+$ becomes vanishingly small in a time of order $\beta^{-1}$. 

\subsection{Linear stability analysis}

The stationary phase transition front has been studied for planar,
cylindrical, and spherical walls (see, e.g., Refs.~\cite{kurkisuonio85deflagration,lm11spherical}).
On the other hand, the hydrodynamic stability
of these solutions has only been studied for planar walls. Below,
we briefly review the treatment and results. Unstable perturbations
in the fluid grow very close to the wall, in a region much shorter
than the extension of the stationary fluid profile. Additionally,
the characteristic wavelength of the unstable modes is much shorter
than the typical bubble radius. 
Therefore, we expect the planar-wall results to be a good approximation.

The standard procedure for linear stability analysis is the following \cite{LandauLifshitz87Fluid}. 
Consider a planar wall perpendicular to the $z$-axis that moves with a constant
velocity $v_{w}$ along this axis. In the wall's reference,
its position is $z=0$. To describe wall perturbations, we can write
the wall's position as $z=\zeta(\mathbf{x}^{\perp},t)$,
where $\mathbf{x}^{\perp}=(x,y)$ is a vector on the plane of the
unperturbed wall. 
The equation for $\zeta$ can be obtained from Eq.~(\ref{eq:walleqgral}).
To linear order, we have \cite{mm14stability} 
\begin{equation}
\sigma_0(\partial_{0}^{2}-\nabla_{\perp}^{2})\zeta
	+\eta\bar{\gamma}\partial_{0}\zeta- \delta F_{\mathrm{fr}}-
{\delta F_{\mathrm{dr}}}=0,\label{walleq}
\end{equation}
where $\nabla_{\perp}=(\partial_{x},\partial_{y})$, 
$\bar\gamma = \frac12(\gamma_+ + \gamma_-)$,
and $\delta F_{\mathrm{fr}}$ and $\delta F_{\mathrm{dr}}$
are the variations of the forces in Eq.~(\ref{eq:eqvw}) due to the fluid perturbations.
The variation of the driving force
depends on the temperature perturbations on each side of the wall,
$\delta T_{\pm}$.
Changes in $p$ and $T$ can be related using the bag EOS.
The variation of the friction force depends on the longitudinal perturbations of the fluid velocity, $\delta v_\pm$.

The matching conditions for the perturbed fluid variables at the interface
can be obtained from Eq.~(\ref{fluideq}).
These conditions are analogous
to Eq.~(\ref{eq:matching1})-(\ref{eq:matching2}) and involve the
perturbation $\delta v$ and the transverse components $\mathbf{v}^{\perp}=(v_{x},v_{y})$.
In addition, the fluid equations (\ref{eq:fluideqsinphi}) must be
considered in each phase. There are four equations for the variables
$\delta p,\delta v,v_{x},v_{y}$. 
We do not write down all the
equations here but only summarize the results. See Ref.~\cite{mm14stability}
for details.

Since the fluid equations have constant coefficients, we consider,
in each phase, a solution of the form 
\begin{equation}
\mathbf{U}\equiv(\delta p,\delta v,\mathbf{v}^{\perp})=\mathbf{L}e^{\Omega t+qz+i\mathbf{\mathbf{k}\cdot\mathbf{x}^{\perp}}}.\label{eq:U}
\end{equation}
The wavevector $\mathbf{k}$ is real, while $q$ and $\Omega$ may
be complex. The boundary conditions far away from the wall impose
the conditions $\mathrm{Re}(q)>0$ for $z<0$ and $\mathrm{Re}(q)<0$
for $z>0$. Once the matching conditions at the interface are imposed,
the existence of solutions with $\mathrm{Re}(\Omega)>0$ implies that
the stationary solution is unstable. 

The vector $\mathbf{v}^{\perp}$ is in the direction of $\mathbf{k}$,
which naturally reduces the degrees of freedom%
\footnote{This is due to symmetry, and can also be derived from one of the fluid equations, $w\gamma^{2}(\partial_{0}+v\partial_{z})\mathbf{v}^{\perp}=-{\nabla}_{\perp}\delta p$,
which implies $\mathbf{v}^{\perp}\propto\mathbf{k}$ for perturbations of
the form (\ref{eq:U}).}.
Defining $v^{\perp}=\mathbf{v}^{\perp}\cdot\hat{k}$ and $x^{\perp}=\mathbf{x}^{\perp}\cdot\hat{k}$,
where $\hat{k}=\mathbf{k}/k$ and $k=|\mathbf{k}|$),
we obtain a system of three equations for the variables 
\begin{equation}
\mathbf{U}^{3}\equiv(\delta p,\delta v,v^{\perp})=\mathbf{L}^{3}e^{\Omega t+qz+ikx^{\perp}}.\label{eq:Ureducido}
\end{equation}
Once we solve for $\mathbf{U}^{3}$, we can obtain $\mathbf{U}$ by replacing $kx^\perp = \mathbf{\mathbf{k}\cdot\mathbf{x}^{\perp}}$
and taking into account that $\mathbf{v}^{\perp} = {v}^{\perp}\hat{k}$.

For a solution of the form (\ref{eq:Ureducido}), the system of differential equations
yields a system of linear equations for the components of the vector
coefficient $\mathbf{L}^{3}$.
Requiring the existence of a non-trivial solution
yields an equation for the parameters $\Omega$, $q$ and $k$,
from which we obtain three dispersion relations $q_{i}(\Omega,k)$.
The corresponding solutions $\mathbf{L}_{i}^{3}(\Omega,k)$ give the relative
amplitudes of the variables $\delta p$, $\delta v$ and $v^{\perp}$ for each
of these three modes.

The general solution is a superposition of these modes.
According to the sign of $q_{i}$, each dispersion relation is only
permitted on one side of the wall.
For $\Omega>0$, there are two modes behind the wall and one in front.
For $\Omega<0$, there are two modes in front and one behind.
The calculation is similar in either case.
We have, e.g.,
\begin{align}
	\mathbf{U}^{3}_+ &= e^{\Omega t+ikx^{\perp}} A \mathbf{L}_{1}^{3}e^{q_{1}z}, \\
	\mathbf{U}^{3}_- &= e^{\Omega t+ikx^{\perp}}\left( B\mathbf{L}_{2}^{3}e^{q_{2}z} + C\mathbf{L}_{3}^{3}e^{q_{3}z} \right).
\end{align}
Continuity at the interface requires that
$\Omega$ and $k$ have the same values on both sides.
Additionally, the wall perturbation must
be of the form 
\begin{equation}
\zeta=De^{\Omega t+ikx^{\perp}}.\label{eq:zetak}
\end{equation}
There are three matching conditions along with the wall equation (\ref{walleq}),
which yields a system of four linear equations for the four amplitudes $A,B,C,D$. 
Again, we require a nontrivial solution, which yields an equation for the quantities $\Omega$ and $k$.

There are only two physical solutions $\Omega_{\pm}(k)$.
The expressions for the quantities $\Omega_{\pm}$ are cumbersome.
For $v_{w}\ll1$ and $\alpha_{T}\ll1$, we have \cite{mm14stability}
\begin{align}
\Omega_{+} & \simeq\frac{3}{2}\alpha_{T}\,v_{w}\left(1-\frac{v_{w}^{2}}{v_{c}^{2}}\right)\left(1-\frac{k}{k_{c}}\right)k,\label{Ommas}\\
\Omega_{-} & \simeq-v_{w}^{-1}d^{-1},\label{Ommen}
\end{align}
where
\begin{equation}
k_{c}\simeq\frac{3}{2}\alpha_{T}\left(1-\frac{v_{w}^{2}}{v_{c}^{2}}\right)\,d^{-1}\label{kc}
\end{equation}
and $d^{-1}\equiv F_{\mathrm{dr}}/\sigma_{0}\simeq Lv_{c}^{2}/\sigma_{0}$.
The quantity $\Omega_{-}$ is always negative. The quantity $\Omega_{+}$
is also negative for $v_{w}>v_{c}$ (in this case, we have $k_{c}<0$).
For $v_{w}<v_{c}$, we have $\Omega_{+}>0$ for $k<k_{c}$. Therefore,
$v_{c}$ is a critical velocity below which perturbations with wavenumbers
below the critical value $k_{c}$ grow exponentially. 

For simplicity, we will use the approximations (\ref{Ommas})-(\ref{kc}),
which describe well the general results. The maximum of $\Omega_{+}$
occurs for $k\sim k_{c}$.
This parameter is of the order of $d^{-1}$.
The quantity $d$ is a length scale that arises from the balance
between the pressure difference and the surface tension. 
Roughly, we have $L\sim T^{4}$ and $\sigma_{0}\sim T^{3}$, and we obtain $d\sim T^{-1}$. 
On the other hand, the final bubble radius $R$ is of the order of $\beta^{-1}$.
For a typical value $\beta\sim100H$, we have, roughly, $d/R\sim T/M_{P}$.
This value is actually amplified by a factor of $(\alpha_{T}v_{c}^{2}v_{w})^{-1}$. 
However, the small ratio of $T/M_{P}$ makes $d\ll R$ for most phase transitions. 

\begin{figure}[hbt]
\includegraphics[width=0.8\columnwidth]{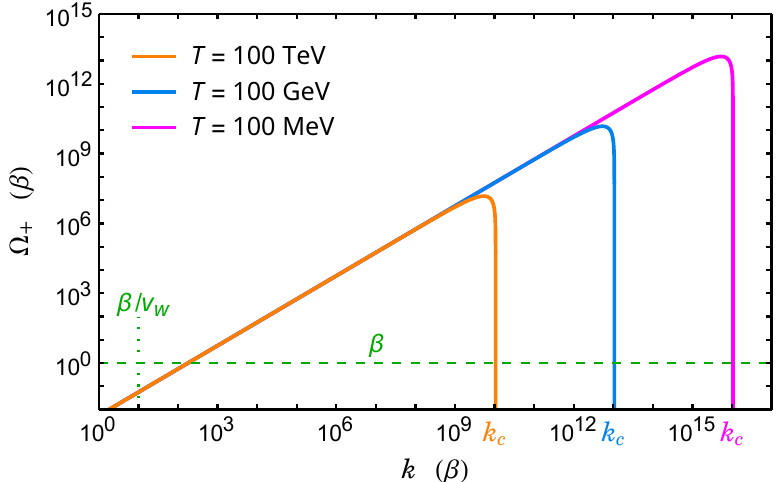}
\caption{The growth rate of the instability as a function of $k$. 
	The values of the parameters are $\beta/H=100$, $\alpha_{T}=0.05$, $v_{c}=0.2$, and $v_{w}=0.1$.\label{figomegamas}}
\end{figure}

In Fig.~\ref{figomegamas} we plot the function $\Omega_{+}(k)$
for phase transitions occurring at different temperature scales.
For the most unstable mode, we have $\Omega_{+}\gg\beta$ and $k\sim k_{c}\gg R^{-1}$.
Therefore, this mode stops its exponential growth after a very
short time, since it becomes nonlinear as soon as its amplitude is
of the order of its wavelength, $k^{-1}\sim k_c^{-1} \ll R$. In contrast, modes with longer wavelengths
grow with a smaller rate $\Omega_{+}$ but may reach sizes of
order $k^{-1}\gg k_{c}^{-1}$ before becoming nonlinear.

\section{Stochastic perturbations}

\label{sec:stochastic}

\subsection{Linear equations with sources}

In the presence of time-dependent sources of perturbations, the equations
no longer admit solutions of the form $e^{\Omega t}$.
However, we can look for solutions of the form 
\begin{equation}
\boldsymbol{\mathbf{U}}=\mathbf{L}(t)e^{i\mathbf{k}\cdot\mathbf{x}^{\perp}+qz},
\quad
\zeta=D(t)e^{i\mathbf{k}\cdot\mathbf{x}^{\perp}}.\label{Uzetakt}
\end{equation}
Therefore, we only need to solve a system of ordinary differential equations
for $\mathbf{L}(t)$ and $D(t)$. 
In particular, Eq.~(\ref{walleq}) yields
\begin{equation}
\sigma_{0}\ddot{D}+\sigma_{0}k^{2}D+\eta\bar{\gamma}\dot{D}+\text{fluid terms}=\text{source},\label{oscarmestable}
\end{equation}
where the fluid terms contain
the fluid variables evaluated at $z=0$.
In the fluid equations, the condition $\mathbf{v}^{\perp} \propto \mathbf{k}$
and the restriction on the sign of $q$ on either side of the wall again
reduce the solutions to a total of three modes.
At the interface, the
matching conditions do not involve derivatives of the fluid variables\footnote{See Eqs.~(46)-(48) of Ref.~\cite{mm14stability}.},
and we can express $\delta v_\pm$ and $\delta p_\pm$ in terms of $\zeta$, $\dot{\zeta}$, and $\ddot{\zeta}$.
Inserting these expressions into Eq.~(\ref{oscarmestable}) results in an equation of the form 
\begin{equation}
\sigma_{0}\ddot{D}+\sigma_{0}\omega_{k}^{2}D+\eta_{k}\dot{D}=\text{source}.\label{oscarm}
\end{equation}
We do not actually need to follow the above steps to derive the coefficients in Eq.~(\ref{oscarm}).
It is sufficient to note that, in the absence of the source term, this equation must produce 
the same result as the linear stability analysis. 

The general solution to	 Eq.~(\ref{oscarm}) is
\begin{equation}
D=D_{+}e^{\Omega_{+}t}+D_{-}e^{\Omega_{-}t},\label{soloscarm}
\end{equation}
where 
\begin{equation}
	\Omega_{\pm}=-\frac{\eta_{k}}{2\sigma_{0}}\pm\sqrt{\frac{\eta_{k}^{2}}{4\sigma_{0}^{2}}-\omega_{k}^{2}} .
	\label{Omegaosc}
\end{equation}
Note that the quantity $\omega_{k}^{2}$ contains a term $k^{2}$
from Eq.~(\ref{oscarmestable}). However, the presence of negative fluid
terms can result in $\omega_{k}^{2}<0$ (and hence, instability) for
small enough $k$. 			
Since we already know the values of $\Omega_{+}$ and $\Omega_{-}$
from the linear stability analysis, Eqs.~(\ref{Ommas})-(\ref{Ommen}), we can obtain the coefficients
$\omega_{k}$ and $\eta_{k}$ by inverting Eq.~(\ref{Omegaosc}), 
\begin{equation}
\eta_{k}=-\sigma_{0}(\Omega_{+}+\Omega_{-}),
\quad
\omega_{k}^{2}=\frac{\eta_{k}^{2}}{4\sigma_{0}^{2}}-\frac{1}{4}\left(\Omega_{+}-\Omega_{-}\right)^{2}.
\end{equation}

The source term in Eq.~(\ref{oscarm}) may represent
thermal fluctuations, bubble collisions, shock fronts coming from other bubbles,
or other sources of perturbations.
We model this term using a random force.
While  Eq.~(\ref{oscarm}) describes corrugation at a specific wave vector $\mathbf{k}$, 
a general wall deformation is a superposition,
\begin{equation}
	\zeta(\mathbf{x}^{\perp},t)=\int\frac{d^{2}k}{(2\pi)^{2}}D_{\mathbf{k}}(t)e^{i\mathbf{k}\cdot\mathbf{x}^{\perp}},
	\label{desazeta}
\end{equation}
with $D_{-\mathbf{k}}=D_{\mathbf{k}}^{*}$.
The Fourier modes obey Eq.~(\ref{oscarm}),
\begin{equation}
	\sigma_{0}\ddot{D}_{\mathbf{k}}+\sigma_{0}\omega_{k}^{2}D_{\mathbf{k}}+\eta_{k}\dot{D}_{\mathbf{k}}=y_{\mathbf{k}}.
	\label{oscarmforzado}
\end{equation}
Since perturbations may occur on all length and time scales, 
we consider a homogeneous white noise force with the properties
\begin{equation}
\langle y_{\mathbf{k}}\rangle=0, ~~ \langle y_{\mathbf{k}}(t)y_{\mathbf{k}'}^{*}(t')\rangle=A\delta(t-t')(2\pi)^{2}\delta(\mathbf{k}-\mathbf{k}'),\label{propsfza}
\end{equation}
where $\langle\,\rangle$ denotes the ensemble average. 
We consider
the case in which the random force corresponds to thermal fluctuations,
for which the amplitude is given by $A=2\eta_{k}T$ (see App.~\ref{app:corrfunc}).
As we will see, the specifics of the initial fluctuations are irrelevant.
This is because the amplitude of relevant deformations is determined by nonlinearity.

For the initial conditions
$D_{\mathbf{k}}(t_{0})=\dot{D}_{\mathbf{k}}(t_{0})=0$,
the solution of Eq.~(\ref{oscarmforzado}) is 
\begin{equation}
D_{\mathbf{k}}(t)=\theta(t-t_{0})\int_{t_{0}}^{t}\frac{e^{\Omega_{+}(t-\tilde{t})}-e^{\Omega_{-}(t-\tilde{t})}}{\sigma_{0}(\Omega_{+}-\Omega_{-})}y_{\mathbf{k}}(\tilde{t})d\tilde{t}.\label{solDk}
\end{equation}
We will determine the initial time $t_{0}$ later. 
Using the properties (\ref{propsfza}) in Eq.~(\ref{solDk}), we obtain $\left\langle D_{\mathbf{k}}(t)\right\rangle =0$
and
\begin{equation}
\left\langle D_{\mathbf{k}}(t)D_{\mathbf{k}'}^{*}(t')\right\rangle =(2\pi)^{2}\delta^{(2)}(\mathbf{k}-\mathbf{k}')\,|D|^{2}(k,t,t'),\label{defD2}
\end{equation}
where
\begin{multline}
|D|^{2}(k,t,t')=D_{0}^{2}\left(1-ce^{-2\Omega_{+}(t_{m}-t_{0})}\right)e^{\Omega_{+}(t+t'-2t_{0})}\\
\times\Theta(t_{m}-t_{0}).
\label{D2}
\end{multline}
The time $t_{m}$ is the minimum of $t$ and $t'$.
Below, we consider the case $t_{m}=t$. The term $e^{\Omega_{-}(t-\tilde{t})}$
in Eq.~(\ref{solDk}) can be neglected, as it decays rapidly (see App.~\ref{app:corrfunc}).
The coefficients in Eq.~(\ref{D2}) are given by 
\begin{equation}
D_{0}^{2}=\frac{|\Omega_{+}+\Omega_{-}|T/\sigma_{0}}{\left(\Omega_{+}-\Omega_{-}\right)^{2}\Omega_{+}},\quad c=\frac{\Omega_{+}-\Omega_{-}}{|\Omega_{+}+\Omega_{-}|}.\label{D0b}
\end{equation}
The coefficient $D_{0}$ diverges for $v_{w}=v_{c}$ and for $k=k_{c}$
due to the factor $\Omega_{+}^{-1}$. However,
for $\Omega_{+}=0$ we have $c=1$, and the factor in parenthesis
in Eq.~(\ref{D2}) vanishes. Taking into account that $\Omega_{+}\ll|\Omega_{-}|$
and using Eqs.~(\ref{Ommas})-(\ref{kc}), we obtain
\begin{equation}
D_{0}^{2}\simeq\frac{5\left(\pi^{2}g_{*}\alpha_{T}^{2}v_{c}^{2}\right)^{-1}}{\left(1-v_{w}^{2}/v_{c}^{2}\right)\left(1-k/k_{c}\right)kT^{3}},\quad c\simeq1.\label{D0bap}
\end{equation}

The correlation function $|D|^{2}$ vanishes continuously as one of
the times $t,t'$ approaches the initial time $t_{0}$. This is because,
near $t_{0}$, the perturbation is still significantly affected by
the fluctuations of the random force, and is therefore not correlated
to later values of its amplitude. When the times $t-t_{0}$ and $t'-t_{0}$
are of order $\Omega_{+}^{-1}$, we have $|D|^{2}\sim D_{0}^{2}$.
This amplitude is of the order of thermal fluctuations. 
However, during
the phase transition, the exponential $e^{\Omega_{+}(t+t'-2t_{0})}$ in Eq.~(\ref{D2}) can become
extremely large since we have 
$\Omega_{+}\beta^{-1}  \sim \alpha_{T}v_{w} k/\beta \sim \alpha_{T}v_{w}10^{-3}(k/T)(M_{P}/T)$.
Even modes with $k/T\ll1$ can grow to significant sizes, provided
$T$ is sufficiently below $M_{P}$. 
Once a mode $k$ becomes relevant, the factor $e^{-2\Omega_{+}(t_{m}-t_{0})}$
in Eq.~(\ref{D2}) is very small, and we have
\begin{equation}
|D|^{2}(k,t,t')\simeq|D|(k,t,t)|D|(k,t',t'),\label{D2DD}
\end{equation}
with 
\begin{equation}
|D|(k,t,t)=D_{0}e^{\Omega_{+}(t-t_{0})}\Theta(t-t_{0}).
\label{Daprox}
\end{equation}
At this stage, the wall perturbations at $t$ and $t'$ are maximally
correlated. This is because, once the amplitude of a perturbation
has grown enough, it is hardly affected by random fluctuations.

\subsection{Spherical bubbles and the Non-linear regime}

\label{subsec:NLregime}

The planar-wall results should be a good approximation as long as
the corrugation wavelength is much smaller than the bubble
radius $R$. 
Conversely, a wavelength $\lambda$ larger than $2\pi R$ cannot even exist on a spherical surface.
Therefore, mode $k$ cannot begin to grow until $R(t)$ is greater than $k^{-1}$.
Furthermore, the initial growth is limited by the bubble expansion. 
The growth rate must exceed the bubble's expansion rate%
\footnote{For a spherical bubble, there is an effective friction term $(3\dot{R}/R)\dot{D}$ in Eq.~(\ref{oscarm}), 
	which has a stabilizing effect for a growing bubble \cite{sharp84overview}. 
	This term can be neglected if $\dot{R}/R\ll\dot{D}/D$, i.e., $v_w/R\ll \Omega_+$.
	A similar condition was used in Ref.~\cite{soj97primordial}.}, 
$\Omega_{+}(k)\geq v_{w}/R$. 
For a bubble nucleated at time $t_{N}$,
this condition gives an activation time of
\begin{equation}
t_{0}-t_{N}=\Omega_{+}(k)^{-1}.\label{t0}
\end{equation}
Although this is a rough estimate, the relevant corrugation amplitudes are
determined by nonlinearity and are rather insensitive to the
initial dynamics.

The linearized equations neglect terms that are quadratic in the derivatives
of the variables, so they are valid for $\partial_{i}\zeta\ll1$,
$\partial_{i}\delta v\ll\delta v$, and so on. All these constraints
are actually equivalent since the variations of all the variables
are related.
For a perturbation mode $\mathbf{k}$, this constraint implies $kD_{\mathbf{k}}\ll1$.
However, in the statistical treatment, we have a continuous
superposition of modes. From Eqs.~(\ref{desazeta}) and (\ref{defD2}),
we have 
\begin{equation}
	\langle(\nabla\zeta)^{2}\rangle=\int\frac{dk}{2\pi}k^{3}\,|D|^{2}(k,t,t).
\end{equation}
We can estimate the contribution of the modes at the scale $k$ as
$
\langle(\nabla\zeta)^{2}\rangle_{k} \sim k\times k^{3}\,|D|^{2}(k,t,t),
$
and assume that the linear approximation is valid for 
$
\langle(\nabla\zeta)^{2}\rangle_{k}\ll 1,
$
which implies%
\footnote{Alternatively, from $\langle\zeta^{2}\rangle=\int_{0}^{\infty}\frac{dk}{2\pi}k\,|D|^{2}(k,t,t)$, 
	we may define the displacement spectrum $\zeta_{k}$ through $\langle\zeta^{2}\rangle\equiv\int\zeta_{k}^{2}\,d\ln k$
	and impose the condition $\zeta_{k}\ll k^{-1}$ (see Ref.~\cite{soj97primordial}).
	We have $\zeta_{k}\propto k|D|(k,t,t)$, so this approach gives 
	the same condition we obtain by considering $\langle(\nabla\zeta)^{2}\rangle$.} 
$|D|(k,t,t)\ll k^{-2}$.

As previously mentioned, we assume that a mode stops growing once it becomes nonlinear.
We manually impose this condition on the quantity 
$|D|(k,t,t)$ at the value $ak^{-2}$, where $a$ is a dimensionless parameter.
Thus, Eq.~(\ref{Daprox}) becomes
\begin{equation}
	|D|(k,t,t)= \Theta(t-t_{0})
	\begin{cases}
		D_{0}e^{\Omega_{+}(t-t_{0})} & t < t_{nl},\\
		ak^{-2} & t_{nl}<t ,
	\end{cases} 
	\label{Daproxsat}
\end{equation}
where $t_{nl}$ is given by the condition $D_{0}e^{\Omega_{+}(t-t_{0})}=ak^{-2}$.
Using Eq.~(\ref{t0}), we obtain
\begin{equation}
	D_{0}(k)e^{\Omega_{+}(k)(t_{nl}-t_{N})-1}=ak^{-2}.\label{defts}
\end{equation}
The constant $a$ is a free parameter in our modeling of the nonlinear dynamics.
Although it could be as large as $a\sim 1$, our perturbative calculation will constrain its value.
This model assumes that the nonlinear coupling between modes and the energy cascade
from shorter to longer wavelengths keeps the average $\langle(\nabla\zeta)^{2}\rangle_{k}$ constant.
However, this assumption is conservative because other factors, such as wall acceleration or cusp formation
may affect the gravitational wave spectrum more strongly.

\begin{figure}[htb]
	\includegraphics[width=\columnwidth]{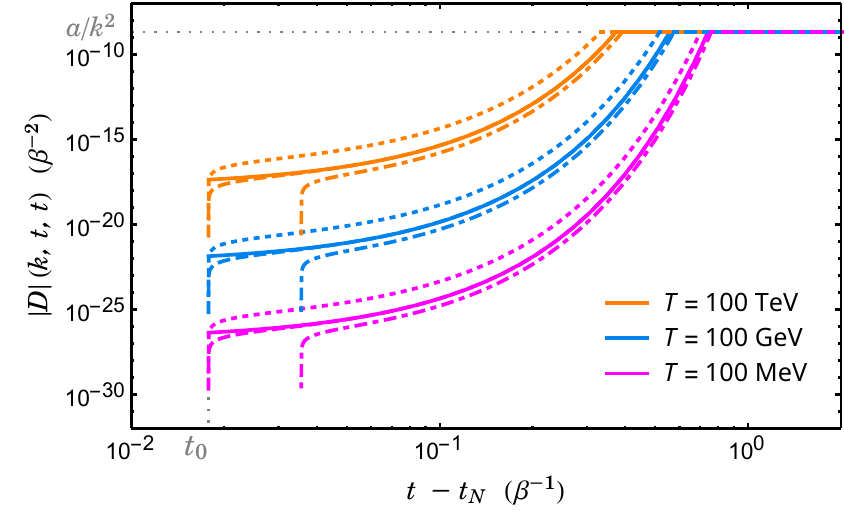}
	\caption{The quantity $|D|(k,t,t)$ for a fixed wavenumber $k=10^{4}\beta$.
		Dashed lines correspond to Eq.~(\ref{D2}), solid lines correspond to Eq.~(\ref{D2DD}), 
		dotted lines correspond to the replacement $A\to 100A$ in Eq.~(\ref{propsfza}),
		and
		dash-dotted lines correspond to the replacement $t_0\to 2t_0$.
		The parameters are $\beta/H=100$, $\alpha_{T}=0.05$, $v_{c}=0.2$, $v_{w}=0.1$, and $a=0.1$.\label{figdkt}}
\end{figure}

We obtained Eq.~(\ref{Daproxsat}) using the approximation (\ref{D2DD})-(\ref{Daprox}), which ignores the decorrelation factor $1-ce^{-2\Omega_{+}(t-t_{0})}$ from Eq.~(\ref{D2}).
The evolution of $|D|(k,t,t)$ is shown in Fig.~\ref{figdkt}.
The solid lines correspond to Eq.~(\ref{Daproxsat}), while the dashed lines correspond to the inclusion of the decorrelation factor.
The approximation does not vanish at the initial time $t_{0}$. 
However, it becomes very accurate once $|D|$ reaches appreciable values.
The dotted lines are obtained by multiplying the amplitude of the initial fluctuations by $100$.
Since the relevant amplitude of $|D|$ is given by its saturation value $a/k^2$, 
the main effect of this change is a slight shift in the saturation time.
The dash-dotted lines are obtained by multiplying the activation time $t_0$ by 2.
This change does not affect significantly the curves in the relevant range.
We have verified that the GW spectrum does not change appreciably under these variations of the initial conditions.
Using Eq.~(\ref{Daproxsat}) in Eq.~(\ref{D2DD}), we obtain, for $t_0<t<t'$,
\begin{equation}
	|D|^{2}(k,t,t')=
	\frac{a^{2}}{k^{4}}\begin{cases}
		e^{\Omega_{+}(t+t'-2t_{nl})} & t<t'<t_{nl},\\
		e^{\Omega_{+}(t-t_{nl})} & t<t_{nl}<t',\\
		1 & t_{nl}<t<t'.
	\end{cases}\label{D2sat}
\end{equation}

\begin{figure}[hbt]
	\includegraphics[width=\columnwidth]{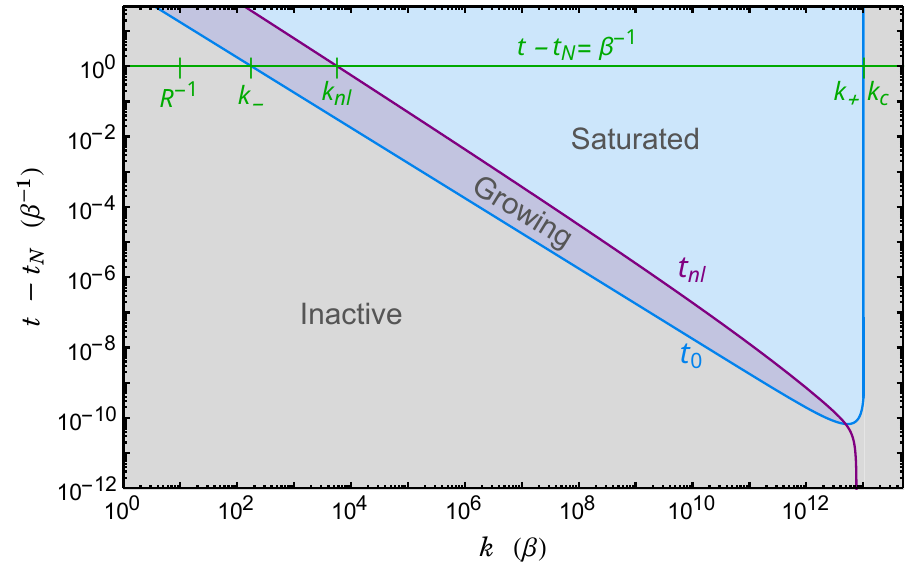}
	\caption{Regions of inactive, growing, and saturated modes, 
		delimited by the curves of $t_{0}(k)$ and $t_{nl}(k)$.
		The values $k=R^{-1}$, $k_-$, $k_{nl}$, and $k_+$ are indicated for $t-t_N=\beta^{-1}$.
		The values of the parameters are the same as in Fig.~\ref{figdkt} and the temperature is $T=100\mathrm{GeV}$.
		The curves are qualitatively similar for other temperatures.}
	\label{figestima}
\end{figure}

In Fig.~\ref{figestima} we consider the times $t_{0}$ and $t_{nl}$ as a function of $k$. 
The time $t_{0}$ diverges at $k=k_{c}$ and has a minimum at $k\sim k_c$.
At a given time $t$, the active modes are in a range $k_{-}(t)<k<k_{+}(t)$
corresponding to the condition $t>t_{0}(k)$. 
From Eqs.~(\ref{t0}) and (\ref{Ommas}), we have
\begin{equation}
\frac{k_{\pm}}{k_{c}}=\frac{1}{2}\pm \frac{1}{2}\sqrt{1-\frac{8}{3\alpha_{T}(1-v_{w}^{2}/v_{c}^{2})k_{c}R}},\label{kinfksup}
\end{equation}
with $R=v_{w}(t-t_{N})$. 
The largest active wavenumber $k_{+}$ is always
of order $k_{c}$, while the smallest active wavenumber $k_{-}$ can
be several orders of magnitude below $k_{c}$. 
At early times, such that $R\lesssim k_c^{-1}$, no mode is active yet. 
At later times,  $t-t_N\sim \beta^{-1}\gg k_c^{-1}$,
we have $k_{+}\simeq k_{c}$ and $k_{-}\simeq\frac23\alpha_T^{-1}(1-v_{w}^{2}/v_{c}^{2})^{-1}R^{-1}\ll k_{c}$.
For $t>t_{nl}(k)$, the growth of mode $k$ saturates.
The value $k_{nl}(t)$ separating growing and saturated modes is given
by the condition 
\begin{equation}
	D_{0}(k_{nl})e^{\Omega_{+}(k_{nl})(t-t_{N})-1}=a k_{nl}^{-2}.\label{defks}
\end{equation}

\begin{figure}[h]
\includegraphics[width=\columnwidth]{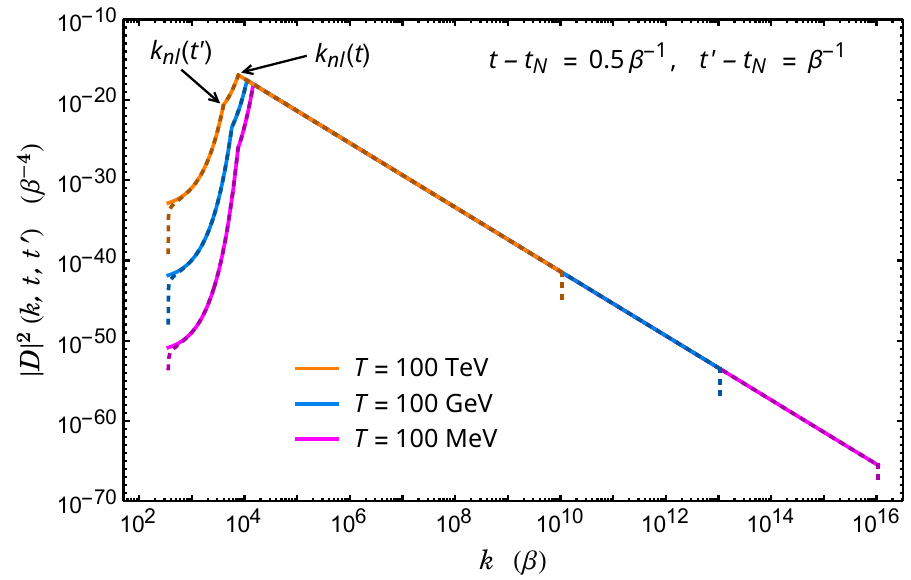}
\caption{The quantity $|D|^2(k,t,t')$ as a function of $k$. 
	Dotted lines correspond to including the factor $1-ce^{-2\Omega_{+}(t-t_{0})}$. 
	The values of the parameters are the same as in previous figures.\label{fig:Dkttp}}
\end{figure}

For $t<t'$, we have $k_-(t')<k_-(t)$ (longer wavelengths become active at later times)
and $k_{nl}(t')<k_{nl}(t)$. 
Expressing the different domains in 
Eq.~(\ref{D2sat}) 
in terms of $k$ instead of $t$, we have
\begin{equation}
|D|^{2}=\frac{a^{2}}{k^{4}}\begin{cases}
e^{\Omega_{+}(t+t'-2t_{nl})} & k_{-}(t)<k<k_{nl}(t'),\\
e^{\Omega_{+}(t-t_{nl})} & k_{nl}(t')<k<k_{nl}(t),\\
1 & k_{nl}(t)<k< k_c.
\end{cases}\label{D2final}
\end{equation}
In Fig.~\ref{fig:Dkttp} we consider this quantity as a function of $k$ for fixed $t$ and $t'$. 
For small $k$, the curves grow exponentially since $\Omega_{+}$ is essentially proportional to $k$. However, for high enough $k$,
the mode has reached saturation, and its amplitude is given by $a/k^{2}$.
The peak (for $t<t'$) is at $k=k_{nl}(t)$.
We see that the approximation of ignoring the factor $1-ce^{-2\Omega_{+}(t-t_{0})}$
only departs from the actual curve for $k$ very close to $k_{\pm}$,
when the perturbations are relatively very small. 
Therefore, we will not consider this factor from now on. 

\begin{figure}[h]
\includegraphics[width=0.8\columnwidth]{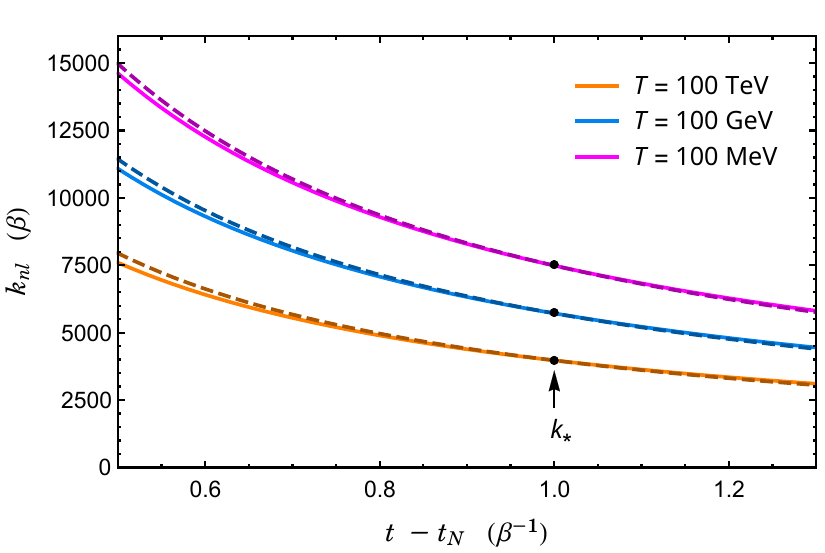}
\caption{The quantity $k_{nl}$ (solid lines) and its approximation (\ref{ksap}) (dashed lines). 
	The values of the parameters are the same as in previous figures.\label{fig:ks}}
\end{figure}

As can be observed in Fig.~\ref{fig:Dkttp}, the relevant wavenumbers fulfill
$k\sim k_{nl}(t)\ll k_{c}$. Therefore, it is a very good approximation
to ignore the factor $(1-k/k_{c})$ in Eqs.~(\ref{Ommas}) and (\ref{D0bap}),
and we have
\begin{equation}
	\Omega_{+}=\hat{\Omega}k,\quad D_{0}^{2}=\hat{D}_{0}^{2}k^{-1},\label{eq:apsOmD0}
\end{equation}
with 
\begin{equation}
	\hat{\Omega}=\frac{3}{2}\alpha_{T}v_{w}(1-v_{w}^{2}/v_{c}^{2}),\quad\hat{D}_{0}^{2}\simeq\frac{5T^{-3}}{\pi^{2}g_{*}\alpha_{T}^{2}(v_{c}^{2}-v_{w}^{2})}.\label{eq:Omhat}
\end{equation}
Using Eqs.~(\ref{eq:apsOmD0}), the solutions of Eqs.~(\ref{defts}) and (\ref{defks}) are
\begin{align}
	t_{nl}(k) &= t_{N} + \frac{\log(ak^{-3/2}/\hat{D}_{0})+1}{\hat{\Omega}k} \label{eq:tssol} \\
	k_{nl}(t) &= \frac{3/2}{\hat{\Omega}(t-t_{N})}\,W\left(\frac{2}{3}(ea)^{2/3}\frac{\hat{\Omega}(t-t_{N})}{\hat{D}_{0}^{2/3}}\right),\label{eq:kssol}
\end{align}
where $W(x)$ is the Lambert function, which is the solution of $We^{W}=x$.
Since deformations of larger size and amplitude are possible at later times,
the relevant scale is given by the parameter 
\begin{equation}
k_{*}\equiv k_{nl}|_{t-t_{N}=\beta^{-1}}.\label{eq:khatdef}
\end{equation}
In Fig.~\ref{fig:ks}, we plot $k_{nl}(t)$ and
indicate the value of $k_{*}$. 

We can obtain a simple approximation for $t_{nl}(k)$ by evaluating the $\log$ (which is slowly varying) at $k=k_*$.
Similarly, the variation of $W(x)$ with $x$ is roughly logarithmic
(see App.~\ref{app:Approximations}),
and we can obtain a simple approximation for $k_{nl}(t)$
by evaluating the argument of $W$ at $t-t_{N}=\beta^{-1}$.
Thus, we obtain
\begin{equation}
k_{nl}(t)=\frac{k_{*}}{\beta(t-t_{N})},\quad t_{nl}(k)-t_{N}=\frac{k_{*}}{\beta k}.\label{ksap}
\end{equation}
The approximation for $k_{nl}$ is shown in Fig.~\ref{fig:ks}. At the electroweak
scale, the error is less than 3\% for $t-t_{N}>0.5\beta^{-1}$ and
less than 1\% for $t-t_{N}>0.8\beta^{-1}$. We will use this approximation
below, since only the later stages of the phase transition are relevant
to GW generation. This is because the amount of wall surface is proportional
to $(t-t_{N})^{2}$.

\subsection{Spacetime correlations}

\label{subsec:correlations}

\begin{figure}[h]
\includegraphics[width=0.8\columnwidth]{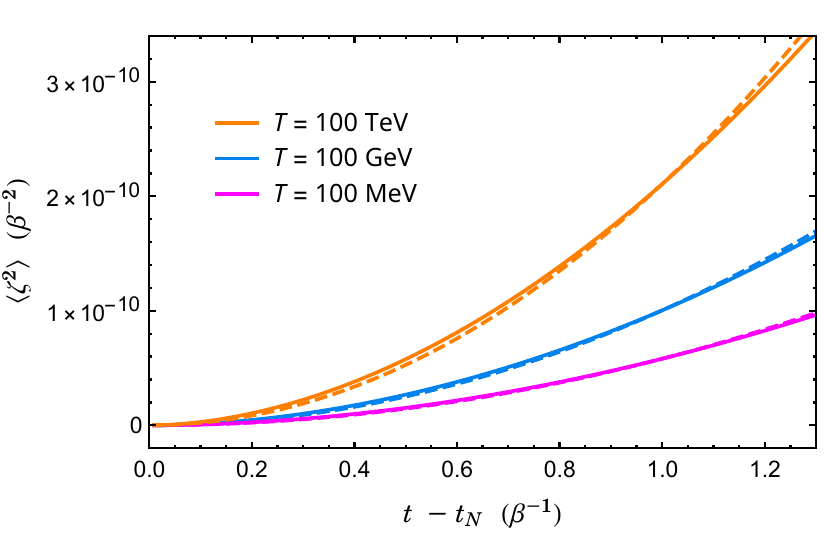}
\caption{The average $\langle\zeta^{2}\rangle$ according to Eq.~(\ref{eq:zeta2medio}) (solid lines)
	and the approximation (\ref{eq:zeta2medioap}) (dashed lines).
	The values of the parameters are the same as in previous figures.\label{fig:z2m}}
\end{figure}

The spacetime correlation function for the deformations can be readily
obtained from Eqs.~(\ref{desazeta}) and (\ref{defD2}),
\begin{equation}
\langle\zeta\zeta'\rangle (x,t,t')=\int\frac{d^{2}k}{(2\pi)^{2}}\,|D|^{2}(k,t,t')e^{i\mathbf{k}\cdot\mathbf{x}},\label{eq:zetazetap}
\end{equation}
where we have used the notation $\zeta\equiv\zeta(\mathbf{x}^{\perp},t)$,
$\zeta'\equiv\zeta(\mathbf{x}^{\prime\perp},t')$, $\mathbf{x}\equiv\mathbf{x}^{\perp}-\mathbf{x}^{\prime\perp}$,
and $x=|\mathbf{x}|$. 
In App.~\ref{app:corrfunc} we analyze the space-time dependence of $\langle\zeta\zeta'\rangle$.
It becomes negligible beyond a small separation $x\ll\beta^{-1}$.
We will not need to use explicitly the function
$\langle\zeta\zeta'\rangle(x,t,t')$ but its Fourier transform $|D|^{2}(k,t,t')$.
We are only interested in the case $x=0$, $t=t'$,
\begin{align}
\langle\zeta^{2}\rangle 
&=\frac{1}{2\pi}\int_{k_{-}}^{k_{+}}kdk|D|^{2}(k,t,t) \\
&= \frac{1}{2\pi}
\left[\frac{a^{2}/k_{nl}^{3}-\hat{D}_{0}^{2}}{2\hat{\Omega}(t-t_{N})} + \frac{a^{2}}{2}\left(\frac{1}{k_{nl}^{2}}-\frac{1}{k_{+}^{2}}\right)\right].\label{eq:zeta2medio}
\end{align}
We show this result in Fig.~\ref{fig:z2m}. The growth of $\langle\zeta^{2}\rangle$
is due to the fact that more and more infrared modes become available
as the bubble grows\footnote{A similar situation occurs for stable fluctuations (which we do not
take into account here). For thermal fluctuations of a plane surface
we have $|D(k,t,t)|^{2}\propto T/k^{2}$ (see App.~\ref{app:corrfunc}).
Therefore, the real-space fluctuations for an infinite plane are logarithmically
divergent \cite{Safran94Statistical}. Using an infrared cut-off
$R$ and an ultraviolet cut-off $\delta$ (e.g., the wall width),
we have $\langle\zeta^{2}\rangle\propto T\log(R/\delta)$. If $R$
is the bubble radius, stable fluctuations grow logarithmically.}. Since we have $\hat{D}_{0}^{2}\ll a^{2}/k_{nl}^{3}$, we can neglect
the quantity $\hat{D}_{0}^{2}$ in this result. This is equivalent
to using the approximation of taking the limit of integration $k_{-}$
to $-\infty$. Similarly, we have
$k_{+}\gg k_{nl}$, and we can neglect the term $k_{+}^{-2}$, which
is equivalent to taking the limit of integration $k_{+}$ to $\infty$.
Using also the approximation (\ref{ksap}), we obtain 
\begin{equation}
\langle\zeta^{2}\rangle=\alpha^{2}(t-t_{N})^{2},\quad\alpha^{2}=\frac{a^{2}}{4\pi}\left(1+\hat{\Omega}^{-1}\frac{\beta}{k_{*}}\right)\frac{\beta^{2}}{k_{*}^{2}}.\label{eq:zeta2medioap}
\end{equation}
This approximation is also shown in Fig.~\ref{fig:z2m}. Its error
is mainly due to the approximation (\ref{ksap}) for $k_{nl}(t)$
and is at most a few percent for $t-t_{N}\sim\beta^{-1}$. 

\begin{figure}[h]
\includegraphics[width=0.8\columnwidth]{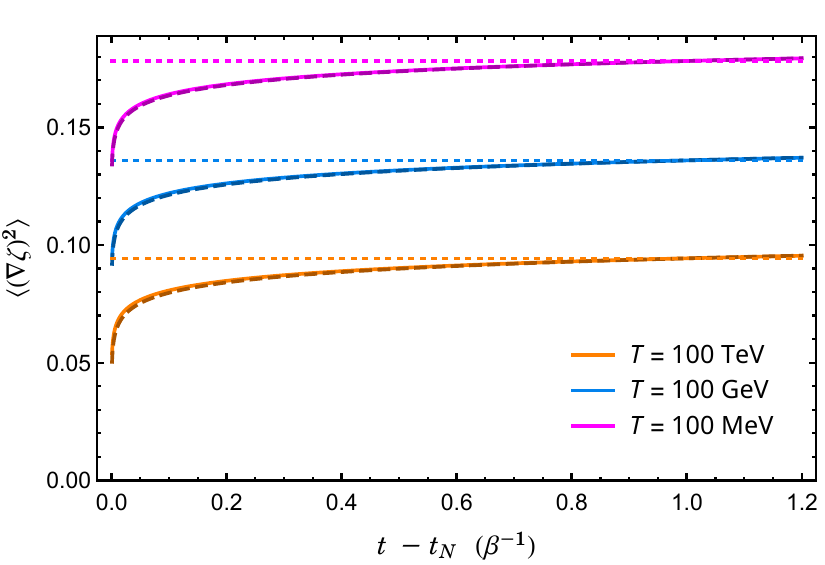}
\caption{The average $\langle(\nabla\zeta)^{2}\rangle$
	according to Eq.~(\ref{gradz2medio}) (solid lines)
	and the approximations (\ref{grad2zmedioap}) (dashed lines) and (\ref{gradz2medioap}) (dotted lines). 
	The values of the parameters are the same as in previous figures.\label{fig:gradz2m}}
\end{figure}

We will also be interested in the quantity
\begin{equation}
\langle(\nabla\zeta)^{2}\rangle=\frac{1}{2\pi}\int_{k_{-}}^{k_{+}}dkk^{3}|D|^{2}(k,t,t),\label{gradz2medio}
\end{equation}
which we already discussed qualitatively above.
Taking into account that we have $\hat{D}_{0}^{2}\ll a^{2}k_{nl}^{-3}$
and $k_{+}\simeq k_{c}$, and using the approximation (\ref{ksap})
for $k_{nl}(t)$, we obtain
\begin{multline}
\langle(\nabla\zeta)^{2}\rangle=\frac{a^{2}}{2\pi}\left[\frac{1-(\hat{\Omega}k_{*}/\beta)^{-1}+\frac{1}{2}(\hat{\Omega}k_{*}/\beta)^{-2}}{2\hat{\Omega}k_{*}/\beta}\right.\\
\left.+\log\left(\frac{k_{c}}{k_{*}}\beta(t-t_{N})\right)\right].\label{grad2zmedioap}
\end{multline}
This quantity diverges logarithmically for $t\to t_{N}$. However,
the amount of wall surface vanishes in this limit. In Fig.~\ref{fig:gradz2m},
we see that Eq.~(\ref{grad2zmedioap}) is a very good approximation
during most of the phase transition. 
For $t-t_{N}\sim\beta^{-1}$,
we can ignore the logarithmic variation as we did before, and
set $t-t_{N}=\beta^{-1}$ inside the $\log$. Taking into account
that the first term is a small constant, we have
\begin{equation}
\langle(\nabla\zeta)^{2}\rangle\simeq\frac{a^{2}}{2\pi}\log(k_{c}/k_{*}).\label{gradz2medioap}
\end{equation}
 This approximation is also shown in Fig.~\ref{fig:gradz2m}.

\section{Gravitational radiation}

\label{sec:gravrad}

For a stochastic source in a large volume $V$, 
the energy density of gravitational waves per logarithmic frequency
interval can be written in the form \cite{mm21gravitational}
\begin{equation}
\frac{d\rho_{GW}}{d\ln\omega}=
\frac{4G\omega^{3}}{\pi}\negthickspace \int_{-\infty}^{\infty}\negthickspace\negthickspace  dt \negmedspace 
\int_{t}^{\infty}\negthickspace \negthickspace dt'
\cos[\omega (t'-t)]\,\Pi(t,t',\omega),\label{rhogw}
\end{equation}
where 
\begin{equation}
\Pi(t,t',\omega)=\frac{1}{V}\Lambda_{ijkl}(\hat{n})\left\langle \tilde{T}_{ij}(t,\omega\hat{n})\tilde{T}_{kl}(t',\omega\hat{n})^{*}\right\rangle .\label{defPi}
\end{equation}
The symbol $\langle\,\rangle$ denotes ensemble average, $\hat{n}$
is the direction of observation, $\Lambda_{ijkl}(\hat{n})$ is the
transverse-traceless projection tensor,
\begin{multline}
\Lambda_{ijkl}=\delta_{ik}\delta_{jl}-\frac{1}{2}\delta_{ij}\delta_{kl}-\hat{n}_{j}\hat{n}_{l}\delta_{ik}-\hat{n}_{i}\hat{n}_{k}\delta_{jl}\\
+\frac{1}{2}\hat{n}_{k}\hat{n}_{l}\delta_{ij}+\frac{1}{2}\hat{n}_{i}\hat{n}_{j}\delta_{kl}+\frac{1}{2}\hat{n}_{i}\hat{n}_{j}\hat{n}_{k}\hat{n}_{l},
\end{multline}
and $\tilde{T}_{ij}$ is the Fourier transform of the energy-momentum
tensor ($i,j$ are space components), 
\begin{equation}
\tilde{T}_{ij}(t,\omega\hat{n})=\int d^{3}xe^{-i\omega\hat{n}\cdot\mathbf{x}}T_{ij}(t,\mathbf{x}).\label{defTijtil}
\end{equation}
If the energy of the bubbles is confined within the
bubble walls or within thin fluid layers adjacent to them,
we can neglect overlapping, and the total
$T_{ij}$ is given by the sum of the contributions from each bubble.

\subsection{Energy-momentum tensor of a bubble}

The energy-momentum tensor of the field is given by
\begin{equation}
T_{\mu\nu}=\partial_{\mu}\phi\partial_{\nu}\phi-\frac{1}{2}g_{\mu\nu}\partial_{\alpha}\phi\partial^{\alpha}\phi+g_{\mu\nu}V(\phi).
\end{equation}
In the thin-wall approximation, we have a field profile $\phi=\phi_{0}(n)$,
as described in Sec.~\ref{sec:instadefla}. Using Eq.~(\ref{fieldeqthinwall}),
we obtain the wall contribution \cite{mm23thin,mm25thin},
\begin{equation}
T_{\mu\nu}=\phi_{0}^{\prime}(n)^{2}(\partial_{\mu}n\partial_{\nu}n+g_{\mu\nu}).\label{TwallGNC}
\end{equation}
Near the hypersurface of the wall, the normal coordinate is given by $n=-N_{\mu}(\xi^{a})(x^{\mu}-X^{\mu}(\xi^{a}))$,
where $X^{\mu}(\xi^{a})$ parametrizes the hypersurface.
The surface of a bubble can be described by $r=R+\zeta(\hat{r})$,
where $R$ is the radius of the bubble without deformations and $\zeta$
is the displacement from the undeformed surface at the position $\mathbf{x}=R\hat{r}$
relative to the bubble's center. 
In spherical coordinates, we have
$N_{\mu}\propto(\dot{R}+\partial_{t}\zeta,\nabla\zeta,-1)$ (the vector
$\nabla\zeta$ is on the tangent plane to the spherical surface).
In the limit of an infinitely thin wall, we have $\phi_{0}^{\prime}(n)^{2}\simeq\sigma_{0}\delta(n)$.
To consider only uncollided bubble walls,
we insert the indicator function $1_{S}(\hat{r})$ for the surface $S$ at time $t$ into Eq.~(\ref{TwallGNC}).
We have $1_{S} = 0$ if the point at $R\hat{r}$ on the surface has collided
and $1_{S}=1$ if the point is uncollided. 
We obtain\footnote{Cf.~Eq.~(6.12) of Ref.~\cite{mm23thin}.}
\begin{align}
T_{00} & =\sigma\,\delta(r-R-\zeta)1_{S}(\hat{r}),\\
T_{ij} & =\frac{(\hat{r}_{i}-\partial_{i}\zeta)(\hat{r}_{j}-\partial_{j}\zeta)-\tilde{\gamma}_{w}^{-2}\delta_{ij}}{1+(\nabla\zeta)^{2}}\,T_{00}
\end{align}
where\footnote{For $\zeta=0$, we obtain the usual result $\sigma=\gamma_{w}\sigma_{0}$.
On the other hand, for constant $R$ and small $\zeta$, we obtain
the energy density of a scalar field living on the surface, $\sigma=\sigma_{0}[1+\frac{1}{2}(\nabla\zeta)^{2}+\frac{1}{2}(\partial_{t}\zeta)^{2}]$. }
\begin{equation}
\sigma=\sigma_{0}[1+(\nabla\zeta)^{2}]\tilde{\gamma}_{w}
\end{equation}
and 
\begin{equation}
\tilde{\gamma}_{w}=[1-(\dot{R}+\partial_{t}\zeta)^{2}+(\nabla\zeta)^{2}]^{-1/2}.
\end{equation}

Only the transverse-traceless projection of $T_{ij}$ is relevant for GWs.
In particular, the term proportional to $\delta_{ij}$
vanishes when contracted with the tensor $\Lambda_{ijkl}$. Therefore,
we drop this term from now on. 
We have
\begin{equation}
T_{ij}=t_{ij}\,\sigma\,\delta(r-R-\zeta)\,1_{S}(\hat{r}),\label{eq:Tijbubble}
\end{equation}
where
\begin{equation}
	t_{ij}=\frac{(\hat{r}_{i}-\partial_{i}\zeta)(\hat{r}_{j}-\partial_{j}\zeta)}{1+(\nabla\zeta)^{2}}.\label{tij}
\end{equation}
For $\zeta=0$, we have $T_{ij}=\hat{r}_{i}\hat{r}_{j}\,\sigma\,\delta(r-R)\,1_{S}(\hat{r})\equiv T_{ij}^{\mathrm{coll}}$,
which is the source of GWs for the bubble collision mechanism in the
envelope approximation. 
The surface energy density $\sigma$ is directly related to the energy
released in the bubble volume. In the absence of plasma, we have $\sigma=\rho_{\mathrm{vac}}R/3$,
where $\rho_{\mathrm{vac}}$ is the false vacuum energy density. 

In the presence of fluid, some of the released
energy is used to reheat the plasma and drive bulk fluid motions.
Only a fraction $\kappa$ goes to the wall. Therefore, we have
\begin{equation}
\sigma=\kappa\rho_{\mathrm{vac}}R/3.\label{sigmaR}
\end{equation}
The bulk fluid motions also produce GWs. 
The fluid is sometimes assumed to occupy a thin layer next to the wall, so that the envelope approximation can be used. However, 
this is generally not a good approximation
for the stationary fluid profile. 
For the unstable perturbations, though, it is a good approximation,
since the fluid perturbation (\ref{eq:U}) occupies a
width $|q|^{-1}\sim k^{-1}$ \cite{mm14stability}, with
$k^{-1}\ll R$ for the relevant modes.
Therefore, we 
use Eqs.~(\ref{eq:Tijbubble})-(\ref{sigmaR}) as an approximation
for the energy-momentum tensor of the wall and the fluid.
Since we are considering perturbations from the stationary case, we assume that
$\kappa$ is given by the efficiency factor calculated in the literature.
The redistribution of this energy among the stationary
fluid profile and the unstable perturbations arises from the evolution
of $T_{ij}$ (see below).

\subsection{GWs from the phase transition}

Due to the surface delta in Eq.~(\ref{eq:Tijbubble}), the contribution of an individual bubble
to Eq.~(\ref{defTijtil}) takes the form
\begin{equation}
	\tilde{T}_{ij}=\sigma R^{2}\negthinspace\int\negthinspace d\hat{r}e^{-i\omega\hat{n}\cdot\hat{r}R}\,e^{-i\omega\hat{n}\cdot\hat{r}\zeta}\,t_{ij} 1_{S}(\hat{r}),\label{Tijtil}
\end{equation}
where $d\hat{r}=\sin\theta d\theta d\phi$.
We have neglected terms of order $\zeta/R$ since we have $|\zeta/R|\ll|k\zeta|\sim|\partial_{i}\zeta|$. 

The total energy-momentum tensor is given by the sum of contributions
from all the bubbles.
Therefore, the product $\tilde{T}_{ij}(t,\omega\hat{n})\tilde{T}_{kl}(t',\omega\hat{n})^{*}$
in Eq.~(\ref{defPi}) is given by a sum of products where the two
factors belong either to the same bubble or to different bubbles.
The average of these two kinds of terms is different, and the quantity
$\Pi$ separates into a single-bubble part $\Pi_{s}$ and a two-bubble
part $\Pi_{d}$ \cite{jt17gravitational,mm21gravitational}. 
Statistically, the sum over bubbles can be expressed as an integral with respect to the nucleation time $t_N$,
taking into account the number of nucleated bubbles, $dt_{N}\Gamma(t_{N})f_+(t_N)$ \cite{mm21gravitational}.
Since the wall deformations are much smaller than the bubble radius, we do not need to take them into account in
the calculation of $f_+$, which is given by Eq.~(\ref{defI}).
Thus, Eq.~(\ref{defPi}) becomes $\Pi=\Pi_{s}+\Pi_{d}$, with (see \cite{mm21gravitational} for details)
\begin{align}
	&\begin{multlined}[b]
		\Pi_{s}=  \int_{-\infty}^{t}\negthickspace\negthickspace dt_{N} \Gamma(t_{N})e^{-I(t_{N})} \sigma R^{2} \sigma' R^{\prime2}
		\times  {}
		\\
		\negmedspace \int \negmedspace d\hat{r} \negmedspace 
		\int\negmedspace d\hat{r}'e^{i\omega\hat{n}\cdot\mathbf{s}}
		\Lambda_{ijkl}\left\langle e^{i\omega\hat{n}\cdot(\zeta'\hat{r}'-\zeta\hat{r})}t_{ij}t_{kl}^{\prime}\,1_{S}1_{S'}\right\rangle ,
	\end{multlined}\label{eq:Pis}
 \\
  &\begin{multlined}[b] 
 	\Pi_{d}=  \int_{-\infty}^{t}\negthickspace\negthickspace dt_{N}\Gamma(t_{N})e^{-I(t_{N})}\negthickspace
 	\int_{-\infty}^{t'}\negthickspace \negthickspace dt_{N}^{\prime}\Gamma(t_{N}^{\prime})e^{-I(t_{N}^{\prime})} 
 	 \\
 	 \times \sigma R^{2} \sigma' R^{\prime2}
 	\negthickspace \int \negmedspace d^{3}l \negthinspace \int \negmedspace  d\hat{r} \negthickspace \int \negmedspace d\hat{r}'
 	e^{i\omega\hat{n}\cdot\mathbf{s}}
 	 \\
 	 \times\Lambda_{ijkl}\left\langle e^{i\omega\hat{n}\cdot(\zeta'\hat{r}'-\zeta\hat{r})}t_{ij}t_{kl}^{\prime}\,1_{S}1_{S'}\right\rangle ,
 \end{multlined}\label{eq:Pid}
\end{align}
where $\sigma$ is given by Eq.~(\ref{sigmaR}), $R=v_w(t-t_N)$, $\zeta=\zeta(\hat{r},t)$,
and similarly for $\sigma'$, $R'$, and $\zeta'$.
In both expressions, the vector $\mathbf{s}$ connects a point
on the surface $S$ of a bubble at time $t$ to a point on the surface
$S'$ of a bubble at time $t'$. We have $\mathbf{s}=\mathbf{l}+R'\hat{r}'-R\hat{r}$,
where $\mathbf{l}$ connects the two nucleation centers (see Ref.~\cite{mm21gravitational} for a figure).
In the single-bubble case, we have $t_N'=t_N$ and $\mathbf{l=0}$.

Since the statistics of bubble collisions and wall deformations are independent of each other, we have 
\begin{equation}
\left\langle 1_{S}(\hat{r})1_{S'}(\hat{r}')\,t_{ij}t_{kl}^{\prime}e^{i\omega\hat{n}\cdot(\zeta'\hat{r}'-\zeta\hat{r})}\right\rangle =P_{u}G_{ijkl},\label{separ}
\end{equation}
where
\begin{equation}
	P_{u}=\left\langle 1_{S}(\hat{r})1_{S'}(\hat{r}')\right\rangle 
\end{equation}
and
\begin{equation}
	G_{ijkl}=\left\langle e^{i\omega\hat{n}\cdot(\zeta'\hat{r}'-\zeta\hat{r})}t_{ij}t_{kl}^{\prime}\right\rangle .
	\label{Gijkldef}
\end{equation}

\subsection{Wall correlations}

The quantity $P_{u}$ is the joint probability that 
point $R\hat{r}$ on surface $S$ and point $R'\hat{r}'$ on surface $S'$ are uncollided at times $t$ and $t'$, respectively. 
The surfaces $S$ and $S'$ may belong to the same or different bubbles.

For the same bubble, $P_{u}$ is of the form \cite{mm20bubble} 
\begin{equation}
P_{u}=e^{-I(t)}e^{-I(t')}e^{I_{\cap}}e^{I(t_{N})}\equiv e^{-J}e^{I(t_{N})}.\label{eq:Pu}
\end{equation}
Here, the factors $e^{-I(t)}$ and $e^{-I(t')}$ give the probabilities that
each point is not collided, while the factor $e^{I_{\cap}}$ takes
into account that these probabilities are not independent. This factor
depends on the distance $s\equiv|\mathbf{s}|$. The factor $e^{I(t_{N})}$
cancels with $e^{-I(t_{N})}$ in Eq.~(\ref{eq:Pis}). The general
expression for the quantity $I_{\cap}$ can be found in Ref.~\cite{mm21gravitational}.
For the exponential nucleation rate (\ref{eq:nuclexp}), we have
\begin{multline}
J=8\pi v_{w}^{3}\beta^{-4}\Gamma_{*}e^{\beta(t_{+}/2-t_{*})}\\
\times\left[2\cosh(\beta t_{-}/2)-e^{-\beta t_{s}/2}\left(\frac{t_{s}^{2}-t_{-}^{2}}{4t_{s}\beta^{-1}}+1\right)\right],\label{eq:Itotexp}
\end{multline}
where we have defined the variables $t_{\pm}=t'\pm t$ and the time $t_{s}=s/v_{w}$.

For two different bubbles, we have \cite{mm20bubble}
\begin{multline}
	P_u = e^{-J}e^{I(t_{N})}  e^{I(t_{N}^{\prime})}	\,
\Theta\big(|\mathbf{l}-R\hat{r}|- v_w (t-t_{N}^{\prime})\big) \\
\times \Theta\big(|\mathbf{l}+R'\hat{r}'|- v_w(t'-t_{N})\big).
\label{eq:Heavi2b}
\end{multline}
The Heaviside functions indicate that the probability vanishes where
the two bubbles overlap. 
We can also write this condition as $\Theta(\hat{r}\cdot\hat{s}+c_{M})\Theta(c_{M}'-\hat{r}'\cdot\hat{s})$,
where $\hat{s}=\mathbf{s}/s$ and
\begin{align}
c_{M} & =\frac{t_{s}^{2}+(t-t_{N})^{2}-(t'-t_{N})^{2}}{2t_{s}(t-t_{N})},\label{cM}\\
c_{M}^{\prime} & =\frac{t_{s}^{2}+(t^{\prime}-t_{N}^{\prime})^{2}-(t-t_{N}^{\prime})^{2}}{2t_{s}(t^{\prime}-t_{N}^{\prime})}.\label{cMp}
\end{align}

\subsection{Correlations of deformations}

To calculate the quantity $G_{ijkl}$, we can expand $t_{ij}$ in powers of $\partial_{i}\zeta$ since,
according to Eq.~(\ref{gradz2medioap}), perturbations are bounded by $\langle(\nabla\zeta)^{2}\rangle\lesssim a^{2}$.
Thus, we write Eq.~(\ref{tij}) as
\begin{equation}
t_{ij}=t_{ij}^{0}+t_{ij}^{1}+t_{ij}^{2}+\cdots,\label{tijpert}
\end{equation}
where
\begin{equation}
	\begin{gathered}
			t_{ij}^{0} =\hat{r}_{i}\hat{r}_{j}, \quad
		t_{ij}^{1}=-(\hat{r}_{i}\partial_{j}\zeta+\hat{r}_{j}\partial_{i}\zeta),
		\\
		t_{ij}^{2} =\partial_{i}\zeta\partial_{j}\zeta-\hat{r}_{i}\hat{r}_{j}(\nabla\zeta)^{2} .
	\end{gathered}	
\end{equation}
Using Eq.~(\ref{tijpert}) in Eq.~(\ref{Gijkldef}), we obtain
\begin{equation}
	G_{ijkl}=\sum_{n,m} \langle e^{i\omega\hat{n}\cdot(\zeta'\hat{r}'-\zeta\hat{r})}t_{ij}^{n}t_{kl}^{\prime m}\rangle
	\equiv\sum_{n,m}G_{ijkl}^{nm}.\label{eq:Gnm}
\end{equation}
Note that we cannot expand the exponential $e^{-i\omega\zeta\hat{n}\cdot\hat{r}}$
in powers of $\zeta$ if we want to consider the high-frequency tail
of the spectrum. 
Assuming Gaussian fluctuations, the averages in Eq.~(\ref{eq:Gnm})
can be written in terms of two-point functions. In particular, we
have
\begin{multline}
G\equiv\langle e^{i\omega\hat{n}\cdot(\zeta'\hat{r}'-\zeta\hat{r})}\rangle=\exp\left\{ -{\textstyle \frac{1}{2}}\omega^{2}\left[(\hat{n}\cdot\hat{r})^{2}\langle\zeta^{2}\rangle\right.\right.\\
\left.\left.+(\hat{n}\cdot\hat{r}')^{2}\langle\zeta^{\prime2}\rangle-2(\hat{n}\cdot\hat{r})(\hat{n}\cdot\hat{r}')\langle\zeta\zeta'\rangle\right]\right\} .\label{eq:G}
\end{multline}
The quantities $G_{ijkl}^{nm}$ are proportional to $G$ and are of order $n+m$.
For this analysis, we will consider terms up to the second order.

\subsubsection{Lowest-order term}

For the lowest-order term $G_{ijkl}^{00}$, we have
\begin{equation}
\Lambda_{ijkl}G_{ijkl}^{00}=\Lambda_{ijkl}\hat{r}_{i}\hat{r}_{j}\hat{r}'_{k}\hat{r}'_{l}\,\langle e^{i\omega\hat{n}\cdot(\zeta'\hat{r}'-\zeta\hat{r})}\rangle=\Lambda G,\label{eq:lowestLG}
\end{equation}
where $\Lambda\equiv\Lambda_{ijkl}\hat{r}_{i}\hat{r}_{j}\hat{r}'_{k}\hat{r}'_{l}$ is given by \cite{mm21gravitational}
\begin{multline}
\Lambda = (\hat{r}\cdot\hat{r}')^{2}-2(\hat{r}\cdot\hat{r}')(\hat{n}\cdot\hat{r})(\hat{n}\cdot\hat{r}')\\
-{\textstyle \frac{1}{2}}+{\textstyle \frac{1}{2}}(\hat{n}\cdot\hat{r}')^{2}+{\textstyle \frac{1}{2}}(\hat{n}\cdot\hat{r})^{2}+{\textstyle \frac{1}{2}}(\hat{n}\cdot\hat{r})^{2}(\hat{n}\cdot\hat{r}')^{2} . \label{eq:Lambda}
\end{multline}
In the absence of deformations, this is the only contribution,
with $G=1$, and we obtain the result of the bubble-collision mechanism \cite{mm21model}. 
The factor $G$ causes a suppression to this result since the exponent
in Eq.~(\ref{eq:G}) is negative. 
However, this suppression only occurs for $\omega\sim k_{*}$, which is a few orders of magnitude
above the scale $\beta$. 
Indeed, the two-point functions in Eq.~(\ref{eq:G})
are of the order of $a^{2}k_{*}^{-2}$ (see Sec.~\ref{sec:stochastic}
and App.~\ref{app:corrfunc}). Therefore, the exponent is of order
$a^2\omega^{2}/k_{*}^{2}$.
For $\omega\sim\beta$, we have $G\simeq1$.

\subsubsection{First-order terms}

The first-order terms are $G_{ijkl}^{01}$ and $G_{ijkl}^{10}$.
We have
\begin{equation}
G_{ijkl}^{10}=-\langle\partial_{i}\zeta e^{i\omega\hat{n}\cdot(\zeta'\hat{r}'-\zeta\hat{r})}\rangle\hat{r}_{j}\hat{r}_{k}'\hat{r}_{l}'+i
\leftrightarrow j .
\label{G10}
\end{equation}
The expression for $G_{ijkl}^{01}$ is similar.
We have \cite{mm21gravitational}
\begin{equation}
	\langle\partial_{i}\zeta e^{i\omega\hat{n}\cdot(\zeta'\hat{r}'-\zeta\hat{r})}\rangle=i(\hat{n}\cdot\hat{r}')\omega\langle\zeta'\partial_{i}\zeta\rangle G.
	\label{zmedio}
\end{equation}
Without the exponential function, Eq.~(\ref{G10}) would be linear in $\zeta$, and the average would vanish.
Dimensionally, the result (\ref{zmedio}) 
is of the order of $\omega k\langle\zeta^{2}\rangle$,
which is of order 1 for relevant deformations and frequencies. 
However, the average  $\langle\zeta'\partial_{i}\zeta\rangle$ is actually much smaller than $k\langle\zeta^{2}\rangle$.
In the first place, it vanishes for the two-bubble contribution since $\zeta$ and $\zeta'$ belong to different bubbles.
For the single-bubble contribution, it vanishes by symmetry at the same point.
Since the correlations rapidly decrease with the point separation, this function is always relatively small.
Therefore, we can neglect these terms.

\subsubsection{Second-order terms}

There are three second-order terms.
One of them is $G_{ijkl}^{20} = \langle e^{i\omega\hat{n}\cdot(\zeta'\hat{r}'-\zeta\hat{r})} t_{ij}^2 t_{kl}^{\prime 0} \rangle$,
where
\begin{equation}
	t_{ij}^2 t_{kl}^{\prime 0} = \partial_{i}\zeta\partial_{j}\zeta \, \hat{r}_{k}'\hat{r}_{l}' - (\nabla\zeta)^{2} \hat{r}_{i}\hat{r}_{j}\hat{r}_{k}'\hat{r}_{l}'
	 .
\end{equation}
This average contains terms proportional to $\langle\zeta'\partial_{i}\zeta\rangle$,
which, as with the first-order contributions, can be neglected
(complete expressions can be found in Ref.~\cite{mm21gravitational}).
Neglecting these terms, we have
\begin{equation}
\langle\partial_{i}\zeta\partial_{j}\zeta e^{i\omega\hat{n}\cdot(\zeta'\hat{r}'-\zeta\hat{r})}\rangle
= \langle\partial_{i}\zeta\partial_{j}\zeta\rangle G.
\end{equation}
By symmetry, we have 
$\langle\partial_{i}\zeta\partial_{j}\zeta\rangle=\frac{1}{2}\langle(\nabla\zeta)^{2}\rangle(\delta_{ij}-\hat{r}_{i}\hat{r}_{j})$.
The calculation of $G_{ijkl}^{02}$ is similar.
Using $\Lambda_{ijkl}\delta_{ij}=0$, we obtain%
\footnote{There is a discrepancy in the numerical factor $3/2$ in Eq.~(\ref{eq:secondLG}) with respect to Ref.~\cite{mm21gravitational},
	which arises from a more correct consideration of the relation between $T_{ij}$ and $\sigma$
	in the present paper, Eq.~(\ref{eq:Tijbubble}).
	See also the discussion in appendix A.2 of Ref.~\cite{mm23thin}.} 
\begin{equation}
\Lambda_{ijkl}(G_{ijkl}^{20}+G_{ijkl}^{02}) =
-\frac{3}{2}\big[\langle(\nabla\zeta)^{2}\rangle+\langle(\nabla\zeta')^{2}\rangle\big] \Lambda G.
\label{eq:secondLG}
\end{equation}

The remaining second-order term is given by
$G^{11}_{ijkl} = \langle e^{i\omega\hat{n}\cdot(\zeta'\hat{r}'-\zeta\hat{r})}t_{ij}^{1}t_{kl}^{\prime 1}\rangle$, with
\begin{multline}
t_{ij}^{1}t_{kl}^{\prime1}=\hat{r}_{i}\hat{r}_{k}'\partial_{j}\zeta\partial_{l}\zeta'+\hat{r}_{i}\hat{r}_{l}'\partial_{j}\zeta\partial_{k}\zeta'\\
+\hat{r}_{j}\hat{r}_{k}'\partial_{i}\zeta\partial_{l}\zeta'+\hat{r}_{j}\hat{r}_{l}'\partial_{i}\zeta\partial_{k}\zeta'.
\end{multline}
Neglecting again terms that contain the average $\langle\zeta'\partial_{i}\zeta\rangle$, we have
\cite{mm21gravitational}
\begin{equation}
\langle\partial_{i}\zeta\partial_{j}\zeta'e^{i\omega\hat{n}\cdot(\zeta'\hat{r}'-\zeta\hat{r})}\rangle
=\langle\partial_{i}\zeta\partial_{j}\zeta'\rangle G,
\end{equation}
This average vanishes for the two-bubble contribution, since $\zeta$ and $\zeta'$ belong to different bubbles%
\footnote{For the previous terms, instead, we had two factors of $\zeta$ evaluated at the same point on the same surface,
	even for the two-bubble contribution.
	For example, in the product $t_{ij}^2 t_{kl}^{\prime 0}$,
	$t_{ij}^2$ contains two factors of $\zeta$, while $t_{kl}^{\prime 0}$ contains no factors of $\zeta'$.}.
For the single-bubble case, we have
$\Lambda_{ijkl} G_{ijkl}^{11} = \Lambda_{ijkl}\langle t_{ij}^{1}t_{kl}^{\prime1}\rangle G$.
Taking into account that $\hat{r}\cdot\nabla\zeta=\hat{r}'\cdot\nabla\zeta'=0$, we obtain 
\begin{multline}
\Lambda_{ijkl}\langle t_{ij}^{1}t_{kl}^{\prime1}\rangle=\big\{2\hat{r}_{j}\hat{r}_{i}^{\prime}-2(\hat{r}_{k}^{\prime}\hat{n}_{k})\hat{n}_{i}\hat{r}_{j}-2(\hat{r}_{k}\hat{n}_{k})\hat{r}_{i}^{\prime}\hat{n}_{j}\\
+2[(\hat{r}_{k}\hat{r}_{k}^{\prime})-(\hat{r}_{k}\hat{n}_{k})(\hat{r}_{l}^{\prime}\hat{n}_{l})](\delta_{ij}-\hat{n}_{i}\hat{n}_{j})\big\}\langle\partial_{i}\zeta\partial_{j}\zeta'\rangle .
\label{eq:L11}
\end{multline}
The main difference from previous contributions is that the two-point
function in Eq.~(\ref{eq:L11}) is evaluated at different points. 
Therefore, it contains the information about the spatial correlation of the deformations. 

\section{The GW spectrum}

Using Eqs.~(\ref{sigmaR}), (\ref{separ}), (\ref{eq:Pu}), (\ref{eq:Heavi2b}), and (\ref{eq:Gnm})
in Eqs.~(\ref{eq:Pis}) and (\ref{eq:Pid}), we obtain $\Pi_{s,d}=\sum_{n,m}\Pi_{s,d}^{nm}$,
with 
\begin{align}
	&\begin{multlined}[b]
		\frac{\Pi_{s}^{nm}}{\kappa^{2}\rho_{\mathrm{vac}}^{2}} 
		=\frac{v_{w}^{6}}{9}\int_{-\infty}^{t}\negthickspace dt_{N}\Gamma(t_{N})(t-t_{N})^{3}(t^{\prime}-t_{N})^{3} 
		\\
		\times\int\negmedspace d\hat{r}\negmedspace\int\negmedspace d\hat{r}'e^{-J}e^{i\omega\hat{n}\cdot\mathbf{s}}\Lambda_{ijkl}G_{ijkl}^{nm},
	\end{multlined}\label{eq:Pisnm}
\\
	&\begin{multlined}[b]
		\frac{\Pi_{d}^{nm}}{\kappa^{2}\rho_{\mathrm{vac}}^{2}}
		=\frac{v_{w}^{6}}{9}\negmedspace\int_{-\infty}^{t}\negthickspace dt_{N}\Gamma(t_{N})\negthickspace\int_{-\infty}^{t'}\negthickspace dt_{N}^{\prime}\Gamma(t_{N}^{\prime}) 
		\\
		\times  (t-t_{N})^{3}(t^{\prime}-t_{N}^{\prime})^{3}\negthickspace\int\negthickspace d^{3}s\negthickspace\int\negthickspace d\hat{r}\negthickspace\int\negthickspace d\hat{r}'e^{-J}e^{i\omega\hat{n}\cdot\mathbf{s}}
		\\
		\times  \Theta(\hat{r}\cdot\hat{s}+c_{M})\Theta(c_{M}'-\hat{r}'\cdot\hat{s}) \Lambda_{ijkl}G_{ijkl}^{nm} .
	\end{multlined}\label{eq:Pidnm}
\end{align}
In Eq.~(\ref{eq:Pisnm}),
the vector $\mathbf{s}=R'\hat{r}'-R\hat{r}$ depends on $\hat{r}$
and $\hat{r}'$. In Eq.~(\ref{eq:Pidnm}), $\mathbf{s}$ is an independent
variable coming from the change of variable $\mathbf{l}\to\mathbf{s}=\mathbf{l}+R'\hat{r}'-R\hat{r}$
in Eq.~(\ref{eq:Pid}). 

\subsection{Collisions of corrugated bubbles}

\label{sec:bbcolldef}

The lowest-order term (\ref{eq:lowestLG}) and the two second-order terms (\ref{eq:secondLG}) are proportional to $\Lambda G$.
Consequently, the calculation of $\Pi^{00}$, $\Pi^{02}$, and $\Pi^{20}$ is essentially the same.
In the exponent of Eq.~(\ref{eq:G}), $\langle\zeta\zeta'\rangle$ vanishes
for the two-bubble contribution.
Therefore, using Eq.~(\ref{eq:zeta2medioap}), we have
\begin{multline}
G=\exp\left\{ -{\textstyle \frac{1}{2}}\omega^{2}\alpha^{2}\left[(\hat{n}\cdot\hat{r})^{2}(t-t_{N})^{2}\right.\right.\\
\left.\left.+(\hat{n}\cdot\hat{r}')^{2}(t'-t'_{N})^{2}\right]\right\} .\label{eq:G00}
\end{multline}
For the single-bubble contribution, $\langle\zeta\zeta'\rangle$ decreases
rapidly as the separation between the points increases (see App.~\ref{app:corrfunc}).
Consequently, it is largely suppressed in the majority of the angular
integral. Thus, we can neglect this term and use Eq.~(\ref{eq:G00})
(with $t_{N}^{\prime}=t_{N}$) for the single-bubble contribution as well.

\subsubsection{Lowest-order contribution}

With $\Lambda_{ijkl}G_{ijkl}^{00}=\Lambda G$,
the expressions for $\Pi^{00}_s$ and $\Pi^{00}_d$
are the same as those for the bubble collision mechanism, except for the factor $G$.
Therefore, the calculation of $\Pi^{00}$ is similar to that case.
We describe the steps below. 
We focus on how to address this factor. 
For more details, see Refs.~\cite{mm21gravitational} and \cite{mm21model}.

Since our stochastic source is homogeneous and isotropic, the GW spectrum
does not depend on the direction of observation. 
Therefore, we can average over $\hat{n}$, which is convenient for calculating $\Pi^{00}$.
In Eqs.~(\ref{eq:Pisnm}) and (\ref{eq:Pidnm}),
only $e^{i\omega\hat{n}\cdot\mathbf{s}}$, $\Lambda$, and $G$ depend
on $\hat{n}$, and we can make the replacement
\begin{equation}
e^{i\omega\hat{n}\cdot\mathbf{s}}\Lambda G\to
\widehat{\Lambda G}\equiv\frac{1}{4\pi}\int d\hat{n}\,e^{i\omega\hat{n}\cdot\mathbf{s}}\Lambda G.\label{eq:LGbar}
\end{equation}
To calculate this integral, we choose coordinates such that $\hat{s}$
is along the $z$-axis and $\hat{r}$ is on the $xz$-plane. We define
the variables $c_{n}=\hat{n}\cdot\hat{s}$, $c_{r}=\hat{r}\cdot\hat{s}$,
$c_{r}'=\hat{r}^{\prime}\cdot\hat{s}$, which are cosines, and we
denote $s_{n}$, $s_{r}$, $s_{r}'$, respectively, the corresponding
sines (e.g., $s_{n}=\sqrt{1-c_{n}^{2}}$), and $\phi_{n},\phi_{r}'$
the azimuthal angles of $\hat{n}$ and $\hat{r}^{\prime}$ (see Ref.~\cite{mm21gravitational}
for a figure). Thus, we have
\begin{align}
\hat{n}\cdot\hat{r} & =s_{r}s_{n}\cos\phi_{n}+c_{r}c_{n},\label{nr}\\
\hat{n}\cdot\hat{r}' & =s_{r}'s_{n}(\cos\phi_{r}'\cos\phi_{n}+\sin\phi_{r}'\sin\phi_{n})+c_{r}'c_{n}.\label{nrp}
\end{align}
The quantity $\Lambda$, Eq.~(\ref{eq:Lambda}), also depends on
\begin{equation}
c\equiv\hat{r}\cdot\hat{r}'=s_{r}s_{r}'\cos\phi_{r}^{\prime}+c_{r}c_{r}'.\label{eq:c_rrp}
\end{equation}
In Eq.~(\ref{eq:LGbar}), we have $d\hat{n}=d\phi_{n}dc_{n}$ and
$\hat{n}\cdot\mathbf{s}=sc_{n}$. 

Integrating by parts $c_{n}$, we have 
\begin{multline}
\int_{-1}^{+1}dc_{n}e^{i\omega sc_{n}}\Lambda G=\left.\frac{e^{i\omega sc_{n}}}{i\omega s}\Lambda G\right|_{-1}^{+1}\\
-\int_{-1}^{+1}dc_{n}\frac{e^{i\omega sc_{n}}}{i\omega s}\partial_{c_{n}}(\Lambda G).\label{eq:intcn}
\end{multline}
The derivative of $G$ with respect to $c_{n}$ introduces factors
of the order of $\omega^{2}\alpha^{2}(t-t_{N})^{2}$. 
Due to the presence of the exponential $e^{-J}$ in Eqs.~(\ref{eq:Pisnm})-(\ref{eq:Pidnm}),
the integrand vanishes beyond a range $t-t_{N}\sim\beta^{-1}$.
Therefore, we have
\begin{equation}
	\partial_{c_{n}}(\Lambda G)=G\partial_{c_{n}}\Lambda + \mathcal{O}(\omega^{2}\alpha^{2}\beta^{-2}) G\Lambda .
	\label{dcnG}
\end{equation}
Using Eq.~(\ref{eq:zeta2medioap}), we have $\omega^{2}\alpha^{2}\beta^{-2} \sim (4\pi)^{-1} a^{2} \omega^{2}{k_{*}^{-2}} $.
For $\omega\sim k_{*}$ or lower, we have $\omega^{2}\alpha^{2}\beta^{-2}\lesssim a^{2}/(4\pi)$.
Therefore, it is a good approximation to drop the second term in (\ref{dcnG}).
This approximation breaks down when $\omega\gg k_{*}$.
However, the exponential suppression of the factor $G$ makes
this contribution irrelevant in this case%
\footnote{If we keep this term and repeat the integration by parts, 
	we obtain an asymptotic expansion in powers of $\alpha^{2}$.}. 
Since $\Lambda$ is a quartic polynomial in $c_{n}$, we can solve
the integral (\ref{eq:intcn}) by repeating the integration by parts.
We obtain
\begin{equation}
\widehat{\Lambda G}=\sum_{k=0}^{2}\frac{j_{n}(\omega s)}{(\omega s)^{n}}C_{n}G,\label{eq:LGmedio}
\end{equation}
where $j_{n}$ are spherical Bessel functions, $C_{n}$ are polynomials
in $c$, $c_{r}$, and $c_{r}'$, which can be found in App.~B of
Ref.~\cite{mm21gravitational}, and $G$ is now given by
\begin{equation}
G=\exp\left\{ -{\textstyle \frac{1}{2}}\omega^{2}\alpha^{2}\left[c_{r}^{2}(t-t_{N})^{2}+c_{r}^{\prime2}(t'-t'_{N})^{2}\right]\right\} \label{G00}
\end{equation}
(with $t_{N}^{\prime}=t_{N}$ in the single-bubble case).

In the last line of Eq.~(\ref{eq:Pisnm}),
after integrating $\hat{r}'$, the result does not depend on
$\hat{r}$, and we have
\begin{equation}
\int\negmedspace d\hat{r}\negmedspace\int\negmedspace d\hat{r}'\,\widehat{\Lambda G}e^{-J}=8\pi^{2}\negthickspace\int_{-1}^{+1}\negthickspace dc\,\widehat{\Lambda G}e^{-J}.
\end{equation}
All the variables in the integrand can be written in terms of $c$.
Indeed, we have $c_{r}=(R'c-R)/s$, $c_{r}'=(R'-Rc)/s$, $\cos\phi_{r}'=(c-c_{r}c_{r}')/(s_{r}s_{r}')$,
and $s^{2}=R^{\prime2}+R^{2}-2RR'c$. We now change variables $c\to s$
and interchange the integrals with respect to $s$ and $t_{N}$ in
Eq.~(\ref{eq:Pisnm}). The quantities $C_{n}$ become functions of
$s$, $t_{-}$ and $T\equiv t_{+}-2t_{N}$. With the change of variable
$t_{N}\to T$, the variable $t_{+}$ disappears almost everywhere.
We obtain
\begin{multline}
\frac{\Pi_{s}^{00}}{\kappa^{2}\rho_{\mathrm{vac}}^{2}}=\frac{\pi^{2}v_{w}^{6}}{72}\Gamma({\textstyle \frac{t_{+}}{2}})\negmedspace\int_{t_{-}}^{\infty}\frac{dt_{s}}{t_{s}^{3}}e^{-J}\\
\times\sum_{n=0}^{2}\frac{j_{n}(\omega v_{w}t_{s})}{(\omega v_{w}t_{s})^{n}}\int_{t_{s}}^{\infty}\negthickspace dTe^{-\frac{1}{2}\beta T}P_{n}G_{s},\label{Pi00sfinal}
\end{multline}
where
\begin{equation}
G_{s}=\exp\left[-{\textstyle \frac{1}{4}}\omega^{2}\alpha^{2}\left(t_{-}^{2}T^{2}/t_{s}^{2}+t_{s}^{2}\right)\right]\label{eq:Gs}
\end{equation}
and
\begin{align}
P_{0}= & \,(t_{s}^{2}-t_{-}^{2})^{2}(t_{s}^{2}-T^{2})^{2},\\
P_{1}= & \,2(t_{s}^{2}-t_{-}^{2})(t_{s}^{2}-T^{2})\nonumber \\
 & \,\times[3t_{s}^{4}+t_{s}^{2}t_{-}^{2}+(t_{s}^{2}-5t_{-}^{2})T^{2}],\\
P_{2}= & \,(3t_{s}^{4}+2t_{s}^{2}t_{-}^{2}+3t_{-}^{4})t_{s}^{4}+(2t_{s}^{4}+12t_{s}^{2}t_{-}^{2}\nonumber \\
 & -30t_{-}^{4})t_{s}^{2}T^{2}+(3t_{s}^{4}-30t_{s}^{2}t_{-}^{2}+35t_{-}^{4})T^{4}.\label{eq:Pn}
\end{align}

In Eq.~(\ref{eq:Pidnm}), $\mathbf{s}$ is an independent variable and we have
\begin{multline}
\int\negmedspace d^{3}s\negmedspace\int\negmedspace d\hat{r}\negmedspace\int\negmedspace d\hat{r}'\,\Theta(c_{r}+c_{M})\Theta(c_{M}'-c_{r}')e^{-J}\,\widehat{\Lambda G}
\\
=\Theta(c_{M}+1)\Theta(c_{M}'+1)\,8\pi^{2}\negthickspace\int_{v_{w}t_{-}}^{\infty}\negthickspace dss^{2} \times {}
\\
\int_{-1}^{\min\{c_{M},1\}}\negthickspace\negthickspace dc_{r}\int_{-1}^{\min\{c_{M}',1\}}\negthickspace \negthickspace dc_{r}'\int_{0}^{2\pi}\negthickspace \negthickspace d\phi_{r}'e^{-J}\,\widehat{\Lambda G}.\label{eq:intangd}
\end{multline}
The integral with respect to $\phi_{r}'$ is straightforward and the
terms with $n=0$ and $n=1$ in Eq.~(\ref{eq:LGmedio}) vanish. The
integrals with respect to $c_{r}$ and $c_{r}'$ are independent,
and the last line in Eq.~(\ref{eq:intangd}) becomes
\begin{equation}
	\frac{j_{2}(\omega s)}{(\omega s)^{2}}e^{-J}I_{c}I_{c}',
\end{equation}
where
\begin{equation}
I_{c}=\int_{-1}^{\min\{c_{M},1\}}\negthickspace dc_{r}(1-3c_{r}^{2})e^{-{\textstyle \frac{1}{2}}\omega^{2}\alpha^{2}c_{r}^{2}(t-t_{N})^{2}},
\end{equation}
and $I_{c}'$ has the same expression as a function of $c_{M}'$ and
$t'-t_{N}'$. 
Integrating by parts and dropping again terms of order $\omega^{2}\alpha^{2}\beta^{-2}\lesssim a^{2}/(4\pi)$, we obtain
\begin{equation}
I_{c}=\Theta(1-c_{M}) (c_{M}-c_{M}^{3}) e^{-{\textstyle \frac{1}{2}}\omega^{2}\alpha^{2}c_{M}^{2}(t-t_{N})^{2}} .
\label{integcr}
\end{equation}
Using Eqs.~(\ref{cM})-(\ref{cMp}), the restrictions on $c_{M}$ and $c_{M}^{\prime}$ in Eqs.~(\ref{eq:intangd}) and (\ref{integcr})
imply $t_{s}\geq t_{-}$, $t_{s}<t_{+}-2t_{N}\equiv T$, and $t_{s}<t_{+}-2t_{N}^{\prime}\equiv T'$.
Using these results in Eq.~(\ref{eq:Pidnm}), interchanging the integrals
with respect to $s$ and $t_{N},t_{N}^{\prime}$, and changing variables
$t_{N}\to T$, $t_{N}'\to T'$, we obtain
\begin{multline}
\frac{\Pi_{d}^{00}}{\kappa^{2}\rho_{\mathrm{vac}}^{2}}=\frac{\pi^{3}v_{w}^{9}}{144}\left[\Gamma({\textstyle \frac{t_{+}}{2}})\right]^{2}\negthickspace\int_{t_{-}}^{\infty}\negthinspace\frac{dt_{s}}{t_{s}^{4}}e^{-J}\frac{j_{2}(\omega v_{w}t_{s})}{(\omega v_{w}t_{s})^{2}}\\
\times\int_{t_{s}}^{\infty}\negthickspace dTe^{-\frac{1}{2}\beta T}QG\negthickspace\int_{t_{s}}^{\infty}\negthickspace dT'e^{-\frac{1}{2}\beta T'}Q'G',\label{Pi00dfinal}
\end{multline}
where 
\begin{equation}
\begin{aligned}G & =\exp[-{\textstyle \frac{1}{8}}\omega^{2}\alpha^{2}(t_{s}-t_{-}T/t_{s})^{2}],\\
G' & =\exp[-{\textstyle \frac{1}{8}}\omega^{2}\alpha^{2}(t_{s}+t_{-}T'/t_{s})^{2}],
\end{aligned}
\label{eq:GGp}
\end{equation}
and
\begin{equation}
\begin{aligned}Q & =(t_{s}^{2}-t_{-}^{2})(T^{2}-t_{s}^{2})(t_{s}^{2}-t_{-}T),\\
Q' & =(t_{s}^{2}-t_{-}^{2})(T^{\prime2}-t_{s}^{2})(t_{s}^{2}+t_{-}T').
\end{aligned}
\label{eq:QQp}
\end{equation}

We now insert the results for $\Pi^{00}$ into Eq.~(\ref{rhogw}). 
Using $\int_{-\infty}^{\infty}dt\int_{t}^{\infty}dt'=\frac{1}{2}\int_{-\infty}^{\infty}dt_{+}\int_{0}^{\infty}dt_{-}$,
the integral with respect to $t_{+}$ is trivial with the change of
variable $t_{+}\to u = \Gamma(\frac{t_+}2) = \Gamma_{*}e^{\beta(t_{+}/2-t_{*})}$.
To avoid writing several constants, it is useful to consider the quantity
\begin{equation}
	\Delta \equiv\frac{3\beta^{2}}{8\pi G(\kappa\rho_{\mathrm{vac}})^{2}}\frac{d\rho_{GW}}{d\ln\omega} .
	\label{eq:Delta_adim}
\end{equation}
The present contribution gives
$\Delta^{00}=\Delta_{s}^{00}+\Delta_{d}^{00}$, with
\begin{align}
	\Delta_{s}^{00} &= \frac{\beta^{5}v_{w}^{3}\omega^{3}}{384\pi}\negthickspace\int_{0}^{\infty}\negthickspace dt_{-}\cos(\omega t_{-})\negthickspace\int_{t_{-}}^{\infty}\frac{dt_{s}}{t_{s}^{3}} 
	\nonumber
	\\
	&\times\frac{e^{\beta t_{s}/2}}{E} \sum_{n=0}^{2}\frac{j_{n}(\omega v_{w}t_{s})}{(\omega v_{w}t_{s})^{n}}\int_{t_{s}}^{\infty}\negthickspace dTe^{-\frac{1}{2}\beta T}P_{n}G_{s},
\label{Delta00sexp}
\\
\Delta_{d}^{00} &= \frac{\beta^{9}v_{w}^{3}\omega^{3}}{6144\pi}\negthickspace\int_{0}^{\infty}\negthickspace dt_{-}
	\cos(\omega t_{-})\negthickspace\int_{t_{-}}^{\infty}\frac{dt_{s}}{t_{s}^{4}}\frac{j_{2}(\omega v_{w}t_{s})}{(\omega v_{w}t_{s})^{2}}
	\nonumber
	\\
	&\times\frac{e^{\beta t_{s}}}{E^{2}} \negthickspace\int_{t_{s}}^{\infty}\negthickspace dTe^{-\frac{1}{2}\beta T}QG\int_{t_{s}}^{\infty}
	\negthickspace dT'e^{-\frac{1}{2}\beta T'}Q'G', 
\label{Delta00dexp}
\end{align}
where 
\begin{equation}
E=2\cosh(\beta t_{-}/2)e^{\beta t_{s}/2}-(t_{s}^{2}-t_{-}^{2})\beta/(4t_{s})-1.
\end{equation}

For $a=0$, we have $G_s=G=G'=1$ and we obtain
the result of the bubble-collision mechanism without corrugations, 
i.e.,
$\Delta^{00}|_{a=0}=\Delta_{\mathrm{coll}}$.
In this case, the integrals with respect to $T$ and $T'$ are trivial
(see Ref.~\cite{mm21model}). 
For $a \neq 0$, these integrals
can still be obtained analytically in terms of the error function.
However, the expressions are rather cumbersome and we do not write
them here. The two remaining integrals with respect to $t_{-}$ and
$t_{s}$ must be calculated numerically. 
Due to the oscillating integrand,
the numerical calculation is difficult for $\omega\gg\beta$.
However, in this case we can use
an asymptotic expansion in powers of $\beta/\omega$ \cite{mm21model}.
Keeping the lowest order in this expansion is a suitable approximation for
$\omega\sim k_{*}$. We discuss this approximation in App.~\ref{app:Approximations}. 

\begin{figure}[h]
\includegraphics[width=0.8\columnwidth]{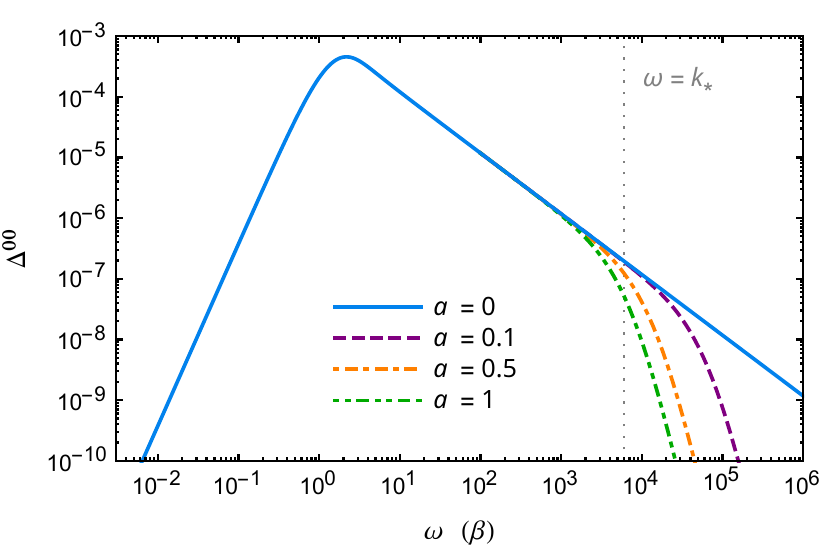}
\caption{The contribution $\Delta^{00}$ for $T=100\mathrm{GeV}$, $\alpha_{T}=0.05$, $v_{c}=0.2$, and $v_{w}=0.1$.\label{fig:D00}}
\end{figure}

In Fig.~\ref{fig:D00} we consider the quantity $\Delta^{00}$ for
various values of the parameter $a$. The case $a=0$
gives the spectrum $\Delta_{\mathrm{coll}}$.
For $a\neq0$, $\Delta^{00}$ essentially coincides with the bubble-collision
spectrum for $\omega\sim \beta$. 
However, for $\omega\sim k_{*}$, $\Delta^{00}$ begins to decrease more rapidly.
In App.~\ref{app:Approximations}, we provide analytic approximations.
For $\omega\gg k_{*}$, we have $\Delta^{00}\propto\omega^{-6}$.

\subsubsection{Second-order equal-point fluctuations}

\begin{figure}[h]
\includegraphics[width=0.8\columnwidth]{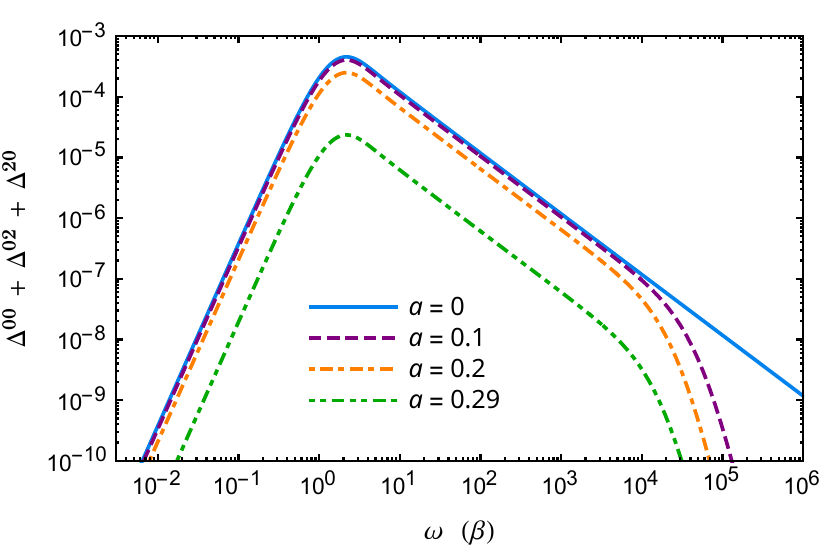}
\caption{The contribution $\Delta^{00}+\Delta^{02}+\Delta^{20}$ for 
	$T=100\mathrm{GeV}$, $\alpha_{T}=0.05$, $v_{c}=0.2$, and $v_{w}=0.1$.\label{fig:D02}}
\end{figure}

Now, let us consider the contributions (\ref{eq:secondLG}).
The quantity $\langle(\nabla\zeta)^{2}\rangle$ is independent of the point on the surface.
Thus, Eqs.~(\ref{eq:Pisnm})-(\ref{eq:Pidnm}) yield
\begin{equation}
\Pi^{20}+\Pi^{02}=-\frac{3}{2}\left[\langle(\nabla\zeta)^{2}\rangle+\langle(\nabla\zeta')^{2}\rangle\right]\Pi^{00}.\label{Pi2002}
\end{equation}
Inserting Eq.~(\ref{Pi2002}) into Eq.~(\ref{rhogw}) and considering the quantity (\ref{eq:Delta_adim}),
we obtain the contribution $\Delta^{02}+\Delta^{20}$, 
which is given by Eqs.~(\ref{Delta00sexp})-(\ref{Delta00dexp}),
with the extra factor $-\frac{3}{2}[\langle(\nabla\zeta)^{2}\rangle+\langle(\nabla\zeta')^{2}\rangle]$
in the integrands. 
As discussed in Sec.~\ref{subsec:correlations},
the quantity $\langle(\nabla\zeta)^{2}\rangle$ depends logarithmically
on $t$, so we can neglect its variation in the relevant time range and
use the approximation (\ref{gradz2medioap}).
Indeed, in Eqs.~(\ref{Delta00sexp})-(\ref{Delta00dexp}), we have factors
$t_{s}^{-3}P_{n}\sim T^{5}$ and $t_{s}^{-4}QQ'\sim T^{8}$ (since
$t_{-}\leq t_{s}\leq T$). 
Therefore, later times $T\sim\beta$ are
much more relevant than earlier times. 
We obtain
\begin{equation}
\Delta^{00}+\Delta^{02}+\Delta^{20}=\left[1-3\frac{a^{2}}{2\pi}\log\left(\frac{k_{c}}{k_{*}}\right)\right]\Delta^{00}.\label{D2002}
\end{equation}
We show this result in Fig.~\ref{fig:D02}. 
The overall suppression to the contribution from bubble collisions occurs because some of the energy that would have gone
into stationary motion now goes into the dynamics of unstable modes.

We introduced the dimensionless parameter $a$ in Eq.~(\ref{Daproxsat})
to model the departure from the linear regime. 
In principle, this parameter could be of order 1.
However,
the value of $\nabla\zeta$ is also constrained by the use of the
expansion (\ref{tijpert}). 
The result (\ref{D2002}) shows
that the perturbative calculation becomes unreliable 
as $\langle(\nabla\zeta)^{2}\rangle$ approaches
the value $1/3$.

\subsection{The instability peak}

\label{sec:Pi11}

Finally, let us consider the term
\begin{multline}
\frac{\Pi^{11}}{\kappa^{2}\rho_{\mathrm{vac}}^{2}}=\frac{1}{9}\int_{-\infty}^{t}\negthickspace dt_{N}\Gamma(t_{N})R^{3}R^{\prime3}\negmedspace\\
\times\int\negmedspace d\hat{r}\negmedspace\int\negmedspace d\hat{r}'e^{i\omega\hat{n}\cdot\mathbf{s}}e^{-J}G\Lambda_{ijkl}\langle t_{ij}^{1}t_{kl}^{\prime1}\rangle.\label{Pi11}
\end{multline}
with $\Lambda_{ijkl}\langle t_{ij}^{1}t_{kl}^{\prime1}\rangle$ given
by Eq.~(\ref{eq:L11}). 
In this case, there is only a single-bubble contribution. 
The quantity $G$ is given by Eq.~(\ref{eq:G}).
For previous contributions, we discarded the last term in the exponent of  Eq.~(\ref{eq:G})
because it is proportional to $\langle\zeta\zeta'\rangle$ and nonvanishing only for small point separations. 
For this contribution, however, the entire integrand in Eq.~(\ref{Pi11})
falls to zero beyond a small distance due to the factor $\langle t_{ij}^{1}t_{kl}^{\prime1}\rangle$.
Therefore, the term $\langle\zeta\zeta'\rangle$
in the exponent of $G$ has the same relevance as the terms $\langle\zeta^{2}\rangle$
and $\langle\zeta^{\prime2}\rangle$. 
For small point separations, we have
$\langle\zeta\zeta'\rangle\simeq\langle\zeta^{2}\rangle$ (for $t<t'$;
see App.~\ref{app:corrfunc}), and we obtain $G\simeq\exp[-\frac{1}{2}\omega^{2}(\hat{r}\cdot\hat{n})^{2}(\langle\zeta^{\prime2}\rangle-\langle\zeta^{2}\rangle)]$.
This factor only matters at high frequency.
However, due to the factor $\cos(\omega t_{-})$ in Eq.~(\ref{rhogw}),
only times $t'\simeq t$ are relevant at high frequency.
This implies $\langle\zeta^{\prime2}\rangle \simeq \langle\zeta^{2}\rangle$ and $G\simeq1$.%
\footnote{We have verified that this is indeed a good approximation.
	Including the function $G$ in the calculations introduces terms of order
	$(a^{2}/4\pi)e^{-k_{*}^{3}/\omega^{3}}k_{*}/\omega$ into the GW spectrum.
	Such terms are irrelevant for $\omega\ll k_{*}$,
	do not change the high-frequency behavior $\Delta\propto\omega^{-1}$,
	and are suppressed by the factor $a^{2}/4\pi$ for $\omega\gtrsim k_*$.}

\begin{figure}[tb]
	\includegraphics[width=0.6\columnwidth]{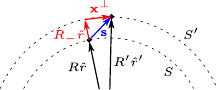}
	\caption{The vectors $\mathbf{s}$, $\mathbf{x}^{\perp}$, and $R_{-}\hat{r}$.}
	\label{fig:sxr}
\end{figure}

In Eq.~(\ref{Pi11}), the vector $\mathbf{s}=R'\hat{r}'-R\hat{r}$ connects two points on
the spherical surface at different times $t,t'$. In order to use
the planar-wall results, we write $\mathbf{s}=R_{-}\hat{r}+\mathbf{x}^{\perp}$,
where $R_{-}=R'-R$ and $\mathbf{x}^{\perp}=R'(\hat{r}'-\hat{r})$
(see Fig.~\ref{fig:sxr}). 
For a given direction $\hat{r}$, the integral with respect
to $\hat{r}'$ effectively extends over a small
patch where $R'\hat{r}'$ is around $R'\hat{r}$. 
Thus, we have $|\mathbf{x}^{\perp}|\ll R$,
$\hat{r}\cdot\mathbf{x}^{\perp} \simeq 0$, $s^{2} \simeq x^{2}+R_{-}^{2}$,
and we can use the planar two-point functions as a function of $x=|\mathbf{x}^{\perp}|$.

Neglecting the curvature of the small patch, we can write $R^{\prime2}d\hat{r}'=d^{2}x^{\perp}$.
Additionally, we have $\hat{n}\cdot\mathbf{x}^{\perp}=\mathbf{n}^{\perp}\cdot\mathbf{x}^{\perp}$,
where $\mathbf{n}^{\perp}$ is the projection of $\hat{n}$ onto the
tangent plane, $\mathbf{n}^{\perp}=\hat{n}-(\hat{n}\cdot\hat{r})\hat{r}$.
The quantity $J$, Eq.~(\ref{eq:Itotexp}), is
approximately constant in this patch, and we can evaluate it at $s=R_{-}$,
\begin{equation}
	J\simeq J|_{t_{s}=t_{-}}=8\pi v_{w}^{3}\beta^{-4}\Gamma_{*}e^{\beta(t'-t_{*})} .
	\label{eq:Itotap}
\end{equation}
Thus, we obtain
\begin{multline}
	\frac{\Pi^{11}}{\kappa^{2}\rho_{\mathrm{vac}}^{2}}=\frac{e^{-J}}{9}\int_{-\infty}^{t}\negthickspace dt_{N}\Gamma(t_{N})R^{3}R^{\prime}\\
	\times\int\negmedspace d\hat{r}e^{i\omega R_{-}\hat{n}\cdot\hat{r}}\negmedspace\int\negthickspace d^{2}x^{\perp}e^{i\omega\mathbf{n}^{\perp}\cdot\mathbf{x}^{\perp}}\Lambda_{ijkl}\langle t_{ij}^{1}t_{kl}^{\prime1}\rangle.\label{Pi11pl}
\end{multline}

The result (\ref{Pi11pl}) can be interpreted as an ensemble of independent deformed planes. 
The rightmost integral
is the Fourier transform of the correlations on the plane, while the
integral $\int\negmedspace d\hat{r}$ corresponds to considering different
plane orientations. 
The factors $\frac{1}{9}\kappa^{2}\rho_{\mathrm{vac}}^{2}R'R$
correspond to the energy injected into the surface, while the factors
$e^{-J}R^{2}$ account for the total uncollided surface area.
The integral with respect to $t_{N}$ considers contributions
from different bubble sizes.

We can use the approximation $\hat{r}'\simeq\hat{r}$ in all the quantities
except for rapidly varying functions such as $\langle\zeta\zeta'\rangle$
and $e^{i\omega\hat{n}\cdot\mathbf{s}}$. In this approximation, $\nabla\zeta$
and $\nabla\zeta'$ are in the same tangent plane and we have $\hat{r}\cdot\nabla\zeta'=\hat{r}'\cdot\nabla\zeta=0$,
$\hat{n}\cdot\nabla\zeta=\mathbf{n}^{\perp}\cdot\nabla\zeta$, $\hat{n}\cdot\nabla\zeta'=\mathbf{n}^{\perp}\cdot\nabla\zeta'$.
Thus, Eq.~(\ref{eq:L11}) becomes
\begin{equation}
	\Lambda_{ijkl}\langle t_{ij}^{1}t_{kl}^{\prime1}\rangle
	=2 n^{\perp2} \left[\delta_{ij}-n_{i}^{\perp}n_{j}^{\perp}\right] \langle\partial_{i}\zeta\partial_{j}\zeta'\rangle.
	\label{Lttpl}
\end{equation}
where $n^{\perp2}=\mathbf{n}^{\perp}\cdot\mathbf{n}^{\perp}=1-(\hat{r}\cdot\hat{n})^{2}$.

Note that the integral with respect to $\mathbf{x}^{\perp}$ in
Eq.~(\ref{Pi11pl}) is the Fourier transform of Eq.~(\ref{Lttpl}) on the plane. 
Let us denote $\tilde{F}(\mathbf{k})\equiv\int d^{2}x^{\perp}\,e^{-i\mathbf{k}\cdot\mathbf{x}^{\perp}}F(\mathbf{x}^{\perp})$
the Fourier transform of a function $F$. According to Eq.~(\ref{eq:zetazetap}),
$\langle\zeta\zeta'\rangle$ is given by the \emph{inverse} transform
of $|D|^{2}$, i.e., we have,
\begin{equation}
\widetilde{\left\langle \zeta\zeta'\right\rangle }=|D|^{2}(k,t,t').
\end{equation}
Similarly, from Eqs.~(\ref{desazeta}) and (\ref{defD2}), we obtain
\begin{equation}
\widetilde{\langle \partial_i\zeta \partial_j \zeta' \rangle} = k_{i}k_{j}\widetilde{\langle\zeta\zeta'\rangle}.\label{eq:FTdzdz}
\end{equation}
Thus, the last integral in Eq.~(\ref{Pi11pl}) is given by
\begin{equation}
\widetilde{\Lambda_{ijkl}\langle t_{ij}^{1}t_{kl}^{\prime1}\rangle}|_{\mathbf{k}=-\omega\mathbf{n}^{\perp}}=2n^{\perp4}c^{2}\omega^{2}|D|^{2}(\omega n^{\perp},t,t'),
\end{equation}
where $c=\hat{r}\cdot\hat{n}$ (note that the cosine $c$ in this
section is not the same as in Sec.~\ref{sec:bbcolldef}) and $n^{\perp2}=1-c^{2}$.

In Eq.~(\ref{Pi11pl}), we have $d\hat{r}=4\pi dc$ and $R=v_{w}T$,
$R'=v_{w}T'$, where $T=t-t_{N}$ and $T'=t'-t_{N}$. Defining also
the times $T_{nl}=t_{nl}-t_{N}$ and $T_{0}=t_{0}-t_{N}$, Eq.~(\ref{D2sat}),
becomes $|D|^{2}=a^{2}\omega^{-4}n^{\perp-4}|\tilde{D}|^{2}$, where
\begin{equation}
|\tilde{D}|^{2}=\begin{cases}
e^{\Omega_{+}(T+T'-2T_{nl})} & T_{0}<T<T'<T_{nl},\\
e^{\Omega_{+}(T-T_{nl})} & T_{0}<T<T_{nl}<T',\\
1 & T_{nl}<T<T',
\end{cases}\label{eq:D2regiones}
\end{equation}
Note that $\Omega_{+}$, $T_{0}$, and $T_{nl}$ depend on $k=\omega s$. 

We now insert these results into Eq.~(\ref{rhogw}) and change
variables $t\to T$ and $t_{N}\to T'$. With the latter change of
variable, $t'$ disappears almost everywhere. Only a factor $e^{\beta(t'-t_{*})}$
remains from Eq.~(\ref{eq:Itotap}) and another factor $e^{\beta(t'-t_{*})}$
arises from $\Gamma(t_{N})$, which is given by Eq.~(\ref{eq:nuclexp}).
We can readily integrate $t'$ using the change of variable $u=\Gamma_{*}e^{\beta(t'-t_{*})}$.
Using the result in Eq.~(\ref{eq:Delta_adim}), we obtain
\begin{multline}
\Delta^{11}=\frac{\beta^{5}a^{2}v_{w}\omega}{12\pi^{2}}\negthickspace\int_{-1}^{+1}\negthickspace dcc^{2}\negmedspace\int_{T_{0}}^{\infty}\negthickspace dT'T^{\prime}e^{-\beta T'}\\
\int_{T_{0}}^{T'}\negthickspace dTT^{3}\cos(\omega T_{-})e^{i\omega v_{w}T_{-}c}|\tilde{D}|^{2},\label{eq:Delta11}
\end{multline}
where $T_{-}=T'-T$. The integrals with respect to $T$ and $T'$
can be done analytically since we only have exponential functions
and powers of these variables. We do not write the complete expression
which is rather cumbersome. The remaining integral with respect to
$c$ must be calculated numerically. We consider analytic approximations
in App.~\ref{app:Approximations}.

\begin{figure}[h]
\includegraphics[width=1\columnwidth]{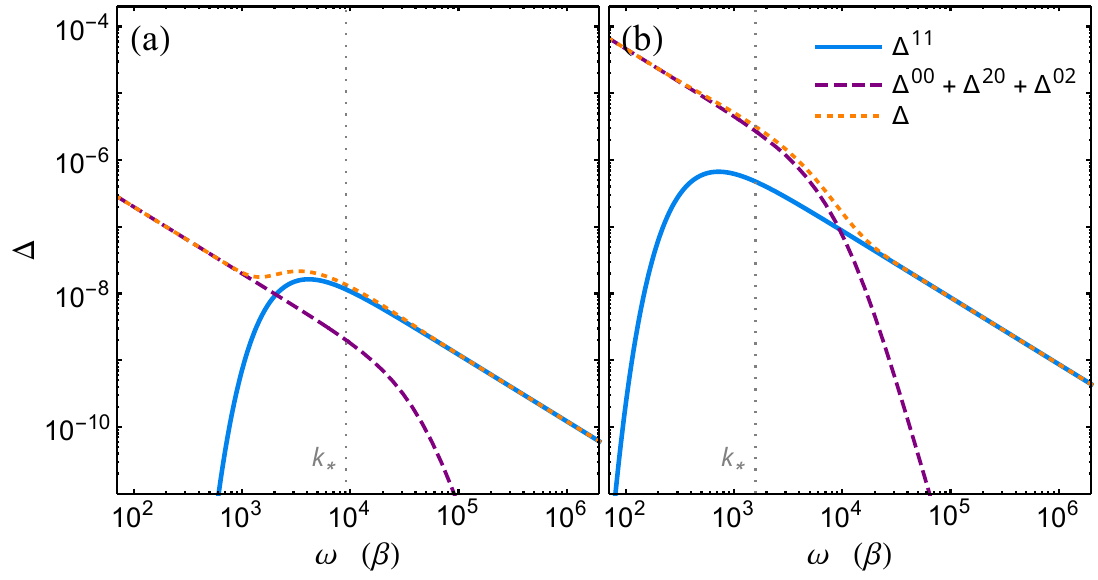}
\caption{Contributions to the GW spectrum for $T=100\mathrm{GeV}$, $\beta/H=100$,
	$\alpha_{T}=0.1$, $a=0.2$, and (a) $v_{c}=0.06$, $v_{w}=0.03$,
	(b) $v_{c}=0.4$, $v_{w}=0.2$.\label{fig:Delta11}}
\end{figure}

The numerical result for $\Delta^{11}$ is shown in Fig.~\ref{fig:Delta11}.
It has a maximum at $\omega\sim k_{*}$. For lower frequencies, it falls
exponentially, while for higher frequencies it falls like $\omega^{-1}$
(see App.~\ref{app:Approximations}). For comparison, we show the
other contributions and the total spectrum. 
The peak of $\Delta^{11}$
dominates for small wall velocities because we have $\Delta^{11}\propto v_{w}$
and $\Delta^{00}\propto v_{w}^{3}$ (see App.~\ref{app:Approximations}).
Above the scale $k_{*}$, $\Delta^{11}$
always dominates since it falls as $\omega^{-1}$ whereas the other
contributions fall as as $\omega^{-6}$. 
Below the scale $k_{*}$,
the other contributions go as $\omega^{-1}$.
Depending on the parameters, the coefficients of the two
power laws $\omega^{-1}$ can differ significantly.

\subsection{Full spectrum}
\label{sec:spectrum}

\begin{figure*}[t]
	\includegraphics[width=\textwidth]{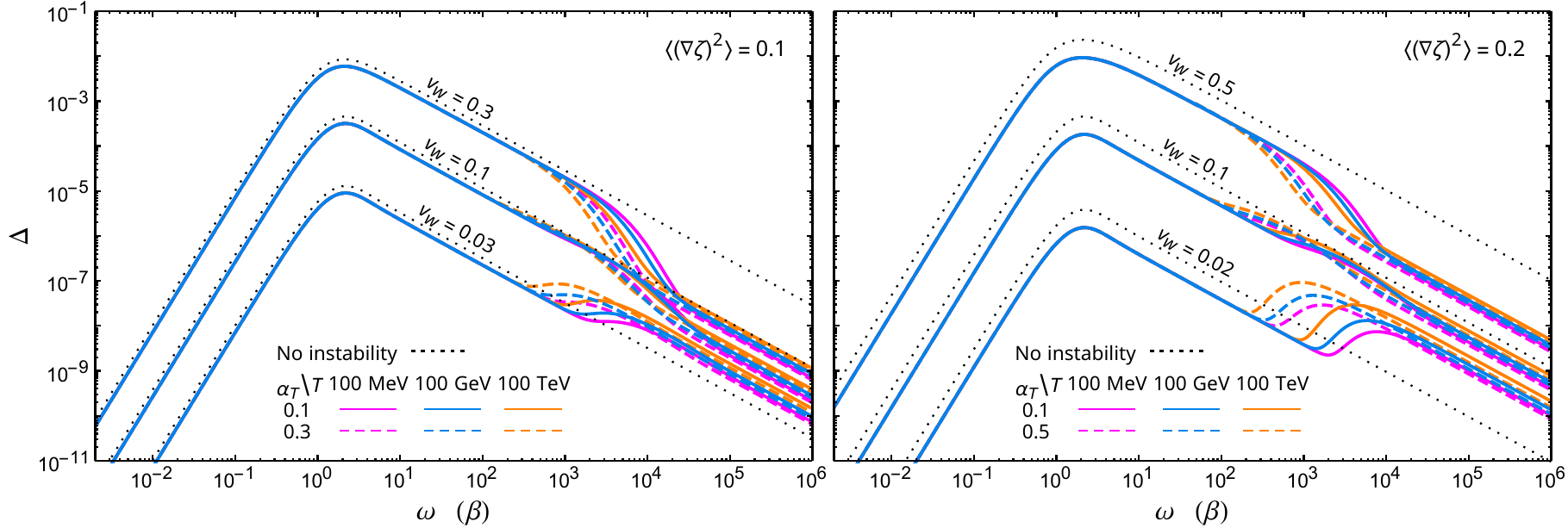}
	\caption{The GW spectrum for $\beta/H=100$ and $v_{c}=2v_{w}$.
		The value of $a$ is chosen so that the value of $\langle(\nabla\zeta)^{2}\rangle$ is as indicated.
		\label{fig:Delta}}
\end{figure*}
Considering all the above contributions, we obtain
\begin{equation}
\Delta=\left[1-3\frac{a^{2}}{2\pi}\log\left(\frac{k_{c}}{k_{*}}\right)\right]\Delta^{00}+\Delta^{11}.\label{eq:Deltatot}
\end{equation}
This quantity is shown in Fig.~\ref{fig:Delta} for various values of the phase transition parameters $T$, $\alpha_{T}$, and $v_w$
and for values of $a$ corresponding to a couple of values of $\langle(\nabla\zeta)^{2}\rangle$.
Some of the parameter values, such as $v_w=0.5$, $\alpha_T=0.5$, and $\langle(\nabla\zeta)^{2}\rangle=0.2$,
are outside the range of validity of our approximations%
\footnote{In particular, for $v_w=0.5$,	 we took $v_c =2v_w=1$. 
	 In this case, Eq.~(\ref{eq:vwvc}) is not valid. 
	 In our approximations, a value $v_c>c_s=1/\sqrt{3}$ represents the case in which every subsonic velocity is unstable, 
	 which occurs for low enough $T_N/T_c$ (see Fig.~11 of Ref.~\cite{mm14stability}).}.
We extrapolated our results to investigate the parameter region where the effect is stronger.
We also consider the case $a=0$, which corresponds to the bubble-collision mechanism without instability.
The latter depends only on $v_w$.
The instability introduces significant changes to the shape of the spectrum at $\omega\sim k_{*}$.
The parameter $k_*$ is roughly given by $k_{*}/\beta\sim (\alpha_{T}v_{w})^{-1}\log(M_{P}/T)$
(see App.~\ref{app:Approximations}). 
Below the scale $k_{*}$, $\Delta$ has the same shape as $\Delta_{\mathrm{coll}}$.
Therefore, this part of the spectrum is indistinguishable from that of the
bubble-collision mechanism.
The peak of the contribution $\Delta^{11}$ is more noticeable at lower wall velocities. 
However, at higher velocities, a departure from the bubble-collision spectrum occurs
due to the change in the power law of $\Delta^{00}$.
The tail $\propto\omega^{-1}$ is due to the contribution $\Delta^{11}$.
This behavior may be due to our abrupt saturation model.

The dynamics of the instability is independent of the nucleation rate. 
However, different nucleation models have different time parameters. 
When defining the characteristic wavenumber $k_*$ in Eq.~(\ref{eq:khatdef}),
we used the timescale $\beta^{-1}$.
For a different time parameter $t_b$,
we can define the characteristic wavenumber $k_{*}^{\prime}=k_{nl}|_{t-t_N=t_{b}}$, which yields
$k_{nl}(t)=k_{*}^{\prime}t_{b}(t-t_N)^{-1}$ and $k_{*}^{\prime}t_{b} = k_{*}\beta^{-1}$.
Similarly, we can define the quantity $\Delta$ in Eq.~(\ref{eq:Delta_adim}) using $t_{b}^{-1}$ instead of $\beta$.
In Ref.~\cite{mm26gravitational}, we consider a nucleation rate of the form
$\Gamma=n_{b}\delta(t-t_{*})$ rather than an exponential rate. 
The parameter $n_{b}$ represents the number density of bubbles nucleating simultaneously at time $t_N = t_{*}$.
The phase transition dynamics is simpler since
the delta function eliminates the
integrals with respect to $t_{N}$ and $t_{N}'$ in Eqs.~(\ref{defI})
and (\ref{eq:Pis})-(\ref{eq:Pid}). 
In this case, the timescale 
is represented by the parameter $t_{b}=R_{b}/v_{w}$, where $R_{b}=\left(4\pi n_{b}/3\right)^{-1/3}$
is the characteristic bubble separation. 
By comparing Fig.~\ref{fig:Delta} with Fig.~1 of Ref.~\cite{mm26gravitational}, we see that the results for the two nucleation models are qualitatively similar.
Below, we consider specific timescales.

\section{Relic gravitational waves}
\label{sec:relic}

The GW spectrum redshifts according to the ratio
\begin{equation}
	a_{*}/a_{0}=(g_{s0}/g_{s*})^{1/3}(T_{0}/T_{*}),\label{eq:scalefactors}
\end{equation}
where $a$ is the scale factor, $g_s$ is the number of entropic degrees of freedom, 
and the subscripts $0$ and $*$ indicate the value of a quantity at present and at the time of GW generation, respectively.
We have $g_{s*}\simeq g_{*}$, $g_{s0}=3.91$, and $T_0$
is related to the energy density of photons by $\rho_{\gamma}=\frac{\pi^{2}}{30}g_{\gamma}T_{0}^{4}$, with $g_\gamma = 2$.

Using Eqs.~(\ref{eq:alphaT}) and (\ref{eq:Friedmann}) in Eq.~(\ref{eq:Delta_adim}), we obtain
\begin{equation}
	\frac{d\rho_{GW}}{d\ln\omega}=\rho_{\mathrm{tot}}\left(\frac{H}{\beta}\right)^{2}\left(\frac{\kappa\alpha_{T}}{1+\alpha_{T}}\right)^{2}
	\Delta.\label{eq:rhoGWast}
\end{equation}
At the end of the phase transition, the total energy
density becomes radiation, $\rho_{\mathrm{tot}}=\frac{\pi^{2}}{30}g_{*}T_{*}^{4}$.
The GW spectrum today is obtained by multiplying Eq.~(\ref{eq:rhoGWast}) by $(a_*/a_0)^4$.
The intensity of the stochastic GW background is characterized by the quantity \cite{maggiore00gravitational}
\begin{equation}
	\Omega_{GW}\equiv\frac{1}{\rho_{c}}\frac{d\rho_{GW}}{d\log f},\label{eq:OmGWtodaydef}
\end{equation}
where $f$ is the frequency and $\rho_{c}$ is the present value of
the critical density.
This quantity is related to the Hubble constant today
by $\rho_{c}=3H_{0}^{2}/(8\pi G)$. 
Using $\rho_{\gamma}/\rho_{c}=2.47\times10^{-5}h^{-2}$ \cite{n+24review},
where $h$ is the dimensionless Hubble parameter, defined by
$H_{0}=100h\mathrm{km}/\mathrm{s}/\mathrm{Mpc}$, we obtain 
\begin{equation}
h^{2}\Omega_{GW}=1.64\times10^{-5}\left(\frac{100}{g_{*}}\right)^{\frac{1}{3}}\left(\frac{H_{*}}{\beta}\right)^{2}
\Big(\frac{\kappa\alpha_{T}}{1+\alpha_{T}}\Big)^{2}\Delta.\label{eq:Omegatoday}
\end{equation}

The efficiency factor $\kappa$ is a function of $\alpha_{T}$ and
$v_{w}$. We use the fit from Ref.~\cite{ekn+10energy},
\begin{equation}
	\kappa=\frac{c_{s}^{11/5}\kappa_{A}\kappa_{B}}{c_{s}^{6/5}\kappa_{A}v_{w}+\kappa_{B}\left(c_{s}^{11/5}-v_{w}^{11/5}\right)},
\end{equation}
where $c_{s}=1/\sqrt{3}$ is the speed of sound and
\begin{align}
	\kappa_{A} & =\frac{6.9\alpha_{T}v_{w}^{6/5}}{\alpha_{T}-0.037\sqrt{\alpha_{T}}+1.36},\\
	\kappa_{B} & =\frac{\alpha_{T}^{2/5}}{(\alpha_{T}+0.997)^{2/5}+0.017}.
\end{align}

The dimensionless quantity $\Delta$ can be expressed as a function
of the dimensionless variable $\omega/\beta$ and the dimensionless
parameter $k_{*}/\beta$. 
The frequency $\omega$ at the time of the phase transition is related to its value today,
$\omega_{0}=2\pi f$, by the factor $a_{*}/a_{0}$.
Using Eqs.~(\ref{eq:Friedmann}) and (\ref{eq:scalefactors}),
with $T_{0}\simeq2.73\mathrm{K}$ and $1\mathrm{eV}\simeq1.5\times10^{15}s^{-1}$,
we obtain
\begin{equation}
\frac{\omega}{\beta}=3.8\times10^{5}\frac{H_{*}}{\beta}\left(\frac{100}{g_{*}}\right)^{1/6}\frac{100\mathrm{GeV}}{T_{*}}
\frac{f}{\mathrm{Hz}}.\label{eq:wtoday}
\end{equation}
 
\begin{figure*}[t]
\includegraphics[width=0.9\textwidth]{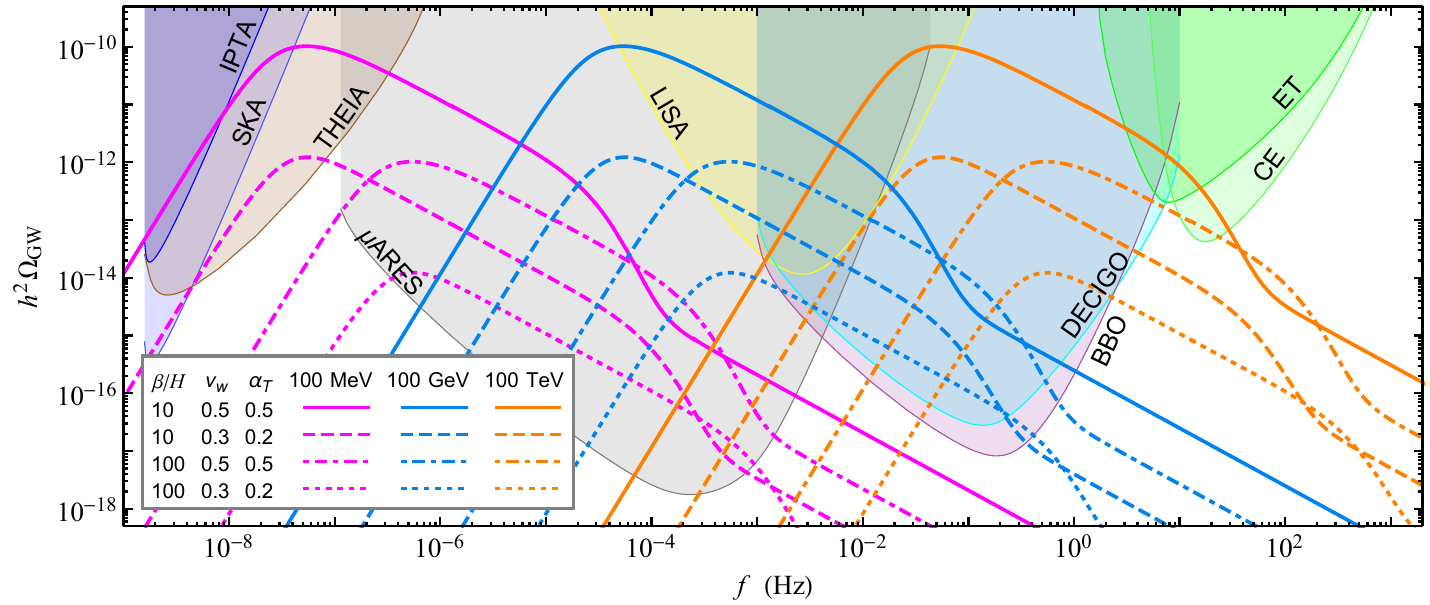}
\caption{\label{fig:spectra}%
	The GW spectrum today for $\langle(\nabla\zeta)^{2}\rangle=0.1$,
	$g_{*}=100$ and $v_{c}=2v_{w}$.
	We include sensitivity curves for future Pulsar Timing Arrays observations, IPTA \cite{haa+10international,manchester13international,vlh+16international}
	and SKA \cite{cr04motivation,jhm+15gravitational,wbc+20fundamental},
	the THEIA \cite{gmw21exploring} proposal, next-generation space-based interferometers $\mu$Ares \cite{s+21unveiling},
	LISA \cite{a+17laser,rcl19construction,c+16science,c+20detecting},
	BBO \cite{cc05lisa,cc06detecting,hfs+06laser} and DECIGO \cite{k+11japanese,ys11detector,inn18multiband},
	and the ground-based interferometers ET \cite{paa+10einstein,m+20science,aaa+26science}
	and CE \cite{aaa+17exploring,r+19cosmic}. 
	Sensitivities are constructed from power-law integrated curves \cite{tr13sensitivity} using signal-to-noise ratios $\mathrm{SNR}$
	and observation times $T_{\mathrm{obs}}$ in years.
	We use $\mathrm{SNR}=1$ and $T_{\mathrm{obs}}=20$ for SKA, IPTA and THEIA;
	$\mathrm{SNR}=1$ and $T_{\mathrm{obs}}=10$ for $\mu$Ares, LISA, BBO and DECIGO;
	$\mathrm{SNR}=1$ and $T_{\mathrm{obs}}=1$ for CE and ET \cite{schmitz21new}.}
\end{figure*}
Figure \ref{fig:spectra} shows the GW spectrum for phase transitions occurring at temperatures between
$100\mathrm{MeV}$ and $100\mathrm{TeV}$, for different sets of phase transition parameters.
We consider $g_{*}=100$ and $v_{c}=2v_{w}$ in all cases,
since the spectrum has a mild dependence on these parameters. 
We use conservative values of the parameter $a$, such that $\langle(\nabla\zeta)^{2}\rangle=0.1$.
We also show the sensitivity curves for future GW experiments.
For all temperatures in this range, the physics at scale $k_{*}$
is potentially observable by some of these detectors.

\begin{figure*}
\includegraphics[width=0.9\textwidth]{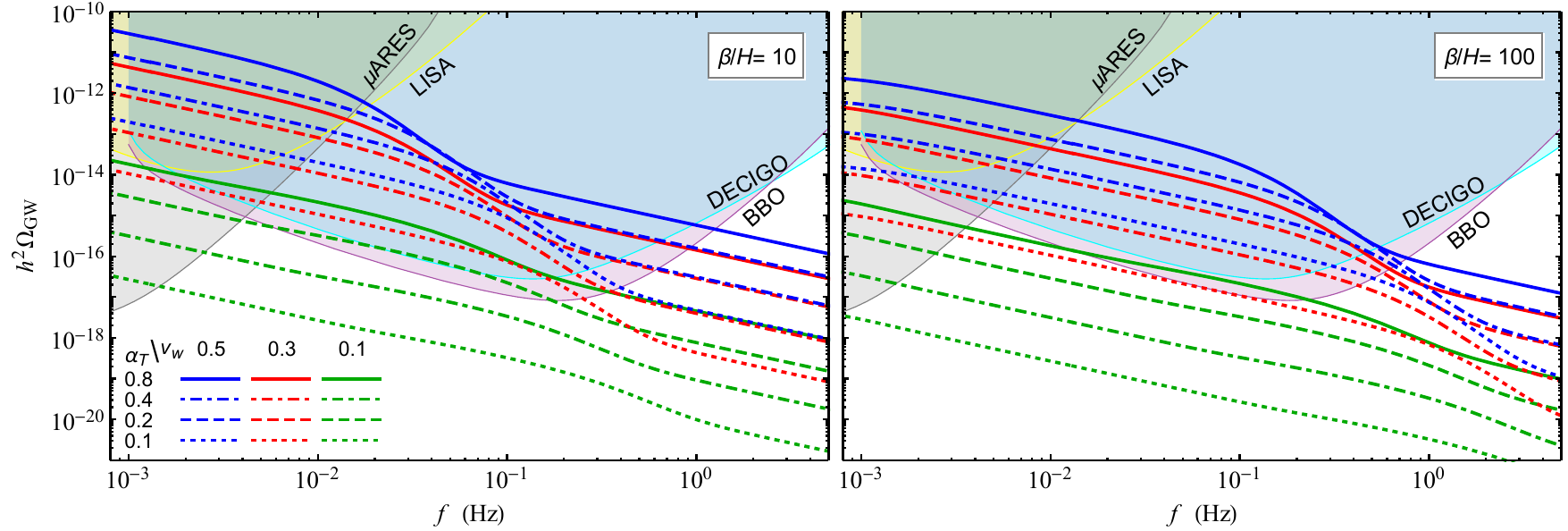}
\caption{The GW spectrum for $T=100\mathrm{GeV}$, $g_{*}=100$, $\langle(\nabla\zeta)^{2}\rangle=0.1$,
	and $v_{c}=2v_{w}$.\label{fig:spectraew}}
\end{figure*}

Figure \ref{fig:spectraew} provides a closer look at a phase transition at the electroweak scale.
Depending on the phase transition parameters,
the knee in the spectrum and the small peak at
$\omega\sim k_{*}$ may be detectable by BBO, DECIGO, and even LISA.
We will discuss specific physical models elsewhere.

As already mentioned, in Ref.~\cite{mm26gravitational} we consider the electroweak phase transition using a simultaneous nucleation model.
To compare the results, we can consider a physical measure of the phase transition duration,
such as the time $\Delta t$ between two values of the fraction of volume, 
$f_{+}= e^{-I_{1}}$ and $f_{+}= e^{-I_{2}}$ \cite{mm21model}. 
For the delta-function rate we
obtain $\Delta t=(I_{2}^{1/3}-I_{1}^{1/3})t_{b}$, while for
the exponential rate we obtain $\Delta t=(\log I_{2}-\log I_{1})\beta^{-1}$.
The values of $\beta$ and $t_{b}$ that give the same
$\Delta t$ are related by 
\begin{equation}
	\beta t_{b}=\frac{\log I_{2}-\log I_{1}}{I_{2}^{1/3}-I_{1}^{1/3}}.\label{eq:betatb}
\end{equation}
For $e^{-I_{1}}=0.99$ and $e^{-I_{2}} =0.01$, we have $\beta t_b \simeq4.23$.
By comparing Fig.~\ref{fig:spectraew} with Fig.~2 of \cite{mm26gravitational}, we see that
the results are qualitatively and quantitatively similar for similar parameter values.

\begin{figure*}
\includegraphics[width=0.9\textwidth]{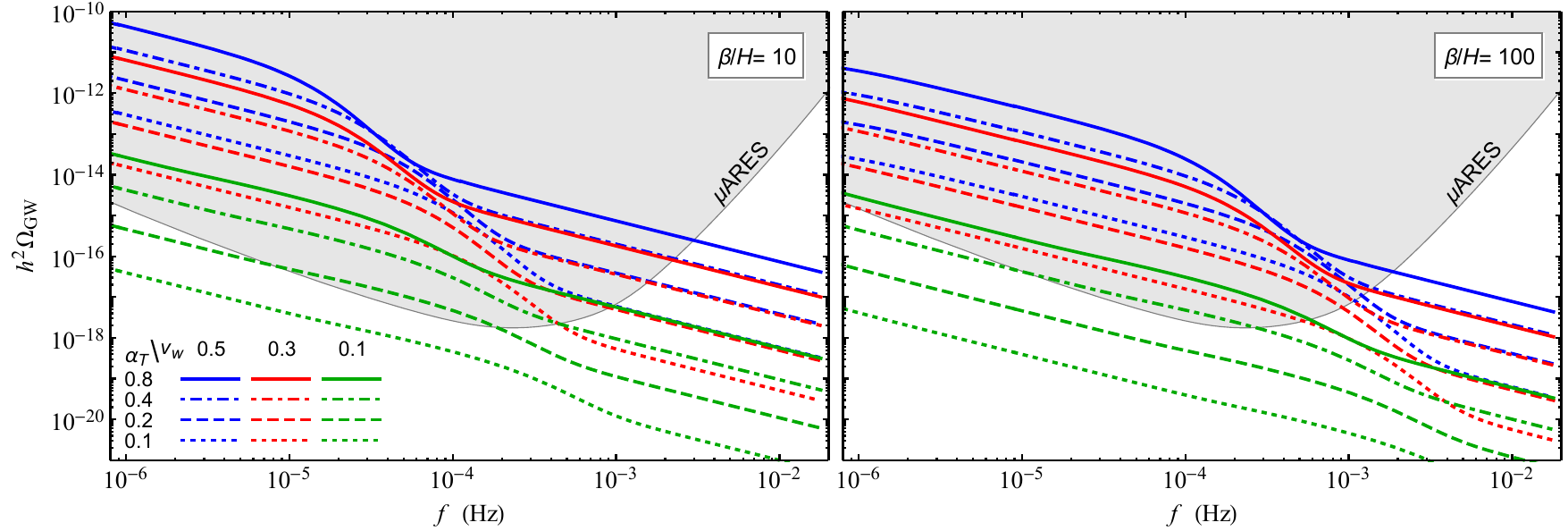}
\caption{The GW spectrum for $T=100\mathrm{MeV}$, $g_{*}=10$, $\langle(\nabla\zeta)^{2}\rangle=0.1$,
	and $v_{c}=2v_{w}$. \label{fig:spectraQCD}}
\end{figure*}

In Fig~\ref{fig:spectraQCD}, we consider a phase transition at the QCD scale in more detail.
Depending on the parameters, the knee in the spectrum and the peak at $\omega\sim k_{*}$ are in
principle detectable by $\mu$ARES.

\section{Conclusions}

\label{sec:conclu}

We have calculated the gravitational radiation produced by 
the exponential growth of perturbations
caused by hydrodynamic instability during a first-order cosmological phase transition.
We focused on the bubble walls and thin fluid shells where the perturbations grow exponentially.
These perturbations also seed magnetohydrodynamic turbulence and long-lasting sound waves.
We will study these sources of gravitational waves elsewhere.

We applied the results from the instability in planar walls,
which should provide a good description since the unstable wavelengths are much smaller than the bubble radius.
We assumed that the growth of the unstable modes ceases once they become nonlinear.  
We introduced a free parameter $a$ through the condition that
the average gradient of the wall displacement $\zeta$ on a length scale $k^{-1}$ is bounded by $a$. 
Although $a$ is in principle a parameter of order 1, our perturbative treatment of stochastic deformations
breaks down as $\langle(\nabla\zeta)^{2}\rangle$ approaches the value $1/3$.

The characteristic scale of the instability is given by the critical wavenumber $k_{c}$.
However, the development of the instability at different length scales depends on the evolution of the phase transition.
The characteristic scale for gravity wave production is given by $k_*\sim\beta (\alpha_{T}v_{w} )^{-1}\log(M_{P}/T)$,
where $\beta$ is the nucleation rate parameter,
$\alpha_{T}$ is the ratio of latent heat to radiation energy density,
$v_{w}$ is the wall velocity,
 $T$ is the temperature, and $M_P$ is the Planck mass.
Typically, $k_*$ is a few orders of magnitude higher than $\beta$.

The GW spectrum has a peak at a frequency $\omega\sim k_*$, which is more noticeable for
wall velocities $v_{w}\lesssim0.1$. 
In any case, the spectrum decreases as $\omega^{-6}$ around $\omega = k_*$, which
distinguishes it from the bubble-collision spectrum. 
These features may be detectable by future GW observatories for a phase transition at the electroweak or the QCD scale.

\section*{Acknowledgments}

This work was supported by UNMdP grant 80020250500055MP.

\appendix

\section{Correlations on the wall surface}

\label{app:corrfunc}

\subsection{The stochastic problem}

The stochastic differential equation for mode $\mathbf{k}$, Eq.~(\ref{oscarmforzado}),
is essentially the same as the equation for a complex driven harmonic oscillator, 
\begin{equation}
m\ddot{x}+\eta\dot{x}+m\omega^{2}x=y. \label{eq:drivenoscarm}
\end{equation}
In this case, the random force satisfies
\begin{equation}
\langle y\rangle=0,\quad\langle y(t)y^{*}(t')\rangle=A\delta(t-t').\label{corrfza}
\end{equation}
Indeed, we obtain Eqs.~(\ref{oscarmforzado})-(\ref{propsfza}) from Eqs.~(\ref{eq:drivenoscarm})-(\ref{corrfza}) 
by substituting
$x$ for $D_{\mathbf{k}}$, $y$ for $y_{\mathbf{k}}$, and
\begin{equation}
m\to\sigma_{0},\:\eta\to\eta_{k},\:\omega\to\omega_{k},\:A\to A(2\pi)^{2}\delta(\mathbf{k}-\mathbf{k}').\label{replacements}
\end{equation}
Therefore, we can solve this one-dimensional problem and make the replacements at the end of the calculation.

The general solution of the homogeneous equation is $x=ae^{\Omega_{+}t}+be^{\Omega_{-}t}$,
with 
\begin{equation}
\Omega_{\pm}=-\eta/2m\pm\sqrt{\eta^{2}/4m^{2}-\omega^{2}}.
\end{equation}
For $\omega^{2}>\eta^{2}/4m^{2}$, we have underdamped oscillations.
For $0<\omega^{2}<\eta^{2}/4m^{2}$ we have overdamped motion. For
$\omega^{2}<0$, we have $\Omega_{+}>0$, which corresponds to the
unstable case. The solution of the non-homogeneous equation can be
obtained, e.g., using the method of variation of constants. For the
initial condition $x(t_{0})=0$, we have 
\begin{equation}
x(t)=\Theta(t-t_{0})\int_{t_{0}}^{t}\frac{e^{\Omega_{+}(t-t')}-e^{\Omega_{-}(t-t')}}{m(\Omega_{+}-\Omega_{-})}y(t')dt'.\label{soloa}
\end{equation}

The correlation function for $x$ readily follows from the expression
(\ref{soloa}) and the property (\ref{corrfza}). Defining the variables
$u=t-t_{0}$, $u'=t'-t_{0}$, and $\tau=t-t'$, we obtain
\begin{multline}
\langle x(t)x(t')^{*}\rangle=\Theta(u)\Theta(u')\frac{A/m^{2}}{|\Omega_{+}-\Omega_{-}|^{2}}\times
\\
\left[
\frac{e^{\Omega_{+}u+\Omega_{+}^{*}u'}}{2\mathrm{Re}(\Omega_{+})}
-\frac{e^{\Omega_{+}u+\Omega_{-}^{*}u'}}{\Omega_{+}+\Omega_{-}^{*}}-\frac{e^{\Omega_{-}u+\Omega_{+}^{*}u'}}{\Omega_{-}+\Omega_{+}^{*}}+\frac{e^{\Omega_{-}u+\Omega_{-}^{*}u'}}{2\mathrm{Re}(\Omega_{-})}
\right.
\\
+\left.\frac{(\Omega_{+}-\Omega_{-})e^{\Omega_{+}^{*}|\tau|}}{(\Omega_{-}+\Omega_{+}^{*})2\mathrm{Re}(\Omega_{+})}+\frac{(\Omega_{-}-\Omega_{+})e^{\Omega_{-}^{*}|\tau|}}{(\Omega_{+}+\Omega_{-}^{*})2\mathrm{Re}(\Omega_{-})}\right] .\label{eq:correloageneral}
\end{multline}

For $\omega^{2}>0$ (stable case), we can take the limit $t_{0}\to-\infty$
since both $\mathrm{Re}(\Omega_{-})$ and $\mathrm{Re}(\Omega_{+})$
are negative, and the first four terms in Eq.~(\ref{eq:correloageneral})
vanish. In particular, for $\omega^{2}>\eta^{2}/4m^{2}$, we obtain
\begin{equation}
\langle x(t)x(t')^{*}\rangle
= \frac{Ae^{-\frac{\eta}{2m}|\tau|}}{2\eta m\omega^{2}}
\Big[\cos(\tilde{\omega}|\tau|)+\frac{\eta \sin(\tilde{\omega}|\tau|)}{2m\tilde{\omega}}\Big].
\label{corroatinf}
\end{equation}
where $\tilde{\omega}^{2}=\omega^{2}-\eta^{2}/4m^{2}$. This is the
result that can be found in standard textbooks \cite{CoffeyKalmykovWaldron04Langevin}.
At equal times, we have $\langle|x(t)|^{2}\rangle=A/2\eta m\omega^{2}$.
This means that the quantity $\langle|x|^{2}\rangle$ reaches a constant
value at long times. If the random force is due to thermal fluctuations,
this value must be given by the equipartition theorem\footnote{With the usual definition $x=(x_{1}+ix_{2})/\sqrt{2}$, the potential
energy is $m\omega^{2}|x|^{2}=\frac{1}{2}m\omega^{2}(x_{1}^{2}+x_{2}^{2})$,
and each of these real degrees of freedom contributes $T/2$ to the
average potential energy.}, 
\begin{equation}
m\omega^{2}\langle|x|^{2}\rangle=T,\label{eqtimeoa}
\end{equation}
which implies $A=2\eta T$. This relation between the size of the
fluctuations and the damping is the simplest form of the fluctuation-dissipation
theorem.

For $\omega^{2}<0$ (unstable case), the limit $t-t_{0}\to\infty$
cannot be taken since we have $\Omega_{+}>0$, and the thermal-equilibrium
value (\ref{eqtimeoa}) is never reached. Nevertheless, we can assume
that the random force is still given by
$A=2\eta T$. Since we have $|\Omega_{-}|\gg\Omega_{+}$, the terms
with $e^{\Omega_{-}u}$, $e^{\Omega_{-}u'}$, and $e^{\Omega_{-}|\tau|}$
decay quickly, and we can keep only the exponentially growing terms.
Thus, Eq.~(\ref{eq:correloageneral}) becomes
\begin{multline}
\langle x(t)x(t')^{*}\rangle=\Theta(t-t_{0})\Theta(t'-t_{0})\\
\times\frac{T}{2m\tilde{\Omega}^{2}}\left[\frac{e^{\Omega_{+}(t+t'-2t_{0})}}{2\Omega_{+}m/\eta}-\frac{\tilde{\Omega}e^{\Omega_{+}|t-t'|}}{\Omega_{+}}\right].\label{corroainest}
\end{multline}
where $\tilde{\Omega}^{2}=\eta^{2}/4m^{2}-\omega^{2}$ (i.e., $\tilde{\Omega}=i\tilde{\omega}$). 

Now, we apply the replacements (\ref{replacements}) to obtain the correlator
for the Fourier modes on the interface. 

For the stable case, Eq.~(\ref{corroatinf}), we have, at equal times,
\begin{equation}
	\left\langle D_{\mathbf{k}}(t)D_{\mathbf{k}'}(t)^{*}\right\rangle \equiv(2\pi)^{2}\delta^{(2)}(\mathbf{k}-\mathbf{k}')T/(\sigma_{0}\omega_{k}^{2}).\label{eq:equipcont}
\end{equation}
If we want to obtain the equipartition condition for a given mode
$\mathbf{k}$, Eq.~(\ref{eq:equipcont}) has an infrared divergence
because there is a continuum of modes. If the surface has finite
area $A$, the modes are discrete, the integral in the Fourier expansion
(\ref{desazeta}) becomes a sum, $\int\frac{d^{2}k}{(2\pi)^{2}}=\frac{1}{A}\sum_{\mathbf{k}}$,
the delta function becomes a Kronecker delta, $(2\pi)^{2}\delta^{(2)}(\mathbf{k}-\mathbf{k}')=A\delta_{\mathbf{k},\mathbf{k}^{\prime}}$,
and we obtain $\sigma_{0}\omega_{k}^{2}\langle|D_{\mathbf{k}}|^{2}\rangle=TA$.
If we neglect the fluid effects in Eq.~(\ref{oscarmestable}), we
have $\omega_{k}^{2}=k^{2}$ and we obtain the standard result for
the fluctuations of a surface \cite{Safran94Statistical}. 
The continuum analogue of Eq.~(\ref{eqtimeoa}) can be expressed
in terms of the quantity $|D|^{2}(k,t,t')$ defined in Eq.~(\ref{defD2}):
$\sigma_{0}\omega_{k}^{2}|D|^{2}(k,t,t)=T$. 

For the unstable case, Eq.~(\ref{corroainest}), we obtain
\begin{multline}
\left\langle D_{\mathbf{k}}(t)D_{\mathbf{k}'}(t')^{*}\right\rangle =(2\pi)^{2}\delta^{(2)}(\mathbf{k}-\mathbf{k}')\Theta(t_{m}-t_{0})\\
\times\frac{T\eta_{k}/\sigma_{0}^{2}}{(2\tilde{\Omega})^{2}\Omega_{+}}\left[e^{\Omega_{+}(t+t'-2t_{0})}-\frac{2\tilde{\Omega}_{k}}{\eta_{k}/\sigma_{0}}e^{\Omega_{+}|t-t'|}\right],\label{eq:D2ap0}
\end{multline}
where $t_{m}=\min\{t,t'\}$, $\Omega_{\pm}=-\eta_{k}/2\sigma_{0}\pm\sqrt{\eta_{k}^{2}/4\sigma_{0}^{2}-\omega_{k}^{2}}$,
and $\tilde{\Omega}_{k}^{2}=\eta_{k}^{2}/4\sigma_{0}^{2}-\omega_{k}^{2}$.
Inverting these relations and using the definition of $|D|^{2}(k,t,t')$,
Eq.~(\ref{defD2}), we obtain Eq.~(\ref{D2}).

\subsection{Spacetime correlations}

Let us now consider the correlation function (\ref{eq:zetazetap}),
\begin{equation}
\langle\zeta\zeta'\rangle(x,t,t')=\frac{1}{2\pi}\int_{k_{-}}^{k_{+}}dk\,k\,|D|^{2}J_{0}(kx),\label{zetazetapspace}
\end{equation}
where $J_{0}$ is the 0th Bessel function of the first kind, $J_{0}(x)=\frac{1}{2\pi}\int_{0}^{2\pi}d\theta e^{ix\cos\theta}$.
For $t\neq t'$, the integral (\ref{zetazetapspace}) separates into
three regions according to Eq.~(\ref{D2final}). For $x=0$, we have
$J_{0}(0)=1$ and the integral in Eq.~(\ref{zetazetapspace}) can
be obtained analytically. For $x\neq0$, there is no simple analytic
approximation. A semi-analytic approximation can be obtained by neglecting
the unsaturated modes, which are exponentially suppressed, and integrating
from $k=k_{nl}$ to $k=\infty$. We obtain
\begin{multline}
\langle\zeta\zeta'\rangle=\frac{a^{2}}{4\pi k_{nl}^{2}}\left\{ 1+\frac{1}{2}\left[\gamma-1+\log(k_{nl}x/2)\right]x^{2}\right.\\
\left.-\frac{1}{4}(k_{nl}x/2)^{4}\,_{2}F_{3}\left(1,1;2,3,3;-(k_{nl}x/2)^{2}\right)\right\} ,\label{zzpsat}
\end{multline}
where $\gamma$ is Euler's constant and $_{p}F_{q}$ is the generalized
hypergeometric function.

\begin{figure}[h]
\includegraphics[width=0.8\columnwidth]{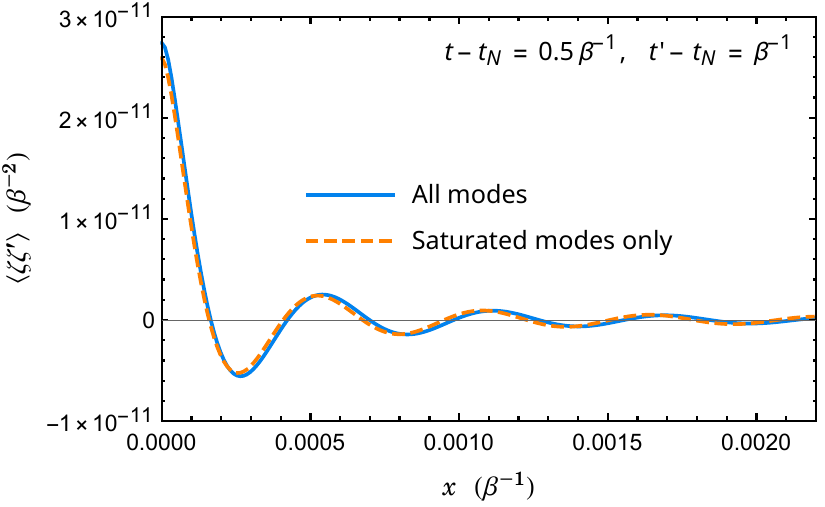}
\caption{The correlation function as a function of the point separation at fixed times.
	The parameters are $T=100\mathrm{GeV}$, $\beta/H=100$, $\alpha_{T}=0.05$, $v_{c}=0.2$, $v_{w}=0.1$, and $a=0.2$.}
\label{figcorrel}
\end{figure}

In Fig.~\ref{figcorrel} we consider the variation of $\langle\zeta\zeta'\rangle$
with the point separation $x$ at fixed times $t,t'$. We see that
the correlation becomes small at $ x \sim k_*^{-1} \ll\beta^{-1}$.
The approximation (\ref{zzpsat}) is shown in dashed lines. 
For other values of $t$ and $t'$ the curve is similar. 
The approximation improves for smaller time differences.

\begin{figure}[h]
\includegraphics[width=0.8\columnwidth]{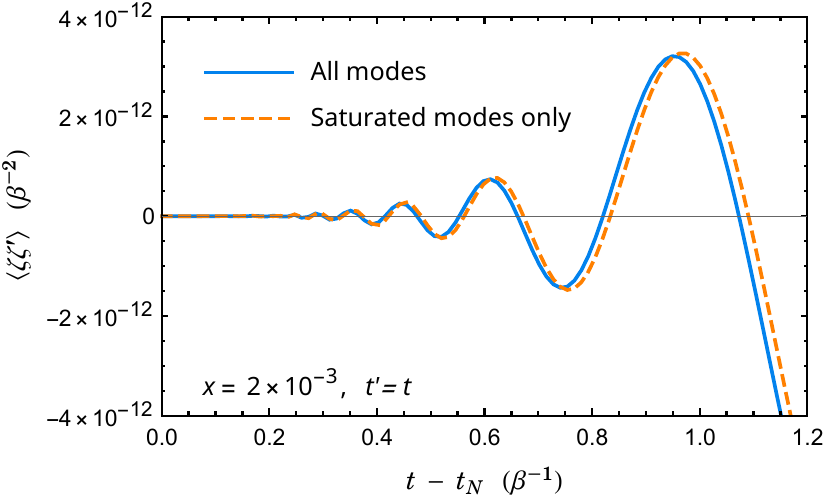}
\caption{Like Fig.~\ref{figcorrel}, but as a function of time for fixed $x$ and equal times.}
\label{figcorrel-1}
\end{figure}

In Fig.~\ref{figcorrel-1}, we consider the equal-time correlation
as a function of time at a given point separation. 
We see that, initially, the
points are not correlated. The correlation increases with time. 

\begin{figure}[h]
\includegraphics[width=0.8\columnwidth]{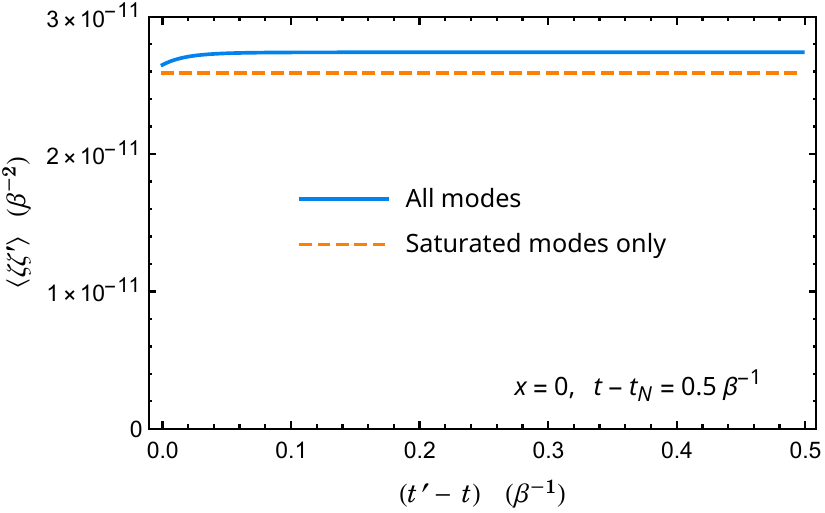}
\caption{Like the previous figures, but as a function of the time difference for $x=0$.}
\label{figcorrel-1-1}
\end{figure}

In Fig.~\ref{figcorrel-1-1} we consider the correlation function
at the same point as a function of the time difference. This quantity
is approximately constant and given by the equal-time value $\langle\zeta^{2}\rangle$
at the smaller time. 

\section{Approximations and fits}

\label{app:Approximations}

In this appendix we discuss various analytical approximations and
fits to the GW spectrum.

\subsection{The instability scale}

The relevant scale in the dynamics of the instability is given by the parameter
$k_{*}$, defined in Eq.~(\ref{eq:khatdef}).
According to Eq.~(\ref{eq:kssol}), we have
\begin{align}
\frac{k_{*}}{\beta} &= \frac{3}{2}\hat{\Omega}^{-1}W\left(\frac{2}{3}a^{2/3}e^{2/3}\hat{\Omega}\hat{D}_{0}^{-2/3}\beta^{-1}\right)
\label{eq:khat}
\\
 &\sim \frac{1}{\alpha_{T}v_{w}}W\left(\frac{a^{2/3}\alpha_{T}^{5/3}v_{c}^{2/3}v_{w}}{g_{*}^{1/6}(\beta/H)}\frac{M_{P}}{T}\right).
\end{align}
The argument of the Lambert function $W$ is of the order of $M_{P}/T$, which is usually very
large. In this case, an approximation can be obtained by using recursively
the equation $W(x)=\log(x)-\log(W(x))$. In particular, taking $W(x)\simeq\log(x)-\log(\log(x))$,
we have an error of 3\% at $x=10^{4}$ and below 1\% for $x\gtrsim10^{8}$.
For a rough estimation, we can retain only the first term, and we
obtain 
\begin{equation}
k_{*}/\beta\sim(\alpha_{T}v_{w})^{-1}\log(M_{P}/T).
\end{equation}
For $\log(M_{P}/T)\sim10$, $\alpha_{T}\lesssim0.1$, and $v_{w}\lesssim0.1$,
we have $k_{*}/\beta\gtrsim10^{3}$.

\subsection{The contribution $\Delta^{00}$}

We consider approximations of $\Delta^{00}$ in different frequency
ranges. 
The approximations for $\omega\ll k_{*}$
and 
$\omega\gg\beta$
are both valid in
the intermediate range $\beta\ll\omega\ll k_{*}$.

\subsubsection{Frequencies $\omega\gg\beta$}

For large $\omega$, most of the functions in Eqs.~(\ref{Delta00sexp})-(\ref{Delta00dexp})
have a very slow variation in comparison with $\cos(\omega t_{-})$
and $j_{n}(\omega v_{w}t_{s})$. Therefore, we can use the change
of variables $t_{-}=\omega^{-1}y$, $t_{s}=\omega^{-1}x$ and expand
the integrand in powers of $\omega^{-1}$ \cite{mm21model}. 
Keeping up to order $\omega^{-1}$ and using the change of variable
$y\to z=y/x$, we obtain
\begin{align}
\Delta_{s,HF}^{00}= & \,\frac{2v_{w}^{3}}{\pi}\negthickspace\int_{0}^{\infty}\negthickspace dxx^{2}\sum_{n=0}^{2}\frac{j_{n}(v_{w}x)}{(v_{w}x)^{n}}\negthickspace\int_{0}^{1}\negthickspace dz\cos(xz)\nonumber \\
 & \times p_{n}(z) \, g_{s} \Big(\frac{\omega\alpha}{2\beta} z \Big) \left[1-\frac{x}{4}(1+z^{2})\frac{\beta}{\omega}\right],
 \label{eq:Delta00HFs}
 \\
\Delta_{d,HF}^{00}= & -\frac{\beta}{\omega}\frac{3v_{w}^{3}}{2\pi}\negthickspace\int_{0}^{\infty}\negthickspace dxx^{3} \frac{j_{2}(v_{w}x)}{(v_{w}x)^{2}}\nonumber \\
 & \times\int_{0}^{1}\negthickspace dz\cos(xz)q(z) \, g_{d} \Big(\frac{\omega\alpha}{2\beta}z \Big),
 \label{eq:Delta00HFd}
\end{align}
where 
$p_{0}=(1-z^{2})^{2}$, $p_{1}=-2(1-z^{2})(1-5z^{2})$, $p_{2}=3-30z^{2}+35z^{4}$,
$q=(1-z^{2})^{2}z^{2}$, 
and
\begin{align}
 & g_{s}(z)=\frac{1}{768}\negmedspace\int_{0}^{\infty}\negthickspace d\tau\tau^{4}e^{-z^{2}\tau^{2}-\frac{1}{2}\tau}\label{eq:gs}\\
 & =\frac{\sqrt{\pi}[48(4z^{4}+z^{2})+1][1-\text{erf}({\textstyle \frac{1}{4z}})]e^{\frac{1}{16z^{2}}}-4(40z^{3}+z)}{768\times512z^{9}},\nonumber \\
 & g_{d}(z)=\left[\frac{1}{96}\int_{0}^{\infty}d\tau\tau^{3}e^{-\frac{1}{2}z^{2}\tau^{2}-\frac{1}{2}\tau}\right]^{2}\label{eq:gd}\\
 & =\frac{\left\{ \sqrt{2\pi}(12z^{2}+1)[1-\text{erf}({\textstyle \frac{1}{2\sqrt{2}z}})]e^{\frac{1}{8z^{2}}}-4(8z^{3}+z)\right\} ^{2}}{96^{2}16^{2}z^{14}}.\nonumber 
\end{align}
The factors in front of the integrals are such that $g_{s}(0)=g_{d}(0)=1$.
In Eqs.~(\ref{eq:Delta00HFs})-(\ref{eq:Delta00HFd}), we have neglected terms proportional to the small constant $\alpha\sim a\beta/k_{*}$.

For $a = 0$, the term of order $(\beta/\omega)^{0}$ in Eq.~(\ref{eq:Delta00HFs})
vanishes upon integration, and we obtain the analytic result
\begin{equation}
\Delta_{HF}^{\mathrm{coll}}=\frac{1}{96\pi}\left[A_{s}(v_{w})+3A_{d}(v_{w})\right]\frac{\beta}{\omega},\label{eq:DeltacollHFexp}
\end{equation}
where
\begin{align}
A_{s} & =2\frac{3-11v_{w}^{2}+69v_{w}^{4}-45v_{w}^{6}}{v_{w}}\nonumber \\
 & + 3 (1-v_{w}^{2})^{2}\frac{15v_{w}^{4}+2v_{w}^{2}-1}{v_{w}^{2}}\log\Big(\frac{1+v_{w}}{1-v_{w}}\Big),\label{eq:As}\\
A_{d} & =(1-v_{w}^{2})\Big[2\frac{15v_{w}^{4}-4v_{w}^{2}-3}{v_{w}} \nonumber \\
 &  {} + 3\frac{1+v_{w}^{2}+3v_{w}^{4}-5v_{w}^{6}}{v_{w}^{2}}\log\Big(\frac{1+v_{w}}{1-v_{w}}\Big)\Big].\label{eq:Ad}
\end{align}

It can be shown that the term of order $(\beta/\omega)^{0}$ in Eq.~(\ref{eq:Delta00HFs})
always vanishes, not only when $a=0$.
This is because the angular integral in Eq.~(\ref{eq:Pisnm}) vanishes for the
contribution $\Pi_{s}^{00}$ if we remove the factor $e^{-J}$ (in
which case the integral becomes quite simpler). 
At order $(\beta/\omega)^0$, we have, precisely, $e^{-J}=1$.

For $a \neq 0$, the integrals with respect to $z$ in Eqs.~(\ref{eq:Delta00HFs})
and (\ref{eq:Delta00HFd}) are of the form $\int_{0}^{1}dzf(z)g(\frac{\omega\alpha}{2\beta}z)$,
where $g=g_{s}$ or $g_{d}$ and $f(z)$ is the rest of the integrand.
Integrating by parts, we obtain
\begin{equation}
F(z) g\Big(\frac{\omega\alpha}{2\beta} z \Big)\Big|_{0}^{1}
-\frac{\omega\alpha}{2\beta} \int_{0}^{1}dzF(z)g^{\prime}\Big( \frac{\omega\alpha}{2\beta}z \Big),\label{eq:intxpartfinal}
\end{equation}
where $F(z)=\int_{0}^{z}f(\zeta)d\zeta$. 
Since $g^{\prime}(\frac{\omega\alpha}{2\beta} z)$ is proportional
to $\frac{\omega\alpha}{2\beta}$
[as can be seen from Eqs.~(\ref{eq:gs}) and (\ref{eq:gd})],
the second term in Eq.~(\ref{eq:intxpartfinal})
is proportional to $(\frac{\omega\alpha}{2\beta})^{2} \sim (16\pi)^{-1} a^{2} (\omega/k_{*})^{2} $.
Therefore, we neglect this term. 
As already discussed, this approximation breaks down
for $\omega\gg k_{*}$, i.e., beyond the relevant frequency range.
Thus, we obtain
\begin{equation}
\Delta_{HF}^{00} = \frac{1}{96\pi}\Big[ A_{s} \, g_{s} \Big(\frac{\alpha}{2}\frac{\omega}{\beta}\big) +
3A_{d} \, g_{d} \Big(\frac{\alpha}{2}\frac{\omega}{\beta}\Big) \Big] \frac{\beta}{\omega}.\label{Delta00HFexp}
\end{equation}
For $\omega\gg k_{*}$, we have 
\begin{equation}
\Delta^{00}=A_{HF}^{00}(\beta/\omega)^{6},\label{eq:D00HF}
\end{equation}
with $A_{HF}^{00}=(6144\sqrt{\pi})^{-1}\alpha^{-5}A_{s}(v_{w})$.

\subsubsection{Frequencies $\omega\ll k_{*}$ ($\Delta^{00}\simeq\Delta_{\mathrm{coll}}$)}

For $\omega\ll k_{*}$, we have $G_{s}\simeq G\simeq G'\simeq 1$, and Eqs.~(\ref{Delta00sexp})-(\ref{Delta00dexp}) yield
$\Delta^{00}\simeq\Delta_{\mathrm{coll}}$. 
Different approximations and fits to $\Delta_{\mathrm{coll}}$ can be found in the literature.
Below we provide an interpolation between the low-frequency and the
high-frequency asymptotic behaviors. The advantage of this approximation
is that it has good precision in the high-frequency range which we
are interested in. 

\paragraph{Case $\omega/\beta\ll1$.}

For very small frequencies, we have $\cos(\omega t_{-})\simeq1$ and
\cite{MiltonAbramowitz70Handbook} 
\begin{equation}
\frac{j_{n}(\omega v_{w}t_{s})}{(\omega v_{w}t_{s})^{n}}\simeq\frac{1}{1\cdot3\cdot5\cdots(2n+1)},
\end{equation}
and Eqs.~(\ref{Delta00sexp})-(\ref{Delta00dexp}) yield (see \cite{mm21model}
for details) 
\begin{equation}
\Delta^{00}\simeq\Delta_{\mathrm{coll}}\simeq A_{LF}^{\mathrm{coll}}(\omega/\beta)^{3},
\end{equation}
with $A_{LF}^{\mathrm{coll}}\simeq0.382v_{w}^{3}$.

\paragraph{Case $\omega/\beta\gg1$.}

The approximation (\ref{Delta00HFexp}) is valid for $\omega\gg\beta$,
with an error of order $\beta/\omega$. In particular, in the range
$\beta\ll\omega\ll k_{*}$, Eq.~(\ref{Delta00HFexp}) becomes Eq.~(\ref{eq:DeltacollHFexp}),
with an error of order $\frac{1}{4\pi}a^{2}(\omega/k_{*})^{2}$, and
we have 
\begin{equation}
\Delta^{00}\simeq\Delta_{\mathrm{coll}}\simeq A_{HF}^{\mathrm{coll}}\beta\omega^{-1},\label{eq:DcollHF}
\end{equation}
with $A_{HF}^{\mathrm{coll}}=\frac{1}{96\pi}(A_{s}+A_{d})$.

\subsubsection{Frequencies $\omega\sim k_{*}$}

\begin{figure}[h]
\includegraphics[width=0.8\columnwidth]{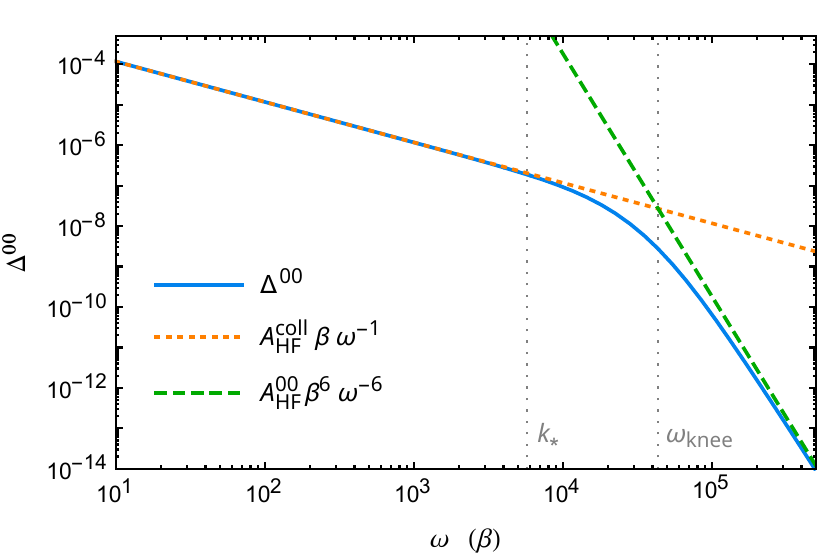}
\caption{Asymptotes of $\Delta^{00}$ for $T=100\mathrm{GeV}$, $\beta/H=100$,
	$\alpha_{T}=0.05$, $v_{w}=0.1$, $v_{c}=0.2$, $a=0.2$.\label{fig:D00asint}}
\end{figure}

Around $\omega\sim k_{*}$, the function $\Delta^{00}$ smoothly changes
between the power laws $\omega^{-1}$ and $\omega^{-6}$ (see Fig.~\ref{fig:D00}).
If we define the position of the knee between these power laws by
the frequency at which the two curves cross, we have
\begin{equation}
\frac{\omega_{\mathrm{knee}}}{\beta}=\frac{1}{2\alpha}\left(\frac{\sqrt{\pi}A_{s}/2}{A_{s}+A_{d}}\right)^{{1}/{5}}.\label{eq:wknee}
\end{equation}
In Fig.~\ref{fig:D00asint} we plot $\Delta^{00}$, the asymptotes
(\ref{eq:DcollHF}) and (\ref{eq:D00HF}), and the position of $\omega_{\mathrm{knee}}$.

\subsubsection{Interpolations }

\begin{figure}[h]
\includegraphics[width=0.8\columnwidth]{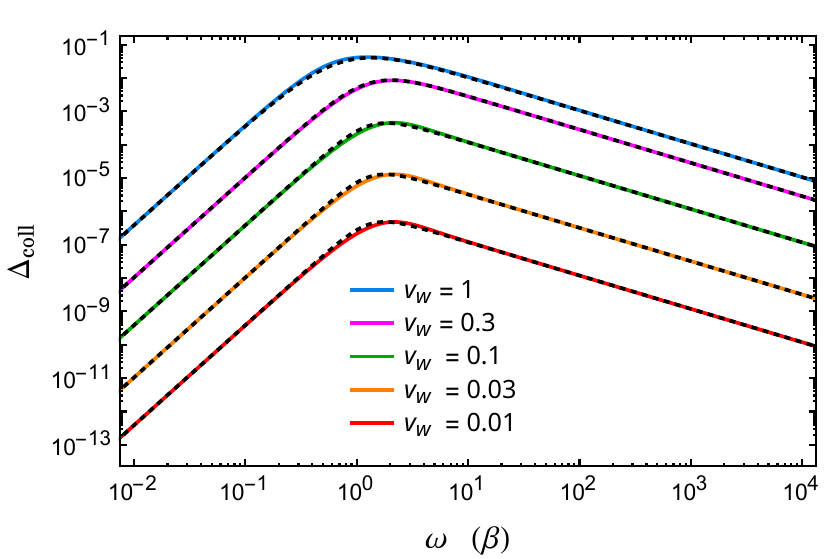}
\caption{The bubble-collision spectrum (solid lines) and the interpolation (\ref{eq:Deltafitcoll}) (dotted lines).\label{fig:fitscoll}}
\end{figure}

For $\omega\ll k_{*}$ we have $\Delta^{00}\simeq\Delta_{\mathrm{coll}}$.
To describe $\Delta_{\mathrm{coll}}$ in the whole frequency range,
we can use an interpolation between the power laws $\omega^{3}$ and
$\omega^{-1}$,
\begin{equation}
\Delta_{\mathrm{coll}}^{\mathrm{fit}}=\Delta_{\mathrm{coll}}^{\mathrm{peak}}[{\textstyle \frac{1}{4}}(\omega_{\mathrm{coll}}^{\mathrm{peak}}/\omega)^{3p}+{\textstyle \frac{3}{4}}(\omega/\omega_{\mathrm{coll}}^{\mathrm{peak}})^{p}]^{-\frac{1}{p}},\label{eq:Deltafitcoll}
\end{equation}
where the peak values are given by
\begin{equation}
\omega_{\mathrm{coll}}^{\mathrm{peak}}=3^{\frac{1}{4p}}\frac{A_{HF}^{\mathrm{coll}\frac{1}{4}}}{A_{LF}^{\mathrm{coll}\frac{1}{4}}},\quad\Delta_{\mathrm{coll}}^{\mathrm{peak}}=\frac{3^{\frac{3}{4p}}}{4^{\frac{1}{p}}}A_{HF}^{\mathrm{coll}\frac{3}{4}}A_{LF}^{\mathrm{coll}\frac{1}{4}}.
\end{equation}
The value of $p$ that best fits the curves is given by
\begin{equation}
p\simeq0.92-0.979v_{w}-0.876v_{w}^{2}+1.93v_{w}^{3}-0.559v_{w}^{4}.
\end{equation}
We show this approximation in Fig.~\ref{fig:fitscoll}. Near $\omega=\beta$
the approximation can exceed a 20\% error, but the error falls below
1\% for $\omega\lesssim10^{-1}\beta$ and $\omega\gtrsim10\beta$.

\begin{figure}[h]
\includegraphics[width=0.8\columnwidth]{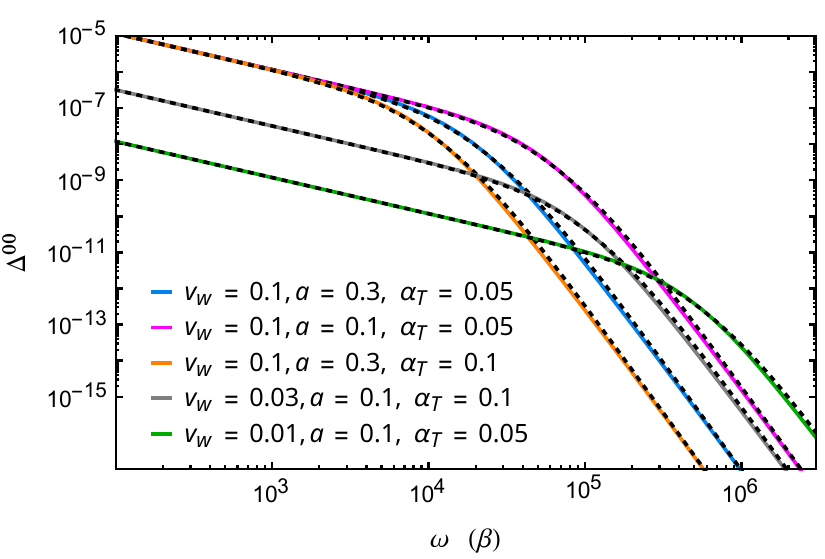}
\caption{Fit to $\Delta^{00}$ for $T=100\,\mathrm{GeV}$, $\beta/H=100$, and $v_c=0.5$.
	\label{fig:D00apps}}
\end{figure}

Similarly, a simple fit for $\omega\gg\beta$ can be obtained as an
interpolation between the power laws $\omega^{-1}$ and $\omega^{-6}$.
In this case, the value $p=1/3$ provides the best fit for any values
of the parameters. An improvement is obtained by adding a term with
a parameter $C$ to fit the value of $\Delta^{00}$ at $\omega=\omega_{\mathrm{knee}}$.
We have
\begin{equation}
\Delta_{\mathrm{fit}}^{00}=A_{HF}^{\mathrm{coll}}\frac{\beta}{\omega}\bigg[1+\frac{\omega^{5/3}}{\omega_{\mathrm{knee}}^{5/3}}+C\Big(1+\frac{\omega_{\mathrm{knee}}^{5}}{\omega^{5}}\Big)^{-1/3}\bigg]^{-3},\label{eq:Delta00fit}
\end{equation}
with $C=0.176$. The error of this fit has a maximum of about 5\%
at $\omega\simeq0.4\omega_{\mathrm{knee}}$ and another maximum of
40\% at $\omega\simeq4\omega_{\mathrm{knee}}$. We show a few examples
in Fig.~\ref{fig:D00apps}. 

\subsection{The contribution $\Delta^{11}$}

Although the integrals with respect to $T$ and $T'$ in Eq.~(\ref{eq:Delta11})
can be evaluated analytically, the result is rather cumbersome. Since the
contribution $\Delta^{11}$ is relevant only for $\omega\sim k_{*} \gg \beta$, 
we can obtain an asymptotic expansion in powers of $\beta/\omega$.
The integrand in Eq.~(\ref{eq:Delta11}) can be written as a sum of terms containing the factors
$e^{i\omega c_{\pm}T'}$ and $e^{-i\omega c_{\pm}T}$, where $c_{\pm}=cv_{w}\pm1$.
These functions vary more rapidly than any other function in the integrand. 
Even the exponential functions $e^{\Omega_{+}T}$ and $e^{\Omega_{+}T'}$
vary more slowly since $\Omega_{+}\sim\hat{\Omega}\omega\ll\omega$.
An asymptotic expansion 
in powers of $\omega^{-1}$ can be obtained by repeatedly using integration by parts, e.g.,
\begin{equation}
\int_{t_{1}}^{t_{2}}\negthickspace dtf(t)e^{i\omega t}=\left.f(t)\frac{e^{i\omega t}}{i\omega}\right|_{t_{1}}^{t_{2}}-\int_{t_{1}}^{t_{2}}\negthickspace dtf'(t)\frac{e^{i\omega t}}{i\omega}.\label{eq:expasint_ixp}
\end{equation}
We apply this procedure to the integrals with respect to $T$ and $T'$
and keep only the lowest order in $\omega^{-1}$,
which has an error of order $\beta/\omega\sim\beta/k_{*}$. 
The resulting expression is still cumbersome, so we also use the approximation
$\Omega_{+}\gg\beta$ (which breaks down for $\omega\lesssim \beta/\hat{\Omega}$).
We obtain
\begin{multline}
\Delta^{11} = \frac{\beta a^{2}v_{w}}{24\pi^{2}\omega} 
\negmedspace \int_{-1}^{+1} \negthickspace dc \frac{c^{2}(1+v_{w}^{2}c^{2})}{(1-v_{w}^{2}c^{2})^{2}}
\Big\{ e^{-\beta T_{nl}}\big[(\beta T_{nl})^{4}
\\
+6(\beta T_{nl})^{3}+18(\beta T_{nl})^{2}+36\beta T_{nl}+36
\\
+3(\beta T_{nl})^{3}\beta T_{0}\big]-2e^{-2(\hat{\Omega}k_{*}/\beta-1)}e^{-\beta T_{0}}(\beta T_{0})^{4}\Big\},
\label{eq:Delta11ap0}
\end{multline}
and, according to Eqs.~(\ref{t0}), (\ref{eq:apsOmD0}), and (\ref{ksap}), we have
\begin{equation}
	\beta T_{nl}=\frac{k_{*}}{\omega\sqrt{1-c^{2}}} , \quad \beta T_{0} = \frac{\hat{\Omega}^{-1} \beta}{\omega\sqrt{1-c^{2}}}.
\end{equation}
The error of this approximation is less than 1\% at the maximum of $\Delta^{11}$, 
and it decreases at higher frequencies. 
At low frequencies, the error increases to 5\% when $\Delta^{11}$
decreases by one order of magnitude and to 10\% when $\Delta^{11}$ decreases
by two orders of magnitude.

At very high frequencies, $T_{nl}$ and $T_{0}$ are very small.
To lowest order in these quantities, Eq.~(\ref{eq:Delta11ap0}) simplifies
significantly, and the integral with respect to $c$ is
straightforward. We obtain
\begin{equation}
\Delta_{HF}^{11}=\frac{3a^{2}}{\pi^{2}v_{w}}\left[\frac{2-v_{w}^{2}}{1-v_{w}^{2}}-\frac{1}{v_{w}}\log\left(\frac{1+v_{w}}{1-v_{w}}\right)\right]\frac{\beta}{\omega}.\label{eq:Delta11HF}
\end{equation}

For $\omega\ll k_{*}$, the quantities $T_{nl}$ and $T_{0}$ are large,
so we only keep the quartic terms in the polynomials of Eq.~(\ref{eq:Delta11ap0}).
It is convenient to make a change of variable from $c$ to $z=(1-c^{2})^{-1/2}$.
The range of this variable is $z>1$, and
the integrand is dominated by $z-1\sim\omega/k_{*}$. Expanding in
powers of $z-1$ we obtain, to lowest order, 
\begin{multline}
\Delta_{LF}^{11} =
\frac{\sqrt{2\pi}a^{2}v_{w}}{24\pi^{2}}
\left[\left( \frac{k_{*}^{5/2}}{\beta^{5/2}}+\frac{3k_{*}^{3/2}}{\beta^{3/2}\hat{\Omega}}\right) e^{-k_{*}/\omega} \right.
\\
\left. - \frac{2e^{-2(\hat{\Omega}k_{*}/\beta-1)}}{\hat{\Omega}^{5/2}} 
e^{-\hat{\Omega}^{-1}\beta/\omega}\right]\left(\frac{\beta}{\omega}\right)^{7/2} .
\label{eq:Delta11LF}
\end{multline}

\begin{figure}[h]
	\includegraphics[width=0.8\columnwidth]{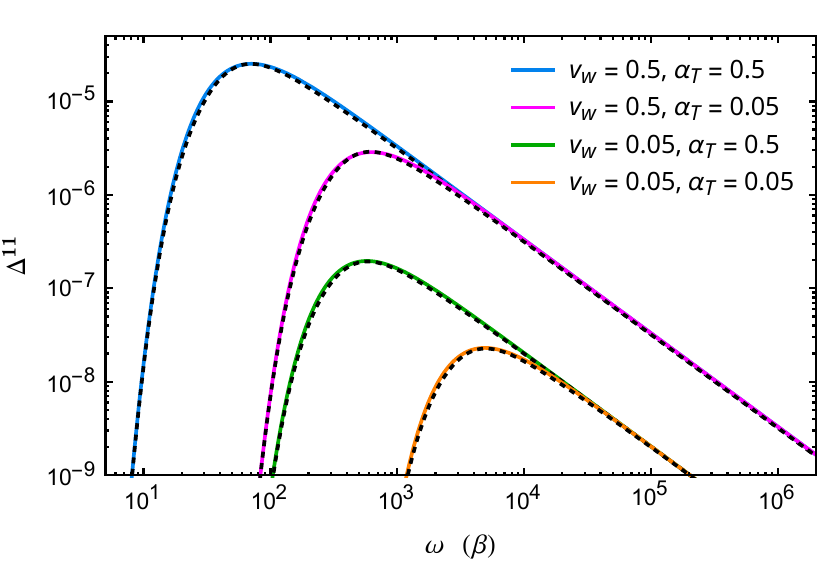}
	\caption{The spectrum contribution $\Delta^{11}$ (solid lines) and its fit (\ref{eq:D11fit}) for $T=100\mathrm{GeV}$, $\beta/H=100$, $a=0.2$,
		and $v_{c}=2v_{w}$.\label{fig:D11fit}}
\end{figure}

A fit to $\Delta^{11}$ is given by the following interpolation
between the high- and low-frequency behaviors:
\begin{equation}
\Delta_{\mathrm{fit}}^{11}=\frac{a^{2}v_{w}\beta}{24\pi^{2}k_{*}}\left[\frac{c_{1}+1}{c_{1}+e^{\frac{k_{*}}{\omega}}}C\frac{k_{*}}{\omega}+c_{2}\frac{k_{*}^{7/2}}{\omega^{7/2}}e^{-\frac{k_{*}}{\omega}}\right],\label{eq:D11fit}
\end{equation}
with $C=24+\frac{216}{5}v_{w}^{2}+\frac{360}{7}v_{w}^{4}$, $c_{1}=4.64+0.142v_{w}^{2}+3.15v_{w}^{4}$,
and $c_{2}=4.12+1.03v_{w}^{2}+6.08v_{w}^{4}$. 
This fit is shown in Fig.~\ref{fig:D11fit}.
The error is below 2\% at the peak but reaches
10\% in some parts of the curve. 

The values of $\omega$ and $\Delta^{11}$
at the peak are approximately given by
\begin{align}
\omega_{p} & =k_{*}\left(0.459+0.122v_{w}^{2}-0.158v_{w}^{4}\right),\label{eq:wpeakfit11}\\
\Delta_{p} & =\frac{a^{2}v_{w}\beta}{24\pi^{2}k_{*}}\left(29+41.5v_{w}^{2}+67v_{w}^{4}\right).\label{eq:hpeakfit11}
\end{align}
The error of both quantities is generally below 2\%.

\subsection{Full spectrum}

In summary, the full GW spectrum can be approximated by
\begin{equation}
\Delta\simeq\begin{cases}
\big[1-\frac{3a^{2}}{2\pi}\log\big(\frac{k_{c}}{k_{*}}\big)\big]\Delta_{\mathrm{coll}}^{\mathrm{fit}} & \omega\ll k_{*},\\
\big[1-\frac{3a^{2}}{2\pi}\log\big(\frac{k_{c}}{k_{*}}\big)\big]\Delta_{\mathrm{fit}}^{00}+\Delta_{\mathrm{fit}}^{11} & \omega\gg\beta.
\end{cases}
\end{equation}
The quantity $\Delta_{\mathrm{coll}}^{\mathrm{fit}}$ is given by
Eq.~(\ref{eq:Deltafitcoll}). It depends only on the parameters $\beta$
and $v_{w}$. The quantity $\Delta_{\mathrm{fit}}^{00}$ is given
by Eq.~(\ref{eq:Delta00fit}). It depends on the parameter $\omega_{\mathrm{knee}}$,
which is given by Eq.~(\ref{eq:wknee}) as a function of the parameter
$\alpha\simeq(2\sqrt{\pi})^{-1}a\beta/k_{*}$. The quantity $\Delta_{\mathrm{fit}}^{11}$
is given by Eq.~(\ref{eq:D11fit}). It depends on the parameters
$v_{w}$, $\beta$, and $k_{*}$. The parameter $k_{*}$ is given
by Eq.~(\ref{eq:khat}), while $k_{c}$ is given by Eq.~(\ref{kc}).

\bibliographystyle{apsrev4-2}
\bibliography{C:/Users/ariel/OneDrive/Biblio/Catalogos/papers,C:/Users/ariel/OneDrive/Biblio/Catalogos/libros-reviews}

\begin{thebibliography}{124}%
\makeatletter
\providecommand \@ifxundefined [1]{%
 \@ifx{#1\undefined}
}%
\providecommand \@ifnum [1]{%
 \ifnum #1\expandafter \@firstoftwo
 \else \expandafter \@secondoftwo
 \fi
}%
\providecommand \@ifx [1]{%
 \ifx #1\expandafter \@firstoftwo
 \else \expandafter \@secondoftwo
 \fi
}%
\providecommand \natexlab [1]{#1}%
\providecommand \enquote  [1]{``#1''}%
\providecommand \bibnamefont  [1]{#1}%
\providecommand \bibfnamefont [1]{#1}%
\providecommand \citenamefont [1]{#1}%
\providecommand \href@noop [0]{\@secondoftwo}%
\providecommand \href [0]{\begingroup \@sanitize@url \@href}%
\providecommand \@href[1]{\@@startlink{#1}\@@href}%
\providecommand \@@href[1]{\endgroup#1\@@endlink}%
\providecommand \@sanitize@url [0]{\catcode `\\12\catcode `\$12\catcode
  `\&12\catcode `\#12\catcode `\^12\catcode `\_12\catcode `\%12\relax}%
\providecommand \@@startlink[1]{}%
\providecommand \@@endlink[0]{}%
\providecommand \url  [0]{\begingroup\@sanitize@url \@url }%
\providecommand \@url [1]{\endgroup\@href {#1}{\urlprefix }}%
\providecommand \urlprefix  [0]{URL }%
\providecommand \Eprint [0]{\href }%
\providecommand \doibase [0]{https://doi.org/}%
\providecommand \selectlanguage [0]{\@gobble}%
\providecommand \bibinfo  [0]{\@secondoftwo}%
\providecommand \bibfield  [0]{\@secondoftwo}%
\providecommand \translation [1]{[#1]}%
\providecommand \BibitemOpen [0]{}%
\providecommand \bibitemStop [0]{}%
\providecommand \bibitemNoStop [0]{.\EOS\space}%
\providecommand \EOS [0]{\spacefactor3000\relax}%
\providecommand \BibitemShut  [1]{\csname bibitem#1\endcsname}%
\let\auto@bib@innerbib\@empty
\bibitem [{\citenamefont {Quiros}(2007)}]{quiros07field}%
  \BibitemOpen
  \bibfield  {author} {\bibinfo {author} {\bibfnamefont {M.}~\bibnamefont
  {Quiros}},\ }\href@noop {} {\bibfield  {journal} {\bibinfo  {journal} {Acta
  Phys. Polon. B}\ }\textbf {\bibinfo {volume} {38}},\ \bibinfo {pages} {3661}
  (\bibinfo {year} {2007})}\BibitemShut {NoStop}%
\bibitem [{\citenamefont {Athron}\ \emph {et~al.}(2024)\citenamefont {Athron},
  \citenamefont {Bal\'azs}, \citenamefont {Fowlie}, \citenamefont {Morris},\
  and\ \citenamefont {Wu}}]{abf+24cosmological}%
  \BibitemOpen
  \bibfield  {author} {\bibinfo {author} {\bibfnamefont {P.}~\bibnamefont
  {Athron}}, \bibinfo {author} {\bibfnamefont {C.}~\bibnamefont {Bal\'azs}},
  \bibinfo {author} {\bibfnamefont {A.}~\bibnamefont {Fowlie}}, \bibinfo
  {author} {\bibfnamefont {L.}~\bibnamefont {Morris}},\ and\ \bibinfo {author}
  {\bibfnamefont {L.}~\bibnamefont {Wu}},\ }\href
  {https://doi.org/10.1016/j.ppnp.2023.104094} {\bibfield  {journal} {\bibinfo
  {journal} {Prog. Part. Nucl. Phys.}\ }\textbf {\bibinfo {volume} {135}},\
  \bibinfo {pages} {104094} (\bibinfo {year} {2024})},\ \Eprint
  {https://arxiv.org/abs/2305.02357} {arXiv:2305.02357 [hep-ph]} \BibitemShut
  {NoStop}%
\bibitem [{\citenamefont {Hogan}(1986)}]{hogan86gravitational}%
  \BibitemOpen
  \bibfield  {author} {\bibinfo {author} {\bibfnamefont {C.~J.}\ \bibnamefont
  {Hogan}},\ }\href@noop {} {\bibfield  {journal} {\bibinfo  {journal} {Mon.
  Not. Roy. Astron. Soc.}\ }\textbf {\bibinfo {volume} {218}},\ \bibinfo
  {pages} {629} (\bibinfo {year} {1986})}\BibitemShut {NoStop}%
\bibitem [{\citenamefont {Turner}\ and\ \citenamefont
  {Wilczek}(1990)}]{tw90relic}%
  \BibitemOpen
  \bibfield  {author} {\bibinfo {author} {\bibfnamefont {M.~S.}\ \bibnamefont
  {Turner}}\ and\ \bibinfo {author} {\bibfnamefont {F.}~\bibnamefont
  {Wilczek}},\ }\href {https://doi.org/10.1103/PhysRevLett.65.3080} {\bibfield
  {journal} {\bibinfo  {journal} {Phys. Rev. Lett.}\ }\textbf {\bibinfo
  {volume} {65}},\ \bibinfo {pages} {3080} (\bibinfo {year}
  {1990})}\BibitemShut {NoStop}%
\bibitem [{\citenamefont {Kosowsky}\ \emph
  {et~al.}(1992{\natexlab{a}})\citenamefont {Kosowsky}, \citenamefont
  {Turner},\ and\ \citenamefont {Watkins}}]{ktw92gravitational}%
  \BibitemOpen
  \bibfield  {author} {\bibinfo {author} {\bibfnamefont {A.}~\bibnamefont
  {Kosowsky}}, \bibinfo {author} {\bibfnamefont {M.~S.}\ \bibnamefont
  {Turner}},\ and\ \bibinfo {author} {\bibfnamefont {R.}~\bibnamefont
  {Watkins}},\ }\href {https://doi.org/10.1103/PhysRevLett.69.2026} {\bibfield
  {journal} {\bibinfo  {journal} {Phys. Rev. Lett.}\ }\textbf {\bibinfo
  {volume} {69}},\ \bibinfo {pages} {2026} (\bibinfo {year}
  {1992}{\natexlab{a}})}\BibitemShut {NoStop}%
\bibitem [{\citenamefont {Kosowsky}\ \emph
  {et~al.}(1992{\natexlab{b}})\citenamefont {Kosowsky}, \citenamefont
  {Turner},\ and\ \citenamefont {Watkins}}]{ktw92gravitationala}%
  \BibitemOpen
  \bibfield  {author} {\bibinfo {author} {\bibfnamefont {A.}~\bibnamefont
  {Kosowsky}}, \bibinfo {author} {\bibfnamefont {M.~S.}\ \bibnamefont
  {Turner}},\ and\ \bibinfo {author} {\bibfnamefont {R.}~\bibnamefont
  {Watkins}},\ }\href {https://doi.org/10.1103/PhysRevD.45.4514} {\bibfield
  {journal} {\bibinfo  {journal} {Phys. Rev. D}\ }\textbf {\bibinfo {volume}
  {45}},\ \bibinfo {pages} {4514} (\bibinfo {year}
  {1992}{\natexlab{b}})}\BibitemShut {NoStop}%
\bibitem [{\citenamefont {Kosowsky}\ and\ \citenamefont
  {Turner}(1993)}]{kt93gravitational}%
  \BibitemOpen
  \bibfield  {author} {\bibinfo {author} {\bibfnamefont {A.}~\bibnamefont
  {Kosowsky}}\ and\ \bibinfo {author} {\bibfnamefont {M.~S.}\ \bibnamefont
  {Turner}},\ }\href {https://doi.org/10.1103/PhysRevD.47.4372} {\bibfield
  {journal} {\bibinfo  {journal} {Phys. Rev.}\ }\textbf {\bibinfo {volume}
  {D47}},\ \bibinfo {pages} {4372} (\bibinfo {year} {1993})},\ \Eprint
  {https://arxiv.org/abs/astro-ph/9211004} {arXiv:astro-ph/9211004 [astro-ph]}
  \BibitemShut {NoStop}%
\bibitem [{\citenamefont {Kamionkowski}\ \emph {et~al.}(1994)\citenamefont
  {Kamionkowski}, \citenamefont {Kosowsky},\ and\ \citenamefont
  {Turner}}]{kkt94gravitational}%
  \BibitemOpen
  \bibfield  {author} {\bibinfo {author} {\bibfnamefont {M.}~\bibnamefont
  {Kamionkowski}}, \bibinfo {author} {\bibfnamefont {A.}~\bibnamefont
  {Kosowsky}},\ and\ \bibinfo {author} {\bibfnamefont {M.~S.}\ \bibnamefont
  {Turner}},\ }\href {https://doi.org/10.1103/PhysRevD.49.2837} {\bibfield
  {journal} {\bibinfo  {journal} {Phys. Rev. D}\ }\textbf {\bibinfo {volume}
  {49}},\ \bibinfo {pages} {2837} (\bibinfo {year} {1994})},\ \Eprint
  {https://arxiv.org/abs/astro-ph/9310044} {arXiv:astro-ph/9310044}
  \BibitemShut {NoStop}%
\bibitem [{\citenamefont {Maggiore}(2000)}]{maggiore00gravitational}%
  \BibitemOpen
  \bibfield  {author} {\bibinfo {author} {\bibfnamefont {M.}~\bibnamefont
  {Maggiore}},\ }\href {https://doi.org/10.1016/S0370-1573(99)00102-7}
  {\bibfield  {journal} {\bibinfo  {journal} {Phys. Rept.}\ }\textbf {\bibinfo
  {volume} {331}},\ \bibinfo {pages} {283} (\bibinfo {year} {2000})},\ \Eprint
  {https://arxiv.org/abs/gr-qc/9909001} {arXiv:gr-qc/9909001} \BibitemShut
  {NoStop}%
\bibitem [{\citenamefont {Amaro-Seoane}\ \emph {et~al.}(2017)\citenamefont
  {Amaro-Seoane} \emph {et~al.}}]{a+17laser}%
  \BibitemOpen
  \bibfield  {author} {\bibinfo {author} {\bibfnamefont {P.}~\bibnamefont
  {Amaro-Seoane}} \emph {et~al.} (\bibinfo {collaboration} {LISA}),\
  }\href@noop {} {\bibinfo {title} {{Laser Interferometer Space Antenna}}}
  (\bibinfo {year} {2017}),\ \Eprint {https://arxiv.org/abs/1702.00786}
  {arXiv:1702.00786 [astro-ph.IM]} \BibitemShut {NoStop}%
\bibitem [{\citenamefont {Janssen}\ \emph {et~al.}(2015)\citenamefont {Janssen}
  \emph {et~al.}}]{jhm+15gravitational}%
  \BibitemOpen
  \bibfield  {author} {\bibinfo {author} {\bibfnamefont {G.}~\bibnamefont
  {Janssen}} \emph {et~al.},\ }\href {https://doi.org/10.22323/1.215.0037}
  {\bibfield  {journal} {\bibinfo  {journal} {PoS}\ }\textbf {\bibinfo {volume}
  {AASKA14}},\ \bibinfo {pages} {037} (\bibinfo {year} {2015})}\BibitemShut
  {NoStop}%
\bibitem [{\citenamefont {Kajantie}\ \emph {et~al.}(1996)\citenamefont
  {Kajantie}, \citenamefont {Laine}, \citenamefont {Rummukainen},\ and\
  \citenamefont {Shaposhnikov}}]{klr+96is}%
  \BibitemOpen
  \bibfield  {author} {\bibinfo {author} {\bibfnamefont {K.}~\bibnamefont
  {Kajantie}}, \bibinfo {author} {\bibfnamefont {M.}~\bibnamefont {Laine}},
  \bibinfo {author} {\bibfnamefont {K.}~\bibnamefont {Rummukainen}},\ and\
  \bibinfo {author} {\bibfnamefont {M.~E.}\ \bibnamefont {Shaposhnikov}},\
  }\href {https://doi.org/10.1103/PhysRevLett.77.2887} {\bibfield  {journal}
  {\bibinfo  {journal} {Phys. Rev. Lett.}\ }\textbf {\bibinfo {volume} {77}},\
  \bibinfo {pages} {2887} (\bibinfo {year} {1996})},\ \Eprint
  {https://arxiv.org/abs/hep-ph/9605288} {arXiv:hep-ph/9605288} \BibitemShut
  {NoStop}%
\bibitem [{\citenamefont {Kajantie}\ \emph {et~al.}(1997)\citenamefont
  {Kajantie}, \citenamefont {Laine}, \citenamefont {Rummukainen},\ and\
  \citenamefont {Shaposhnikov}}]{klr+97nonperturbative}%
  \BibitemOpen
  \bibfield  {author} {\bibinfo {author} {\bibfnamefont {K.}~\bibnamefont
  {Kajantie}}, \bibinfo {author} {\bibfnamefont {M.}~\bibnamefont {Laine}},
  \bibinfo {author} {\bibfnamefont {K.}~\bibnamefont {Rummukainen}},\ and\
  \bibinfo {author} {\bibfnamefont {M.~E.}\ \bibnamefont {Shaposhnikov}},\
  }\href {https://doi.org/10.1016/S0550-3213(97)00164-8} {\bibfield  {journal}
  {\bibinfo  {journal} {Nucl. Phys. B}\ }\textbf {\bibinfo {volume} {493}},\
  \bibinfo {pages} {413} (\bibinfo {year} {1997})},\ \Eprint
  {https://arxiv.org/abs/hep-lat/9612006} {arXiv:hep-lat/9612006} \BibitemShut
  {NoStop}%
\bibitem [{\citenamefont {Aoki}\ \emph {et~al.}(2006)\citenamefont {Aoki},
  \citenamefont {Endrodi}, \citenamefont {Fodor}, \citenamefont {Katz},\ and\
  \citenamefont {Szabo}}]{aef+06order}%
  \BibitemOpen
  \bibfield  {author} {\bibinfo {author} {\bibfnamefont {Y.}~\bibnamefont
  {Aoki}}, \bibinfo {author} {\bibfnamefont {G.}~\bibnamefont {Endrodi}},
  \bibinfo {author} {\bibfnamefont {Z.}~\bibnamefont {Fodor}}, \bibinfo
  {author} {\bibfnamefont {S.~D.}\ \bibnamefont {Katz}},\ and\ \bibinfo
  {author} {\bibfnamefont {K.~K.}\ \bibnamefont {Szabo}},\ }\href
  {https://doi.org/10.1038/nature05120} {\bibfield  {journal} {\bibinfo
  {journal} {Nature}\ }\textbf {\bibinfo {volume} {443}},\ \bibinfo {pages}
  {675} (\bibinfo {year} {2006})},\ \Eprint
  {https://arxiv.org/abs/hep-lat/0611014} {arXiv:hep-lat/0611014} \BibitemShut
  {NoStop}%
\bibitem [{\citenamefont {Apreda}\ \emph {et~al.}(2001)\citenamefont {Apreda},
  \citenamefont {Maggiore}, \citenamefont {Nicolis},\ and\ \citenamefont
  {Riotto}}]{amn+01supersymmetric}%
  \BibitemOpen
  \bibfield  {author} {\bibinfo {author} {\bibfnamefont {R.}~\bibnamefont
  {Apreda}}, \bibinfo {author} {\bibfnamefont {M.}~\bibnamefont {Maggiore}},
  \bibinfo {author} {\bibfnamefont {A.}~\bibnamefont {Nicolis}},\ and\ \bibinfo
  {author} {\bibfnamefont {A.}~\bibnamefont {Riotto}},\ }\href
  {https://doi.org/10.1088/0264-9381/18/23/101} {\bibfield  {journal} {\bibinfo
   {journal} {Class. Quant. Grav.}\ }\textbf {\bibinfo {volume} {18}},\
  \bibinfo {pages} {L155} (\bibinfo {year} {2001})},\ \Eprint
  {https://arxiv.org/abs/hep-ph/0102140} {arXiv:hep-ph/0102140} \BibitemShut
  {NoStop}%
\bibitem [{\citenamefont {Espinosa}\ \emph {et~al.}(2008)\citenamefont
  {Espinosa}, \citenamefont {Konstandin}, \citenamefont {No},\ and\
  \citenamefont {Quiros}}]{ekn+08some}%
  \BibitemOpen
  \bibfield  {author} {\bibinfo {author} {\bibfnamefont {J.~R.}\ \bibnamefont
  {Espinosa}}, \bibinfo {author} {\bibfnamefont {T.}~\bibnamefont
  {Konstandin}}, \bibinfo {author} {\bibfnamefont {J.~M.}\ \bibnamefont {No}},\
  and\ \bibinfo {author} {\bibfnamefont {M.}~\bibnamefont {Quiros}},\ }\href
  {https://doi.org/10.1103/PhysRevD.78.123528} {\bibfield  {journal} {\bibinfo
  {journal} {Phys. Rev. D}\ }\textbf {\bibinfo {volume} {78}},\ \bibinfo
  {pages} {123528} (\bibinfo {year} {2008})},\ \Eprint
  {https://arxiv.org/abs/0809.3215} {arXiv:0809.3215 [hep-ph]} \BibitemShut
  {NoStop}%
\bibitem [{\citenamefont {Huber}\ and\ \citenamefont
  {Konstandin}(2008{\natexlab{a}})}]{hk08production}%
  \BibitemOpen
  \bibfield  {author} {\bibinfo {author} {\bibfnamefont {S.~J.}\ \bibnamefont
  {Huber}}\ and\ \bibinfo {author} {\bibfnamefont {T.}~\bibnamefont
  {Konstandin}},\ }\href {https://doi.org/10.1088/1475-7516/2008/05/017}
  {\bibfield  {journal} {\bibinfo  {journal} {JCAP}\ }\textbf {\bibinfo
  {volume} {2008}}\bibfield  {number} {\bibinfo  {number} { (05)},\ \bibinfo
  {pages} {017}},\ }\Eprint {https://arxiv.org/abs/0709.2091} {arXiv:0709.2091
  [hep-ph]} \BibitemShut {NoStop}%
\bibitem [{\citenamefont {Leitao}\ \emph {et~al.}(2012)\citenamefont {Leitao},
  \citenamefont {Megevand},\ and\ \citenamefont
  {Sanchez}}]{lms12gravitational}%
  \BibitemOpen
  \bibfield  {author} {\bibinfo {author} {\bibfnamefont {L.}~\bibnamefont
  {Leitao}}, \bibinfo {author} {\bibfnamefont {A.}~\bibnamefont {Megevand}},\
  and\ \bibinfo {author} {\bibfnamefont {A.~D.}\ \bibnamefont {Sanchez}},\
  }\href {https://doi.org/10.1088/1475-7516/2012/10/024} {\bibfield  {journal}
  {\bibinfo  {journal} {JCAP}\ }\textbf {\bibinfo {volume} {2012}}\bibfield
  {number} {\bibinfo  {number} { (10)},\ \bibinfo {pages} {024}},\ }\Eprint
  {https://arxiv.org/abs/1205.3070} {arXiv:1205.3070 [astro-ph.CO]}
  \BibitemShut {NoStop}%
\bibitem [{\citenamefont {Dorsch}\ \emph {et~al.}(2014)\citenamefont {Dorsch},
  \citenamefont {Huber},\ and\ \citenamefont {No}}]{dhn14cosmological}%
  \BibitemOpen
  \bibfield  {author} {\bibinfo {author} {\bibfnamefont {G.~C.}\ \bibnamefont
  {Dorsch}}, \bibinfo {author} {\bibfnamefont {S.~J.}\ \bibnamefont {Huber}},\
  and\ \bibinfo {author} {\bibfnamefont {J.~M.}\ \bibnamefont {No}},\ }\href
  {https://doi.org/10.1103/PhysRevLett.113.121801} {\bibfield  {journal}
  {\bibinfo  {journal} {Phys. Rev. Lett.}\ }\textbf {\bibinfo {volume} {113}},\
  \bibinfo {pages} {121801} (\bibinfo {year} {2014})},\ \Eprint
  {https://arxiv.org/abs/1403.5583} {arXiv:1403.5583 [hep-ph]} \BibitemShut
  {NoStop}%
\bibitem [{\citenamefont {Kakizaki}\ \emph {et~al.}(2015)\citenamefont
  {Kakizaki}, \citenamefont {Kanemura},\ and\ \citenamefont
  {Matsui}}]{kkm15gravitational}%
  \BibitemOpen
  \bibfield  {author} {\bibinfo {author} {\bibfnamefont {M.}~\bibnamefont
  {Kakizaki}}, \bibinfo {author} {\bibfnamefont {S.}~\bibnamefont {Kanemura}},\
  and\ \bibinfo {author} {\bibfnamefont {T.}~\bibnamefont {Matsui}},\ }\href
  {https://doi.org/10.1103/PhysRevD.92.115007} {\bibfield  {journal} {\bibinfo
  {journal} {Phys. Rev. D}\ }\textbf {\bibinfo {volume} {92}},\ \bibinfo
  {pages} {115007} (\bibinfo {year} {2015})},\ \Eprint
  {https://arxiv.org/abs/1509.08394} {arXiv:1509.08394 [hep-ph]} \BibitemShut
  {NoStop}%
\bibitem [{\citenamefont {Huber}\ \emph {et~al.}(2016)\citenamefont {Huber},
  \citenamefont {Konstandin}, \citenamefont {Nardini},\ and\ \citenamefont
  {Rues}}]{hkn+16detectable}%
  \BibitemOpen
  \bibfield  {author} {\bibinfo {author} {\bibfnamefont {S.~J.}\ \bibnamefont
  {Huber}}, \bibinfo {author} {\bibfnamefont {T.}~\bibnamefont {Konstandin}},
  \bibinfo {author} {\bibfnamefont {G.}~\bibnamefont {Nardini}},\ and\ \bibinfo
  {author} {\bibfnamefont {I.}~\bibnamefont {Rues}},\ }\href
  {https://doi.org/10.1088/1475-7516/2016/03/036} {\bibfield  {journal}
  {\bibinfo  {journal} {JCAP}\ }\textbf {\bibinfo {volume} {2016}}\bibfield
  {number} {\bibinfo  {number} { (03)},\ \bibinfo {pages} {036}},\ }\Eprint
  {https://arxiv.org/abs/1512.06357} {arXiv:1512.06357 [hep-ph]} \BibitemShut
  {NoStop}%
\bibitem [{\citenamefont {Leitao}\ and\ \citenamefont
  {Megevand}(2016{\natexlab{a}})}]{lm16gravitational}%
  \BibitemOpen
  \bibfield  {author} {\bibinfo {author} {\bibfnamefont {L.}~\bibnamefont
  {Leitao}}\ and\ \bibinfo {author} {\bibfnamefont {A.}~\bibnamefont
  {Megevand}},\ }\href {https://doi.org/10.1088/1475-7516/2016/05/037}
  {\bibfield  {journal} {\bibinfo  {journal} {JCAP}\ }\textbf {\bibinfo
  {volume} {2016}}\bibfield  {number} {\bibinfo  {number} { (05)},\ \bibinfo
  {pages} {037}},\ }\Eprint {https://arxiv.org/abs/1512.08962}
  {arXiv:1512.08962 [astro-ph.CO]} \BibitemShut {NoStop}%
\bibitem [{\citenamefont {Garcia-Pepin}\ and\ \citenamefont
  {Quiros}(2016)}]{gq16strong}%
  \BibitemOpen
  \bibfield  {author} {\bibinfo {author} {\bibfnamefont {M.}~\bibnamefont
  {Garcia-Pepin}}\ and\ \bibinfo {author} {\bibfnamefont {M.}~\bibnamefont
  {Quiros}},\ }\href {https://doi.org/10.1007/JHEP05(2016)177} {\bibfield
  {journal} {\bibinfo  {journal} {JHEP}\ }\textbf {\bibinfo {volume}
  {2016}}\bibfield  {number} {\bibinfo  {number} { (05)},\ \bibinfo {pages}
  {177}},\ }\Eprint {https://arxiv.org/abs/1602.01351} {arXiv:1602.01351
  [hep-ph]} \BibitemShut {NoStop}%
\bibitem [{\citenamefont {Tsumura}\ \emph {et~al.}(2017)\citenamefont
  {Tsumura}, \citenamefont {Yamada},\ and\ \citenamefont
  {Yamaguchi}}]{tyy17gravitational}%
  \BibitemOpen
  \bibfield  {author} {\bibinfo {author} {\bibfnamefont {K.}~\bibnamefont
  {Tsumura}}, \bibinfo {author} {\bibfnamefont {M.}~\bibnamefont {Yamada}},\
  and\ \bibinfo {author} {\bibfnamefont {Y.}~\bibnamefont {Yamaguchi}},\ }\href
  {https://doi.org/10.1088/1475-7516/2017/07/044} {\bibfield  {journal}
  {\bibinfo  {journal} {JCAP}\ }\textbf {\bibinfo {volume} {2017}}\bibfield
  {number} {\bibinfo  {number} { (07)},\ \bibinfo {pages} {044}},\ }\Eprint
  {https://arxiv.org/abs/1704.00219} {arXiv:1704.00219 [hep-ph]} \BibitemShut
  {NoStop}%
\bibitem [{\citenamefont {Dorsch}\ \emph {et~al.}(2017)\citenamefont {Dorsch},
  \citenamefont {Huber}, \citenamefont {Konstandin},\ and\ \citenamefont
  {No}}]{dhk+17second}%
  \BibitemOpen
  \bibfield  {author} {\bibinfo {author} {\bibfnamefont {G.~C.}\ \bibnamefont
  {Dorsch}}, \bibinfo {author} {\bibfnamefont {S.~J.}\ \bibnamefont {Huber}},
  \bibinfo {author} {\bibfnamefont {T.}~\bibnamefont {Konstandin}},\ and\
  \bibinfo {author} {\bibfnamefont {J.~M.}\ \bibnamefont {No}},\ }\href
  {https://doi.org/10.1088/1475-7516/2017/05/052} {\bibfield  {journal}
  {\bibinfo  {journal} {JCAP}\ }\textbf {\bibinfo {volume} {2017}}\bibfield
  {number} {\bibinfo  {number} { (05)},\ \bibinfo {pages} {052}},\ }\Eprint
  {https://arxiv.org/abs/1611.05874} {arXiv:1611.05874 [hep-ph]} \BibitemShut
  {NoStop}%
\bibitem [{\citenamefont {Cai}\ \emph {et~al.}(2017)\citenamefont {Cai},
  \citenamefont {Sasaki},\ and\ \citenamefont {Wang}}]{csw17gravitational}%
  \BibitemOpen
  \bibfield  {author} {\bibinfo {author} {\bibfnamefont {R.-G.}\ \bibnamefont
  {Cai}}, \bibinfo {author} {\bibfnamefont {M.}~\bibnamefont {Sasaki}},\ and\
  \bibinfo {author} {\bibfnamefont {S.-J.}\ \bibnamefont {Wang}},\ }\href
  {https://doi.org/10.1088/1475-7516/2017/08/004} {\bibfield  {journal}
  {\bibinfo  {journal} {JCAP}\ }\textbf {\bibinfo {volume} {2017}}\bibfield
  {number} {\bibinfo  {number} { (08)},\ \bibinfo {pages} {004}},\ }\Eprint
  {https://arxiv.org/abs/1707.03001} {arXiv:1707.03001 [astro-ph.CO]}
  \BibitemShut {NoStop}%
\bibitem [{\citenamefont {Aoki}\ \emph {et~al.}(2017)\citenamefont {Aoki},
  \citenamefont {Goto},\ and\ \citenamefont {Kubo}}]{agk17gravitational}%
  \BibitemOpen
  \bibfield  {author} {\bibinfo {author} {\bibfnamefont {M.}~\bibnamefont
  {Aoki}}, \bibinfo {author} {\bibfnamefont {H.}~\bibnamefont {Goto}},\ and\
  \bibinfo {author} {\bibfnamefont {J.}~\bibnamefont {Kubo}},\ }\href
  {https://doi.org/10.1103/PhysRevD.96.075045} {\bibfield  {journal} {\bibinfo
  {journal} {Phys. Rev. D}\ }\textbf {\bibinfo {volume} {96}},\ \bibinfo
  {pages} {075045} (\bibinfo {year} {2017})},\ \Eprint
  {https://arxiv.org/abs/1709.07572} {arXiv:1709.07572 [hep-ph]} \BibitemShut
  {NoStop}%
\bibitem [{\citenamefont {von Harling}\ and\ \citenamefont
  {Servant}(2018)}]{hs18qcd}%
  \BibitemOpen
  \bibfield  {author} {\bibinfo {author} {\bibfnamefont {B.}~\bibnamefont {von
  Harling}}\ and\ \bibinfo {author} {\bibfnamefont {G.}~\bibnamefont
  {Servant}},\ }\href {https://doi.org/10.1007/JHEP01(2018)159} {\bibfield
  {journal} {\bibinfo  {journal} {JHEP}\ }\textbf {\bibinfo {volume}
  {2018}}\bibfield  {number} {\bibinfo  {number} { (01)},\ \bibinfo {pages}
  {159}},\ }\Eprint {https://arxiv.org/abs/1711.11554} {arXiv:1711.11554
  [hep-ph]} \BibitemShut {NoStop}%
\bibitem [{\citenamefont {Meg\'\i{}as}\ \emph {et~al.}(2018)\citenamefont
  {Meg\'\i{}as}, \citenamefont {Nardini},\ and\ \citenamefont
  {Quir\'os}}]{mnq18cosmological}%
  \BibitemOpen
  \bibfield  {author} {\bibinfo {author} {\bibfnamefont {E.}~\bibnamefont
  {Meg\'\i{}as}}, \bibinfo {author} {\bibfnamefont {G.}~\bibnamefont
  {Nardini}},\ and\ \bibinfo {author} {\bibfnamefont {M.}~\bibnamefont
  {Quir\'os}},\ }\href {https://doi.org/10.1007/JHEP09(2018)095} {\bibfield
  {journal} {\bibinfo  {journal} {JHEP}\ }\textbf {\bibinfo {volume}
  {2018}}\bibfield  {number} {\bibinfo  {number} { (09)},\ \bibinfo {pages}
  {095}},\ }\Eprint {https://arxiv.org/abs/1806.04877} {arXiv:1806.04877
  [hep-ph]} \BibitemShut {NoStop}%
\bibitem [{\citenamefont {Caprini}\ \emph {et~al.}(2025)\citenamefont
  {Caprini}, \citenamefont {Pujol{\`a}s}, \citenamefont {Quelquejay-Leclere},
  \citenamefont {Rompineve},\ and\ \citenamefont {Steer}}]{cpq+25primordial}%
  \BibitemOpen
  \bibfield  {author} {\bibinfo {author} {\bibfnamefont {C.}~\bibnamefont
  {Caprini}}, \bibinfo {author} {\bibfnamefont {O.}~\bibnamefont
  {Pujol{\`a}s}}, \bibinfo {author} {\bibfnamefont {H.}~\bibnamefont
  {Quelquejay-Leclere}}, \bibinfo {author} {\bibfnamefont {F.}~\bibnamefont
  {Rompineve}},\ and\ \bibinfo {author} {\bibfnamefont {D.~A.}\ \bibnamefont
  {Steer}},\ }\href {https://doi.org/10.1088/1361-6382/ad9a48} {\bibfield
  {journal} {\bibinfo  {journal} {Class. Quant. Grav.}\ }\textbf {\bibinfo
  {volume} {42}},\ \bibinfo {pages} {045015} (\bibinfo {year} {2025})},\
  \Eprint {https://arxiv.org/abs/2406.02359} {arXiv:2406.02359 [astro-ph.CO]}
  \BibitemShut {NoStop}%
\bibitem [{\citenamefont {Huber}\ and\ \citenamefont
  {Konstandin}(2008{\natexlab{b}})}]{hk08gravitational}%
  \BibitemOpen
  \bibfield  {author} {\bibinfo {author} {\bibfnamefont {S.~J.}\ \bibnamefont
  {Huber}}\ and\ \bibinfo {author} {\bibfnamefont {T.}~\bibnamefont
  {Konstandin}},\ }\href {https://doi.org/10.1088/1475-7516/2008/09/022}
  {\bibfield  {journal} {\bibinfo  {journal} {JCAP}\ }\textbf {\bibinfo
  {volume} {2008}}\bibfield  {number} {\bibinfo  {number} { (09)},\ \bibinfo
  {pages} {022}},\ }\Eprint {https://arxiv.org/abs/0806.1828} {arXiv:0806.1828
  [hep-ph]} \BibitemShut {NoStop}%
\bibitem [{\citenamefont {Megevand}\ and\ \citenamefont
  {Ramirez}(2017)}]{mr17bubble}%
  \BibitemOpen
  \bibfield  {author} {\bibinfo {author} {\bibfnamefont {A.}~\bibnamefont
  {Megevand}}\ and\ \bibinfo {author} {\bibfnamefont {S.}~\bibnamefont
  {Ramirez}},\ }\href {https://doi.org/10.1016/j.nuclphysb.2017.03.009}
  {\bibfield  {journal} {\bibinfo  {journal} {Nucl. Phys. B}\ }\textbf
  {\bibinfo {volume} {919}},\ \bibinfo {pages} {74} (\bibinfo {year} {2017})},\
  \Eprint {https://arxiv.org/abs/1611.05853} {arXiv:1611.05853 [astro-ph.CO]}
  \BibitemShut {NoStop}%
\bibitem [{\citenamefont {Ellis}\ \emph
  {et~al.}(2019{\natexlab{a}})\citenamefont {Ellis}, \citenamefont {Lewicki},
  \citenamefont {No},\ and\ \citenamefont {Vaskonen}}]{eln+19gravitational}%
  \BibitemOpen
  \bibfield  {author} {\bibinfo {author} {\bibfnamefont {J.}~\bibnamefont
  {Ellis}}, \bibinfo {author} {\bibfnamefont {M.}~\bibnamefont {Lewicki}},
  \bibinfo {author} {\bibfnamefont {J.~M.}\ \bibnamefont {No}},\ and\ \bibinfo
  {author} {\bibfnamefont {V.}~\bibnamefont {Vaskonen}},\ }\href
  {https://doi.org/10.1088/1475-7516/2019/06/024} {\bibfield  {journal}
  {\bibinfo  {journal} {JCAP}\ }\textbf {\bibinfo {volume} {2019}}\bibfield
  {number} {\bibinfo  {number} { (06)},\ \bibinfo {pages} {024}},\ }\Eprint
  {https://arxiv.org/abs/1903.09642} {arXiv:1903.09642 [hep-ph]} \BibitemShut
  {NoStop}%
\bibitem [{\citenamefont {Ellis}\ \emph
  {et~al.}(2019{\natexlab{b}})\citenamefont {Ellis}, \citenamefont {Lewicki},\
  and\ \citenamefont {No}}]{eln19maximal}%
  \BibitemOpen
  \bibfield  {author} {\bibinfo {author} {\bibfnamefont {J.}~\bibnamefont
  {Ellis}}, \bibinfo {author} {\bibfnamefont {M.}~\bibnamefont {Lewicki}},\
  and\ \bibinfo {author} {\bibfnamefont {J.~M.}\ \bibnamefont {No}},\ }\href
  {https://doi.org/10.1088/1475-7516/2019/04/003} {\bibfield  {journal}
  {\bibinfo  {journal} {JCAP}\ }\textbf {\bibinfo {volume} {2019}}\bibfield
  {number} {\bibinfo  {number} { (04)},\ \bibinfo {pages} {003}},\ }\Eprint
  {https://arxiv.org/abs/1809.08242} {arXiv:1809.08242 [hep-ph]} \BibitemShut
  {NoStop}%
\bibitem [{\citenamefont {Ellis}\ \emph {et~al.}(2020)\citenamefont {Ellis},
  \citenamefont {Lewicki},\ and\ \citenamefont {Vaskonen}}]{elv20updated}%
  \BibitemOpen
  \bibfield  {author} {\bibinfo {author} {\bibfnamefont {J.}~\bibnamefont
  {Ellis}}, \bibinfo {author} {\bibfnamefont {M.}~\bibnamefont {Lewicki}},\
  and\ \bibinfo {author} {\bibfnamefont {V.}~\bibnamefont {Vaskonen}},\ }\href
  {https://doi.org/10.1088/1475-7516/2020/11/020} {\bibfield  {journal}
  {\bibinfo  {journal} {JCAP}\ }\textbf {\bibinfo {volume} {2020}}\bibfield
  {number} {\bibinfo  {number} { (11)},\ \bibinfo {pages} {020}},\ }\Eprint
  {https://arxiv.org/abs/2007.15586} {arXiv:2007.15586 [astro-ph.CO]}
  \BibitemShut {NoStop}%
\bibitem [{\citenamefont {Espinosa}\ \emph {et~al.}(2010)\citenamefont
  {Espinosa}, \citenamefont {Konstandin}, \citenamefont {No},\ and\
  \citenamefont {Servant}}]{ekn+10energy}%
  \BibitemOpen
  \bibfield  {author} {\bibinfo {author} {\bibfnamefont {J.~R.}\ \bibnamefont
  {Espinosa}}, \bibinfo {author} {\bibfnamefont {T.}~\bibnamefont
  {Konstandin}}, \bibinfo {author} {\bibfnamefont {J.~M.}\ \bibnamefont {No}},\
  and\ \bibinfo {author} {\bibfnamefont {G.}~\bibnamefont {Servant}},\ }\href
  {https://doi.org/10.1088/1475-7516/2010/06/028} {\bibfield  {journal}
  {\bibinfo  {journal} {JCAP}\ }\textbf {\bibinfo {volume} {2010}}\bibfield
  {number} {\bibinfo  {number} { (06)},\ \bibinfo {pages} {028}},\ }\Eprint
  {https://arxiv.org/abs/1004.4187} {arXiv:1004.4187 [hep-ph]} \BibitemShut
  {NoStop}%
\bibitem [{\citenamefont {M\'egevand}(2013)}]{megevand13friction}%
  \BibitemOpen
  \bibfield  {author} {\bibinfo {author} {\bibfnamefont {A.}~\bibnamefont
  {M\'egevand}},\ }\href {https://doi.org/10.1088/1475-7516/2013/07/045}
  {\bibfield  {journal} {\bibinfo  {journal} {JCAP}\ }\textbf {\bibinfo
  {volume} {2013}}\bibfield  {number} {\bibinfo  {number} { (07)},\ \bibinfo
  {pages} {045}},\ }\Eprint {https://arxiv.org/abs/1303.4233} {arXiv:1303.4233
  [astro-ph.CO]} \BibitemShut {NoStop}%
\bibitem [{\citenamefont {Leitao}\ and\ \citenamefont
  {Megevand}(2016{\natexlab{b}})}]{lm16hydrodynamics}%
  \BibitemOpen
  \bibfield  {author} {\bibinfo {author} {\bibfnamefont {L.}~\bibnamefont
  {Leitao}}\ and\ \bibinfo {author} {\bibfnamefont {A.}~\bibnamefont
  {Megevand}},\ }\href {https://doi.org/10.1016/j.nuclphysb.2016.02.009}
  {\bibfield  {journal} {\bibinfo  {journal} {Nucl. Phys.}\ }\textbf {\bibinfo
  {volume} {B905}},\ \bibinfo {pages} {45} (\bibinfo {year}
  {2016}{\natexlab{b}})},\ \Eprint {https://arxiv.org/abs/1510.07747}
  {arXiv:1510.07747 [astro-ph.CO]} \BibitemShut {NoStop}%
\bibitem [{\citenamefont {DeGrand}\ and\ \citenamefont
  {Kajantie}(1984)}]{dk84supercooling}%
  \BibitemOpen
  \bibfield  {author} {\bibinfo {author} {\bibfnamefont {T.~A.}\ \bibnamefont
  {DeGrand}}\ and\ \bibinfo {author} {\bibfnamefont {K.}~\bibnamefont
  {Kajantie}},\ }\href {https://doi.org/10.1016/0370-2693(84)90115-1}
  {\bibfield  {journal} {\bibinfo  {journal} {Phys. Lett. B}\ }\textbf
  {\bibinfo {volume} {147}},\ \bibinfo {pages} {273} (\bibinfo {year}
  {1984})}\BibitemShut {NoStop}%
\bibitem [{\citenamefont {Enqvist}\ \emph {et~al.}(1992)\citenamefont
  {Enqvist}, \citenamefont {Ignatius}, \citenamefont {Kajantie},\ and\
  \citenamefont {Rummukainen}}]{eik+92nucleation}%
  \BibitemOpen
  \bibfield  {author} {\bibinfo {author} {\bibfnamefont {K.}~\bibnamefont
  {Enqvist}}, \bibinfo {author} {\bibfnamefont {J.}~\bibnamefont {Ignatius}},
  \bibinfo {author} {\bibfnamefont {K.}~\bibnamefont {Kajantie}},\ and\
  \bibinfo {author} {\bibfnamefont {K.}~\bibnamefont {Rummukainen}},\ }\href
  {https://doi.org/10.1103/PhysRevD.45.3415} {\bibfield  {journal} {\bibinfo
  {journal} {Phys. Rev. D}\ }\textbf {\bibinfo {volume} {45}},\ \bibinfo
  {pages} {3415} (\bibinfo {year} {1992})}\BibitemShut {NoStop}%
\bibitem [{\citenamefont {Kurki-Suonio}\ and\ \citenamefont
  {Laine}(1996)}]{kl96bubble}%
  \BibitemOpen
  \bibfield  {author} {\bibinfo {author} {\bibfnamefont {H.}~\bibnamefont
  {Kurki-Suonio}}\ and\ \bibinfo {author} {\bibfnamefont {M.}~\bibnamefont
  {Laine}},\ }\href {https://doi.org/10.1103/PhysRevD.54.7163} {\bibfield
  {journal} {\bibinfo  {journal} {Phys. Rev. D}\ }\textbf {\bibinfo {volume}
  {54}},\ \bibinfo {pages} {7163} (\bibinfo {year} {1996})},\ \Eprint
  {https://arxiv.org/abs/hep-ph/9512202} {arXiv:hep-ph/9512202} \BibitemShut
  {NoStop}%
\bibitem [{\citenamefont {Megevand}(2008)}]{megevand08gravitational}%
  \BibitemOpen
  \bibfield  {author} {\bibinfo {author} {\bibfnamefont {A.}~\bibnamefont
  {Megevand}},\ }\href {https://doi.org/10.1103/PhysRevD.78.084003} {\bibfield
  {journal} {\bibinfo  {journal} {Phys. Rev. D}\ }\textbf {\bibinfo {volume}
  {78}},\ \bibinfo {pages} {084003} (\bibinfo {year} {2008})},\ \Eprint
  {https://arxiv.org/abs/0804.0391} {arXiv:0804.0391 [astro-ph]} \BibitemShut
  {NoStop}%
\bibitem [{\citenamefont {Megevand}\ and\ \citenamefont
  {Sanchez}(2010)}]{ms10velocity}%
  \BibitemOpen
  \bibfield  {author} {\bibinfo {author} {\bibfnamefont {A.}~\bibnamefont
  {Megevand}}\ and\ \bibinfo {author} {\bibfnamefont {A.~D.}\ \bibnamefont
  {Sanchez}},\ }\href {https://doi.org/10.1016/j.nuclphysb.2009.09.019}
  {\bibfield  {journal} {\bibinfo  {journal} {Nucl. Phys. B}\ }\textbf
  {\bibinfo {volume} {825}},\ \bibinfo {pages} {151} (\bibinfo {year}
  {2010})},\ \Eprint {https://arxiv.org/abs/0908.3663} {arXiv:0908.3663
  [hep-ph]} \BibitemShut {NoStop}%
\bibitem [{\citenamefont {Leitao}\ and\ \citenamefont
  {Megevand}(2015)}]{lm15hydrodynamics}%
  \BibitemOpen
  \bibfield  {author} {\bibinfo {author} {\bibfnamefont {L.}~\bibnamefont
  {Leitao}}\ and\ \bibinfo {author} {\bibfnamefont {A.}~\bibnamefont
  {Megevand}},\ }\href {https://doi.org/10.1016/j.nuclphysb.2014.12.008}
  {\bibfield  {journal} {\bibinfo  {journal} {Nucl. Phys. B}\ }\textbf
  {\bibinfo {volume} {891}},\ \bibinfo {pages} {159} (\bibinfo {year}
  {2015})},\ \Eprint {https://arxiv.org/abs/1410.3875} {arXiv:1410.3875
  [hep-ph]} \BibitemShut {NoStop}%
\bibitem [{\citenamefont {M\'egevand}\ and\ \citenamefont
  {Ram\'{\i}rez}(2018)}]{mr18bubble}%
  \BibitemOpen
  \bibfield  {author} {\bibinfo {author} {\bibfnamefont {A.}~\bibnamefont
  {M\'egevand}}\ and\ \bibinfo {author} {\bibfnamefont {S.}~\bibnamefont
  {Ram\'{\i}rez}},\ }\href {https://doi.org/10.1016/j.nuclphysb.2018.01.012}
  {\bibfield  {journal} {\bibinfo  {journal} {Nucl. Phys.}\ }\textbf {\bibinfo
  {volume} {B928}},\ \bibinfo {pages} {38} (\bibinfo {year} {2018})},\ \Eprint
  {https://arxiv.org/abs/1710.06279} {arXiv:1710.06279 [astro-ph.CO]}
  \BibitemShut {NoStop}%
\bibitem [{\citenamefont {Hindmarsh}\ \emph {et~al.}(2015)\citenamefont
  {Hindmarsh}, \citenamefont {Huber}, \citenamefont {Rummukainen},\ and\
  \citenamefont {Weir}}]{hhr+15numerical}%
  \BibitemOpen
  \bibfield  {author} {\bibinfo {author} {\bibfnamefont {M.}~\bibnamefont
  {Hindmarsh}}, \bibinfo {author} {\bibfnamefont {S.~J.}\ \bibnamefont
  {Huber}}, \bibinfo {author} {\bibfnamefont {K.}~\bibnamefont {Rummukainen}},\
  and\ \bibinfo {author} {\bibfnamefont {D.~J.}\ \bibnamefont {Weir}},\ }\href
  {https://doi.org/10.1103/PhysRevD.92.123009} {\bibfield  {journal} {\bibinfo
  {journal} {Phys. Rev. D}\ }\textbf {\bibinfo {volume} {92}},\ \bibinfo
  {pages} {123009} (\bibinfo {year} {2015})},\ \Eprint
  {https://arxiv.org/abs/1504.03291} {arXiv:1504.03291 [astro-ph.CO]}
  \BibitemShut {NoStop}%
\bibitem [{\citenamefont {Hindmarsh}\ \emph {et~al.}(2014)\citenamefont
  {Hindmarsh}, \citenamefont {Huber}, \citenamefont {Rummukainen},\ and\
  \citenamefont {Weir}}]{hhr+14gravitational}%
  \BibitemOpen
  \bibfield  {author} {\bibinfo {author} {\bibfnamefont {M.}~\bibnamefont
  {Hindmarsh}}, \bibinfo {author} {\bibfnamefont {S.~J.}\ \bibnamefont
  {Huber}}, \bibinfo {author} {\bibfnamefont {K.}~\bibnamefont {Rummukainen}},\
  and\ \bibinfo {author} {\bibfnamefont {D.~J.}\ \bibnamefont {Weir}},\ }\href
  {https://doi.org/10.1103/PhysRevLett.112.041301} {\bibfield  {journal}
  {\bibinfo  {journal} {Phys. Rev. Lett.}\ }\textbf {\bibinfo {volume} {112}},\
  \bibinfo {pages} {041301} (\bibinfo {year} {2014})},\ \Eprint
  {https://arxiv.org/abs/1304.2433} {arXiv:1304.2433 [hep-ph]} \BibitemShut
  {NoStop}%
\bibitem [{\citenamefont {Hindmarsh}\ \emph {et~al.}(2017)\citenamefont
  {Hindmarsh}, \citenamefont {Huber}, \citenamefont {Rummukainen},\ and\
  \citenamefont {Weir}}]{hhr+17shape}%
  \BibitemOpen
  \bibfield  {author} {\bibinfo {author} {\bibfnamefont {M.}~\bibnamefont
  {Hindmarsh}}, \bibinfo {author} {\bibfnamefont {S.~J.}\ \bibnamefont
  {Huber}}, \bibinfo {author} {\bibfnamefont {K.}~\bibnamefont {Rummukainen}},\
  and\ \bibinfo {author} {\bibfnamefont {D.~J.}\ \bibnamefont {Weir}},\ }\href
  {https://doi.org/10.1103/PhysRevD.96.103520} {\bibfield  {journal} {\bibinfo
  {journal} {Phys. Rev. D}\ }\textbf {\bibinfo {volume} {96}},\ \bibinfo
  {pages} {103520} (\bibinfo {year} {2017})},\ \bibinfo {note} {[Erratum:
  Phys.Rev.D 101, 089902 (2020)]},\ \Eprint {https://arxiv.org/abs/1704.05871}
  {arXiv:1704.05871 [astro-ph.CO]} \BibitemShut {NoStop}%
\bibitem [{\citenamefont {Hindmarsh}(2018)}]{hindmarsh18sound}%
  \BibitemOpen
  \bibfield  {author} {\bibinfo {author} {\bibfnamefont {M.}~\bibnamefont
  {Hindmarsh}},\ }\href {https://doi.org/10.1103/PhysRevLett.120.071301}
  {\bibfield  {journal} {\bibinfo  {journal} {Phys. Rev. Lett.}\ }\textbf
  {\bibinfo {volume} {120}},\ \bibinfo {pages} {071301} (\bibinfo {year}
  {2018})},\ \Eprint {https://arxiv.org/abs/1608.04735} {arXiv:1608.04735
  [astro-ph.CO]} \BibitemShut {NoStop}%
\bibitem [{\citenamefont {Hindmarsh}\ and\ \citenamefont
  {Hijazi}(2019)}]{hh19gravitational}%
  \BibitemOpen
  \bibfield  {author} {\bibinfo {author} {\bibfnamefont {M.}~\bibnamefont
  {Hindmarsh}}\ and\ \bibinfo {author} {\bibfnamefont {M.}~\bibnamefont
  {Hijazi}},\ }\href {https://doi.org/10.1088/1475-7516/2019/12/062} {\bibfield
   {journal} {\bibinfo  {journal} {JCAP}\ }\textbf {\bibinfo {volume}
  {2019}}\bibfield  {number} {\bibinfo  {number} { (12)},\ \bibinfo {pages}
  {062}},\ }\Eprint {https://arxiv.org/abs/1909.10040} {arXiv:1909.10040
  [astro-ph.CO]} \BibitemShut {NoStop}%
\bibitem [{\citenamefont {Dolgov}\ \emph {et~al.}(2002)\citenamefont {Dolgov},
  \citenamefont {Grasso},\ and\ \citenamefont {Nicolis}}]{dgn02relic}%
  \BibitemOpen
  \bibfield  {author} {\bibinfo {author} {\bibfnamefont {A.~D.}\ \bibnamefont
  {Dolgov}}, \bibinfo {author} {\bibfnamefont {D.}~\bibnamefont {Grasso}},\
  and\ \bibinfo {author} {\bibfnamefont {A.}~\bibnamefont {Nicolis}},\ }\href
  {https://doi.org/10.1103/PhysRevD.66.103505} {\bibfield  {journal} {\bibinfo
  {journal} {Phys. Rev. D}\ }\textbf {\bibinfo {volume} {66}},\ \bibinfo
  {pages} {103505} (\bibinfo {year} {2002})},\ \Eprint
  {https://arxiv.org/abs/astro-ph/0206461} {arXiv:astro-ph/0206461}
  \BibitemShut {NoStop}%
\bibitem [{\citenamefont {Kosowsky}\ \emph {et~al.}(2002)\citenamefont
  {Kosowsky}, \citenamefont {Mack},\ and\ \citenamefont
  {Kahniashvili}}]{kmk02gravitational}%
  \BibitemOpen
  \bibfield  {author} {\bibinfo {author} {\bibfnamefont {A.}~\bibnamefont
  {Kosowsky}}, \bibinfo {author} {\bibfnamefont {A.}~\bibnamefont {Mack}},\
  and\ \bibinfo {author} {\bibfnamefont {T.}~\bibnamefont {Kahniashvili}},\
  }\href {https://doi.org/10.1103/PhysRevD.66.024030} {\bibfield  {journal}
  {\bibinfo  {journal} {Phys. Rev. D}\ }\textbf {\bibinfo {volume} {66}},\
  \bibinfo {pages} {024030} (\bibinfo {year} {2002})},\ \Eprint
  {https://arxiv.org/abs/astro-ph/0111483} {arXiv:astro-ph/0111483}
  \BibitemShut {NoStop}%
\bibitem [{\citenamefont {Caprini}\ and\ \citenamefont
  {Durrer}(2006)}]{cd06gravitational}%
  \BibitemOpen
  \bibfield  {author} {\bibinfo {author} {\bibfnamefont {C.}~\bibnamefont
  {Caprini}}\ and\ \bibinfo {author} {\bibfnamefont {R.}~\bibnamefont
  {Durrer}},\ }\href {https://doi.org/10.1103/PhysRevD.74.063521} {\bibfield
  {journal} {\bibinfo  {journal} {Phys. Rev. D}\ }\textbf {\bibinfo {volume}
  {74}},\ \bibinfo {pages} {063521} (\bibinfo {year} {2006})},\ \Eprint
  {https://arxiv.org/abs/astro-ph/0603476} {arXiv:astro-ph/0603476}
  \BibitemShut {NoStop}%
\bibitem [{\citenamefont {Caprini}\ \emph {et~al.}(2009)\citenamefont
  {Caprini}, \citenamefont {Durrer},\ and\ \citenamefont
  {Servant}}]{cds09stochastic}%
  \BibitemOpen
  \bibfield  {author} {\bibinfo {author} {\bibfnamefont {C.}~\bibnamefont
  {Caprini}}, \bibinfo {author} {\bibfnamefont {R.}~\bibnamefont {Durrer}},\
  and\ \bibinfo {author} {\bibfnamefont {G.}~\bibnamefont {Servant}},\ }\href
  {https://doi.org/10.1088/1475-7516/2009/12/024} {\bibfield  {journal}
  {\bibinfo  {journal} {JCAP}\ }\textbf {\bibinfo {volume} {2009}}\bibfield
  {number} {\bibinfo  {number} { (12)},\ \bibinfo {pages} {024}},\ }\Eprint
  {https://arxiv.org/abs/0909.0622} {arXiv:0909.0622 [astro-ph.CO]}
  \BibitemShut {NoStop}%
\bibitem [{\citenamefont {Kahniashvili}\ \emph {et~al.}(2010)\citenamefont
  {Kahniashvili}, \citenamefont {Kisslinger},\ and\ \citenamefont
  {Stevens}}]{kks10gravitational}%
  \BibitemOpen
  \bibfield  {author} {\bibinfo {author} {\bibfnamefont {T.}~\bibnamefont
  {Kahniashvili}}, \bibinfo {author} {\bibfnamefont {L.}~\bibnamefont
  {Kisslinger}},\ and\ \bibinfo {author} {\bibfnamefont {T.}~\bibnamefont
  {Stevens}},\ }\href {https://doi.org/10.1103/PhysRevD.81.023004} {\bibfield
  {journal} {\bibinfo  {journal} {Phys. Rev. D}\ }\textbf {\bibinfo {volume}
  {81}},\ \bibinfo {pages} {023004} (\bibinfo {year} {2010})},\ \Eprint
  {https://arxiv.org/abs/0905.0643} {arXiv:0905.0643 [astro-ph.CO]}
  \BibitemShut {NoStop}%
\bibitem [{\citenamefont {Gogoberidze}\ \emph {et~al.}(2007)\citenamefont
  {Gogoberidze}, \citenamefont {Kahniashvili},\ and\ \citenamefont
  {Kosowsky}}]{gkk07spectrum}%
  \BibitemOpen
  \bibfield  {author} {\bibinfo {author} {\bibfnamefont {G.}~\bibnamefont
  {Gogoberidze}}, \bibinfo {author} {\bibfnamefont {T.}~\bibnamefont
  {Kahniashvili}},\ and\ \bibinfo {author} {\bibfnamefont {A.}~\bibnamefont
  {Kosowsky}},\ }\href {https://doi.org/10.1103/PhysRevD.76.083002} {\bibfield
  {journal} {\bibinfo  {journal} {Phys. Rev. D}\ }\textbf {\bibinfo {volume}
  {76}},\ \bibinfo {pages} {083002} (\bibinfo {year} {2007})},\ \Eprint
  {https://arxiv.org/abs/0705.1733} {arXiv:0705.1733 [astro-ph]} \BibitemShut
  {NoStop}%
\bibitem [{\citenamefont {Roper~Pol}\ \emph {et~al.}(2020)\citenamefont
  {Roper~Pol}, \citenamefont {Mandal}, \citenamefont {Brandenburg},
  \citenamefont {Kahniashvili},\ and\ \citenamefont
  {Kosowsky}}]{rmb+20numerical}%
  \BibitemOpen
  \bibfield  {author} {\bibinfo {author} {\bibfnamefont {A.}~\bibnamefont
  {Roper~Pol}}, \bibinfo {author} {\bibfnamefont {S.}~\bibnamefont {Mandal}},
  \bibinfo {author} {\bibfnamefont {A.}~\bibnamefont {Brandenburg}}, \bibinfo
  {author} {\bibfnamefont {T.}~\bibnamefont {Kahniashvili}},\ and\ \bibinfo
  {author} {\bibfnamefont {A.}~\bibnamefont {Kosowsky}},\ }\href
  {https://doi.org/10.1103/PhysRevD.102.083512} {\bibfield  {journal} {\bibinfo
   {journal} {Phys. Rev. D}\ }\textbf {\bibinfo {volume} {102}},\ \bibinfo
  {pages} {083512} (\bibinfo {year} {2020})},\ \Eprint
  {https://arxiv.org/abs/1903.08585} {arXiv:1903.08585 [astro-ph.CO]}
  \BibitemShut {NoStop}%
\bibitem [{\citenamefont {Landau}\ and\ \citenamefont
  {Lifshitz}(1987)}]{LandauLifshitz87Fluid}%
  \BibitemOpen
  \bibfield  {author} {\bibinfo {author} {\bibfnamefont {L.~D.}\ \bibnamefont
  {Landau}}\ and\ \bibinfo {author} {\bibfnamefont {E.~M.}\ \bibnamefont
  {Lifshitz}},\ }\href@noop {} {\emph {\bibinfo {title} {Fluid Mechanics}}},\
  \bibinfo {edition} {2nd}\ ed.\ (\bibinfo  {publisher} {Pergamon Press},\
  \bibinfo {address} {London},\ \bibinfo {year} {1987})\BibitemShut {NoStop}%
\bibitem [{\citenamefont {Freese}\ and\ \citenamefont
  {Adams}(1990)}]{fa90hadron}%
  \BibitemOpen
  \bibfield  {author} {\bibinfo {author} {\bibfnamefont {K.}~\bibnamefont
  {Freese}}\ and\ \bibinfo {author} {\bibfnamefont {F.~C.}\ \bibnamefont
  {Adams}},\ }\href {https://doi.org/10.1103/PhysRevD.41.2449} {\bibfield
  {journal} {\bibinfo  {journal} {Phys. Rev. D}\ }\textbf {\bibinfo {volume}
  {41}},\ \bibinfo {pages} {2449} (\bibinfo {year} {1990})}\BibitemShut
  {NoStop}%
\bibitem [{\citenamefont {Link}(1992)}]{link92deflagration}%
  \BibitemOpen
  \bibfield  {author} {\bibinfo {author} {\bibfnamefont {B.}~\bibnamefont
  {Link}},\ }\href {https://doi.org/10.1103/PhysRevLett.68.2425} {\bibfield
  {journal} {\bibinfo  {journal} {Phys. Rev. Lett.}\ }\textbf {\bibinfo
  {volume} {68}},\ \bibinfo {pages} {2425} (\bibinfo {year}
  {1992})}\BibitemShut {NoStop}%
\bibitem [{\citenamefont {Kamionkowski}\ and\ \citenamefont
  {Freese}(1992)}]{kf92instability}%
  \BibitemOpen
  \bibfield  {author} {\bibinfo {author} {\bibfnamefont {M.}~\bibnamefont
  {Kamionkowski}}\ and\ \bibinfo {author} {\bibfnamefont {K.}~\bibnamefont
  {Freese}},\ }\href {https://doi.org/10.1103/PhysRevLett.69.2743} {\bibfield
  {journal} {\bibinfo  {journal} {Phys. Rev. Lett.}\ }\textbf {\bibinfo
  {volume} {69}},\ \bibinfo {pages} {2743} (\bibinfo {year} {1992})},\ \Eprint
  {https://arxiv.org/abs/hep-ph/9208202} {arXiv:hep-ph/9208202} \BibitemShut
  {NoStop}%
\bibitem [{\citenamefont {Huet}\ \emph {et~al.}(1993)\citenamefont {Huet},
  \citenamefont {Kajantie}, \citenamefont {Leigh}, \citenamefont {Liu},\ and\
  \citenamefont {McLerran}}]{hkl+93hydrodynamic}%
  \BibitemOpen
  \bibfield  {author} {\bibinfo {author} {\bibfnamefont {P.~Y.}\ \bibnamefont
  {Huet}}, \bibinfo {author} {\bibfnamefont {K.}~\bibnamefont {Kajantie}},
  \bibinfo {author} {\bibfnamefont {R.~G.}\ \bibnamefont {Leigh}}, \bibinfo
  {author} {\bibfnamefont {B.-H.}\ \bibnamefont {Liu}},\ and\ \bibinfo {author}
  {\bibfnamefont {L.~D.}\ \bibnamefont {McLerran}},\ }\href
  {https://doi.org/10.1103/PhysRevD.48.2477} {\bibfield  {journal} {\bibinfo
  {journal} {Phys. Rev.}\ }\textbf {\bibinfo {volume} {D48}},\ \bibinfo {pages}
  {2477} (\bibinfo {year} {1993})},\ \Eprint
  {https://arxiv.org/abs/hep-ph/9212224} {arXiv:hep-ph/9212224 [hep-ph]}
  \BibitemShut {NoStop}%
\bibitem [{\citenamefont {Fragile}\ and\ \citenamefont
  {Anninos}(2003)}]{fa03hydrodynamic}%
  \BibitemOpen
  \bibfield  {author} {\bibinfo {author} {\bibfnamefont {P.~C.}\ \bibnamefont
  {Fragile}}\ and\ \bibinfo {author} {\bibfnamefont {P.}~\bibnamefont
  {Anninos}},\ }\href {https://doi.org/10.1103/PhysRevD.67.103010} {\bibfield
  {journal} {\bibinfo  {journal} {Phys. Rev. D}\ }\textbf {\bibinfo {volume}
  {67}},\ \bibinfo {pages} {103010} (\bibinfo {year} {2003})},\ \Eprint
  {https://arxiv.org/abs/gr-qc/0303018} {arXiv:gr-qc/0303018} \BibitemShut
  {NoStop}%
\bibitem [{\citenamefont {Megevand}\ and\ \citenamefont
  {Membiela}(2014)}]{mm14stability}%
  \BibitemOpen
  \bibfield  {author} {\bibinfo {author} {\bibfnamefont {A.}~\bibnamefont
  {Megevand}}\ and\ \bibinfo {author} {\bibfnamefont {F.~A.}\ \bibnamefont
  {Membiela}},\ }\href {https://doi.org/10.1103/PhysRevD.89.103507} {\bibfield
  {journal} {\bibinfo  {journal} {Phys. Rev.}\ }\textbf {\bibinfo {volume}
  {D89}},\ \bibinfo {pages} {103507} (\bibinfo {year} {2014})},\ \Eprint
  {https://arxiv.org/abs/1311.2453} {arXiv:1311.2453 [astro-ph.CO]}
  \BibitemShut {NoStop}%
\bibitem [{\citenamefont {Megevand}\ \emph {et~al.}(2015)\citenamefont
  {Megevand}, \citenamefont {Membiela},\ and\ \citenamefont
  {Sanchez}}]{mms15lower}%
  \BibitemOpen
  \bibfield  {author} {\bibinfo {author} {\bibfnamefont {A.}~\bibnamefont
  {Megevand}}, \bibinfo {author} {\bibfnamefont {F.~A.}\ \bibnamefont
  {Membiela}},\ and\ \bibinfo {author} {\bibfnamefont {A.~D.}\ \bibnamefont
  {Sanchez}},\ }\href {https://doi.org/10.1088/1475-7516/2015/03/051}
  {\bibfield  {journal} {\bibinfo  {journal} {JCAP}\ }\textbf {\bibinfo
  {volume} {2015}}\bibfield  {number} {\bibinfo  {number} { (03)},\ \bibinfo
  {pages} {051}},\ }\Eprint {https://arxiv.org/abs/1412.8064} {arXiv:1412.8064
  [hep-ph]} \BibitemShut {NoStop}%
\bibitem [{\citenamefont {Megevand}\ and\ \citenamefont
  {Membiela}(2021{\natexlab{a}})}]{mm21gravitational}%
  \BibitemOpen
  \bibfield  {author} {\bibinfo {author} {\bibfnamefont {A.}~\bibnamefont
  {Megevand}}\ and\ \bibinfo {author} {\bibfnamefont {F.~A.}\ \bibnamefont
  {Membiela}},\ }\href {https://doi.org/10.1088/1475-7516/2021/10/073}
  {\bibfield  {journal} {\bibinfo  {journal} {JCAP}\ }\textbf {\bibinfo
  {volume} {2021}}\bibfield  {number} {\bibinfo  {number} { (10)},\ \bibinfo
  {pages} {073}},\ }\Eprint {https://arxiv.org/abs/2108.05510}
  {arXiv:2108.05510 [astro-ph.CO]} \BibitemShut {NoStop}%
\bibitem [{\citenamefont {Bychkov}\ and\ \citenamefont
  {Liberman}(2000)}]{bl00dynamics}%
  \BibitemOpen
  \bibfield  {author} {\bibinfo {author} {\bibfnamefont {V.}~\bibnamefont
  {Bychkov}}\ and\ \bibinfo {author} {\bibfnamefont {M.}~\bibnamefont
  {Liberman}},\ }\href
  {https://doi.org/https://doi.org/10.1016/S0370-1573(99)00081-2} {\bibfield
  {journal} {\bibinfo  {journal} {Physics Reports}\ }\textbf {\bibinfo {volume}
  {325}},\ \bibinfo {pages} {115} (\bibinfo {year} {2000})}\BibitemShut
  {NoStop}%
\bibitem [{\citenamefont {M{\'e}gevand}\ and\ \citenamefont
  {Membiela}(2026)}]{mm26gravitational}%
  \BibitemOpen
  \bibfield  {author} {\bibinfo {author} {\bibfnamefont {A.}~\bibnamefont
  {M{\'e}gevand}}\ and\ \bibinfo {author} {\bibfnamefont {F.~A.}\ \bibnamefont
  {Membiela}},\ }\href@noop {} {\bibinfo {title} {{Gravitational waves from the
  instability of cosmological phase transition fronts}}} (\bibinfo {year}
  {2026}),\ \Eprint {https://arxiv.org/abs/2611.35408} {arXiv:2611.35408
  [astro-ph.CO]} \BibitemShut {NoStop}%
\bibitem [{\citenamefont {Coleman}(1977)}]{coleman77fate}%
  \BibitemOpen
  \bibfield  {author} {\bibinfo {author} {\bibfnamefont {S.~R.}\ \bibnamefont
  {Coleman}},\ }\href {https://doi.org/10.1103/PhysRevD.16.1248} {\bibfield
  {journal} {\bibinfo  {journal} {Phys. Rev. D}\ }\textbf {\bibinfo {volume}
  {15}},\ \bibinfo {pages} {2929} (\bibinfo {year} {1977})},\ \bibinfo {note}
  {[Erratum: Phys.Rev.D 16, 1248 (1977)]}\BibitemShut {NoStop}%
\bibitem [{\citenamefont {Callan}\ and\ \citenamefont
  {Coleman}(1977)}]{cc77fate}%
  \BibitemOpen
  \bibfield  {author} {\bibinfo {author} {\bibfnamefont {J.}~\bibnamefont
  {Callan}, \bibfnamefont {Curtis~G.}}\ and\ \bibinfo {author} {\bibfnamefont
  {S.~R.}\ \bibnamefont {Coleman}},\ }\href
  {https://doi.org/10.1103/PhysRevD.16.1762} {\bibfield  {journal} {\bibinfo
  {journal} {Phys. Rev. D}\ }\textbf {\bibinfo {volume} {16}},\ \bibinfo
  {pages} {1762} (\bibinfo {year} {1977})}\BibitemShut {NoStop}%
\bibitem [{\citenamefont {Linde}(1981)}]{linde81fate}%
  \BibitemOpen
  \bibfield  {author} {\bibinfo {author} {\bibfnamefont {A.~D.}\ \bibnamefont
  {Linde}},\ }\href {https://doi.org/10.1016/0370-2693(81)90281-1} {\bibfield
  {journal} {\bibinfo  {journal} {Phys. Lett. B}\ }\textbf {\bibinfo {volume}
  {100}},\ \bibinfo {pages} {37} (\bibinfo {year} {1981})}\BibitemShut
  {NoStop}%
\bibitem [{\citenamefont {Linde}(1983)}]{linde83decay}%
  \BibitemOpen
  \bibfield  {author} {\bibinfo {author} {\bibfnamefont {A.~D.}\ \bibnamefont
  {Linde}},\ }\href {https://doi.org/10.1016/0550-3213(83)90072-X} {\bibfield
  {journal} {\bibinfo  {journal} {Nucl. Phys. B}\ }\textbf {\bibinfo {volume}
  {216}},\ \bibinfo {pages} {421} (\bibinfo {year} {1983})},\ \bibinfo {note}
  {[Erratum: Nucl.Phys.B 223, 544 (1983)]}\BibitemShut {NoStop}%
\bibitem [{\citenamefont {M\'egevand}\ and\ \citenamefont
  {Membiela}(2023)}]{mm23thin}%
  \BibitemOpen
  \bibfield  {author} {\bibinfo {author} {\bibfnamefont {A.}~\bibnamefont
  {M\'egevand}}\ and\ \bibinfo {author} {\bibfnamefont {F.~A.}\ \bibnamefont
  {Membiela}},\ }\href {https://doi.org/10.1088/1475-7516/2023/06/007}
  {\bibfield  {journal} {\bibinfo  {journal} {JCAP}\ }\textbf {\bibinfo
  {volume} {2023}}\bibfield  {number} {\bibinfo  {number} { (06)},\ \bibinfo
  {pages} {007}},\ }\Eprint {https://arxiv.org/abs/2302.13349}
  {arXiv:2302.13349 [gr-qc]} \BibitemShut {NoStop}%
\bibitem [{\citenamefont {M\'egevand}\ and\ \citenamefont
  {Membiela}(2024)}]{mm24thinb}%
  \BibitemOpen
  \bibfield  {author} {\bibinfo {author} {\bibfnamefont {A.}~\bibnamefont
  {M\'egevand}}\ and\ \bibinfo {author} {\bibfnamefont {F.~A.}\ \bibnamefont
  {Membiela}},\ }\href {https://doi.org/10.1088/1475-7516/2024/04/003}
  {\bibfield  {journal} {\bibinfo  {journal} {JCAP}\ }\textbf {\bibinfo
  {volume} {2024}}\bibfield  {number} {\bibinfo  {number} { (04)},\ \bibinfo
  {pages} {003}},\ }\bibinfo {note} {arXiv:2401.03102 [astro-ph]}\BibitemShut
  {NoStop}%
\bibitem [{\citenamefont {M{\'e}gevand}\ and\ \citenamefont
  {Membiela}(2025)}]{mm25thin}%
  \BibitemOpen
  \bibfield  {author} {\bibinfo {author} {\bibfnamefont {A.}~\bibnamefont
  {M{\'e}gevand}}\ and\ \bibinfo {author} {\bibfnamefont {F.~A.}\ \bibnamefont
  {Membiela}},\ }\href {https://doi.org/10.1088/1475-7516/2025/05/076}
  {\bibfield  {journal} {\bibinfo  {journal} {JCAP}\ }\textbf {\bibinfo
  {volume} {2025}}\bibfield  {number} {\bibinfo  {number} { (05)},\ \bibinfo
  {pages} {076}},\ }\Eprint {https://arxiv.org/abs/2501.05612}
  {arXiv:2501.05612 [gr-qc]} \BibitemShut {NoStop}%
\bibitem [{\citenamefont {Gyulassy}\ \emph {et~al.}(1984)\citenamefont
  {Gyulassy}, \citenamefont {Kajantie}, \citenamefont {Kurki-Suonio},\ and\
  \citenamefont {McLerran}}]{gkk+84deflagrations}%
  \BibitemOpen
  \bibfield  {author} {\bibinfo {author} {\bibfnamefont {M.}~\bibnamefont
  {Gyulassy}}, \bibinfo {author} {\bibfnamefont {K.}~\bibnamefont {Kajantie}},
  \bibinfo {author} {\bibfnamefont {H.}~\bibnamefont {Kurki-Suonio}},\ and\
  \bibinfo {author} {\bibfnamefont {L.~D.}\ \bibnamefont {McLerran}},\ }\href
  {https://doi.org/10.1016/0550-3213(84)90004-X} {\bibfield  {journal}
  {\bibinfo  {journal} {Nucl. Phys. B}\ }\textbf {\bibinfo {volume} {237}},\
  \bibinfo {pages} {477} (\bibinfo {year} {1984})}\BibitemShut {NoStop}%
\bibitem [{\citenamefont {Kurki-Suonio}(1985)}]{kurkisuonio85deflagration}%
  \BibitemOpen
  \bibfield  {author} {\bibinfo {author} {\bibfnamefont {H.}~\bibnamefont
  {Kurki-Suonio}},\ }\href {https://doi.org/10.1016/0550-3213(85)90135-X}
  {\bibfield  {journal} {\bibinfo  {journal} {Nucl. Phys. B}\ }\textbf
  {\bibinfo {volume} {255}},\ \bibinfo {pages} {231} (\bibinfo {year}
  {1985})}\BibitemShut {NoStop}%
\bibitem [{\citenamefont {Kajantie}\ and\ \citenamefont
  {Kurki-Suonio}(1986)}]{kk86bubble}%
  \BibitemOpen
  \bibfield  {author} {\bibinfo {author} {\bibfnamefont {K.}~\bibnamefont
  {Kajantie}}\ and\ \bibinfo {author} {\bibfnamefont {H.}~\bibnamefont
  {Kurki-Suonio}},\ }\href {https://doi.org/10.1103/PhysRevD.34.1719}
  {\bibfield  {journal} {\bibinfo  {journal} {Phys. Rev. D}\ }\textbf {\bibinfo
  {volume} {34}},\ \bibinfo {pages} {1719} (\bibinfo {year}
  {1986})}\BibitemShut {NoStop}%
\bibitem [{\citenamefont {Kurki-Suonio}\ and\ \citenamefont
  {Laine}(1995)}]{kl95supersonic}%
  \BibitemOpen
  \bibfield  {author} {\bibinfo {author} {\bibfnamefont {H.}~\bibnamefont
  {Kurki-Suonio}}\ and\ \bibinfo {author} {\bibfnamefont {M.}~\bibnamefont
  {Laine}},\ }\href {https://doi.org/10.1103/PhysRevD.51.5431} {\bibfield
  {journal} {\bibinfo  {journal} {Phys. Rev. D}\ }\textbf {\bibinfo {volume}
  {51}},\ \bibinfo {pages} {5431} (\bibinfo {year} {1995})},\ \Eprint
  {https://arxiv.org/abs/hep-ph/9501216} {arXiv:hep-ph/9501216} \BibitemShut
  {NoStop}%
\bibitem [{\citenamefont {Liu}\ \emph {et~al.}(1992)\citenamefont {Liu},
  \citenamefont {McLerran},\ and\ \citenamefont {Turok}}]{lmt92bubble}%
  \BibitemOpen
  \bibfield  {author} {\bibinfo {author} {\bibfnamefont {B.-H.}\ \bibnamefont
  {Liu}}, \bibinfo {author} {\bibfnamefont {L.~D.}\ \bibnamefont {McLerran}},\
  and\ \bibinfo {author} {\bibfnamefont {N.}~\bibnamefont {Turok}},\ }\href
  {https://doi.org/10.1103/PhysRevD.46.2668} {\bibfield  {journal} {\bibinfo
  {journal} {Phys. Rev. D}\ }\textbf {\bibinfo {volume} {46}},\ \bibinfo
  {pages} {2668} (\bibinfo {year} {1992})}\BibitemShut {NoStop}%
\bibitem [{\citenamefont {Turok}(1992)}]{turok92electroweak}%
  \BibitemOpen
  \bibfield  {author} {\bibinfo {author} {\bibfnamefont {N.}~\bibnamefont
  {Turok}},\ }\href {https://doi.org/10.1103/PhysRevLett.68.1803} {\bibfield
  {journal} {\bibinfo  {journal} {Phys. Rev. Lett.}\ }\textbf {\bibinfo
  {volume} {68}},\ \bibinfo {pages} {1803} (\bibinfo {year}
  {1992})}\BibitemShut {NoStop}%
\bibitem [{\citenamefont {Dine}\ \emph {et~al.}(1992)\citenamefont {Dine},
  \citenamefont {Leigh}, \citenamefont {Huet}, \citenamefont {Linde},\ and\
  \citenamefont {Linde}}]{dlh+92towards}%
  \BibitemOpen
  \bibfield  {author} {\bibinfo {author} {\bibfnamefont {M.}~\bibnamefont
  {Dine}}, \bibinfo {author} {\bibfnamefont {R.~G.}\ \bibnamefont {Leigh}},
  \bibinfo {author} {\bibfnamefont {P.~Y.}\ \bibnamefont {Huet}}, \bibinfo
  {author} {\bibfnamefont {A.~D.}\ \bibnamefont {Linde}},\ and\ \bibinfo
  {author} {\bibfnamefont {D.~A.}\ \bibnamefont {Linde}},\ }\href
  {https://doi.org/10.1103/PhysRevD.46.550} {\bibfield  {journal} {\bibinfo
  {journal} {Phys. Rev. D}\ }\textbf {\bibinfo {volume} {46}},\ \bibinfo
  {pages} {550} (\bibinfo {year} {1992})},\ \Eprint
  {https://arxiv.org/abs/hep-ph/9203203} {arXiv:hep-ph/9203203} \BibitemShut
  {NoStop}%
\bibitem [{\citenamefont {Khlebnikov}(1992)}]{khlebnikov92fluctuation}%
  \BibitemOpen
  \bibfield  {author} {\bibinfo {author} {\bibfnamefont {S.~Y.}\ \bibnamefont
  {Khlebnikov}},\ }\href {https://doi.org/10.1103/PhysRevD.46.R3223} {\bibfield
   {journal} {\bibinfo  {journal} {Phys. Rev. D}\ }\textbf {\bibinfo {volume}
  {46}},\ \bibinfo {pages} {3223} (\bibinfo {year} {1992})}\BibitemShut
  {NoStop}%
\bibitem [{\citenamefont {Arnold}(1993)}]{arnold93one}%
  \BibitemOpen
  \bibfield  {author} {\bibinfo {author} {\bibfnamefont {P.~B.}\ \bibnamefont
  {Arnold}},\ }\href {https://doi.org/10.1103/PhysRevD.48.1539} {\bibfield
  {journal} {\bibinfo  {journal} {Phys. Rev. D}\ }\textbf {\bibinfo {volume}
  {48}},\ \bibinfo {pages} {1539} (\bibinfo {year} {1993})},\ \Eprint
  {https://arxiv.org/abs/hep-ph/9302258} {arXiv:hep-ph/9302258} \BibitemShut
  {NoStop}%
\bibitem [{\citenamefont {Moore}\ and\ \citenamefont
  {Prokopec}(1995{\natexlab{a}})}]{mp95bubble}%
  \BibitemOpen
  \bibfield  {author} {\bibinfo {author} {\bibfnamefont {G.~D.}\ \bibnamefont
  {Moore}}\ and\ \bibinfo {author} {\bibfnamefont {T.}~\bibnamefont
  {Prokopec}},\ }\href {https://doi.org/10.1103/PhysRevLett.75.777} {\bibfield
  {journal} {\bibinfo  {journal} {Phys. Rev. Lett.}\ }\textbf {\bibinfo
  {volume} {75}},\ \bibinfo {pages} {777} (\bibinfo {year}
  {1995}{\natexlab{a}})},\ \Eprint {https://arxiv.org/abs/hep-ph/9503296}
  {arXiv:hep-ph/9503296} \BibitemShut {NoStop}%
\bibitem [{\citenamefont {Moore}\ and\ \citenamefont
  {Prokopec}(1995{\natexlab{b}})}]{mp95how}%
  \BibitemOpen
  \bibfield  {author} {\bibinfo {author} {\bibfnamefont {G.~D.}\ \bibnamefont
  {Moore}}\ and\ \bibinfo {author} {\bibfnamefont {T.}~\bibnamefont
  {Prokopec}},\ }\href {https://doi.org/10.1103/PhysRevD.52.7182} {\bibfield
  {journal} {\bibinfo  {journal} {Phys. Rev. D}\ }\textbf {\bibinfo {volume}
  {52}},\ \bibinfo {pages} {7182} (\bibinfo {year} {1995}{\natexlab{b}})},\
  \Eprint {https://arxiv.org/abs/hep-ph/9506475} {arXiv:hep-ph/9506475}
  \BibitemShut {NoStop}%
\bibitem [{\citenamefont {Barrow}\ \emph {et~al.}(2000)\citenamefont {Barrow},
  \citenamefont {Bean},\ and\ \citenamefont {Magueijo}}]{bbm00can}%
  \BibitemOpen
  \bibfield  {author} {\bibinfo {author} {\bibfnamefont {J.}~\bibnamefont
  {Barrow}}, \bibinfo {author} {\bibfnamefont {R.}~\bibnamefont {Bean}},\ and\
  \bibinfo {author} {\bibfnamefont {J.}~\bibnamefont {Magueijo}},\ }\href
  {https://doi.org/10.1046/j.1365-8711.2000.03778.x} {\bibfield  {journal}
  {\bibinfo  {journal} {Mon. Not. Roy. Astron. Soc.}\ }\textbf {\bibinfo
  {volume} {316}},\ \bibinfo {pages} {L41} (\bibinfo {year} {2000})},\ \Eprint
  {https://arxiv.org/abs/astro-ph/0004321} {arXiv:astro-ph/0004321}
  \BibitemShut {NoStop}%
\bibitem [{\citenamefont {Bodeker}\ and\ \citenamefont
  {Moore}(2017)}]{bm17electroweak}%
  \BibitemOpen
  \bibfield  {author} {\bibinfo {author} {\bibfnamefont {D.}~\bibnamefont
  {Bodeker}}\ and\ \bibinfo {author} {\bibfnamefont {G.~D.}\ \bibnamefont
  {Moore}},\ }\href {https://doi.org/10.1088/1475-7516/2017/05/025} {\bibfield
  {journal} {\bibinfo  {journal} {JCAP}\ }\textbf {\bibinfo {volume}
  {2017}}\bibfield  {number} {\bibinfo  {number} { (05)},\ \bibinfo {pages}
  {025}},\ }\Eprint {https://arxiv.org/abs/1703.08215} {arXiv:1703.08215
  [hep-ph]} \BibitemShut {NoStop}%
\bibitem [{\citenamefont {Ignatius}\ \emph {et~al.}(1994)\citenamefont
  {Ignatius}, \citenamefont {Kajantie}, \citenamefont {Kurki-Suonio},\ and\
  \citenamefont {Laine}}]{ikk+94growth}%
  \BibitemOpen
  \bibfield  {author} {\bibinfo {author} {\bibfnamefont {J.}~\bibnamefont
  {Ignatius}}, \bibinfo {author} {\bibfnamefont {K.}~\bibnamefont {Kajantie}},
  \bibinfo {author} {\bibfnamefont {H.}~\bibnamefont {Kurki-Suonio}},\ and\
  \bibinfo {author} {\bibfnamefont {M.}~\bibnamefont {Laine}},\ }\href
  {https://doi.org/10.1103/PhysRevD.49.3854} {\bibfield  {journal} {\bibinfo
  {journal} {Phys. Rev. D}\ }\textbf {\bibinfo {volume} {49}},\ \bibinfo
  {pages} {3854} (\bibinfo {year} {1994})},\ \Eprint
  {https://arxiv.org/abs/astro-ph/9309059} {arXiv:astro-ph/9309059}
  \BibitemShut {NoStop}%
\bibitem [{\citenamefont {Leitao}\ and\ \citenamefont
  {Megevand}(2011)}]{lm11spherical}%
  \BibitemOpen
  \bibfield  {author} {\bibinfo {author} {\bibfnamefont {L.}~\bibnamefont
  {Leitao}}\ and\ \bibinfo {author} {\bibfnamefont {A.}~\bibnamefont
  {Megevand}},\ }\href {https://doi.org/10.1016/j.nuclphysb.2010.11.012}
  {\bibfield  {journal} {\bibinfo  {journal} {Nucl. Phys. B}\ }\textbf
  {\bibinfo {volume} {844}},\ \bibinfo {pages} {450} (\bibinfo {year}
  {2011})},\ \Eprint {https://arxiv.org/abs/1010.2134} {arXiv:1010.2134
  [astro-ph.CO]} \BibitemShut {NoStop}%
\bibitem [{\citenamefont {Heckler}(1995)}]{heckler95effects}%
  \BibitemOpen
  \bibfield  {author} {\bibinfo {author} {\bibfnamefont {A.~F.}\ \bibnamefont
  {Heckler}},\ }\href {https://doi.org/10.1103/PhysRevD.51.405} {\bibfield
  {journal} {\bibinfo  {journal} {Phys. Rev. D}\ }\textbf {\bibinfo {volume}
  {51}},\ \bibinfo {pages} {405} (\bibinfo {year} {1995})},\ \Eprint
  {https://arxiv.org/abs/astro-ph/9407064} {arXiv:astro-ph/9407064}
  \BibitemShut {NoStop}%
\bibitem [{\citenamefont {Guth}\ and\ \citenamefont
  {Weinberg}(1981)}]{gw81cosmological}%
  \BibitemOpen
  \bibfield  {author} {\bibinfo {author} {\bibfnamefont {A.~H.}\ \bibnamefont
  {Guth}}\ and\ \bibinfo {author} {\bibfnamefont {E.~J.}\ \bibnamefont
  {Weinberg}},\ }\href {https://doi.org/10.1103/PhysRevD.23.876} {\bibfield
  {journal} {\bibinfo  {journal} {Phys. Rev.}\ }\textbf {\bibinfo {volume}
  {D23}},\ \bibinfo {pages} {876} (\bibinfo {year} {1981})}\BibitemShut
  {NoStop}%
\bibitem [{\citenamefont {Sharp}(1984)}]{sharp84overview}%
  \BibitemOpen
  \bibfield  {author} {\bibinfo {author} {\bibfnamefont {D.~H.}\ \bibnamefont
  {Sharp}},\ }\href
  {https://doi.org/https://doi.org/10.1016/0167-2789(84)90510-4} {\bibfield
  {journal} {\bibinfo  {journal} {Physica D: Nonlinear Phenomena}\ }\textbf
  {\bibinfo {volume} {12}},\ \bibinfo {pages} {3 } (\bibinfo {year}
  {1984})}\BibitemShut {NoStop}%
\bibitem [{\citenamefont {Sigl}\ \emph {et~al.}(1997)\citenamefont {Sigl},
  \citenamefont {Olinto},\ and\ \citenamefont {Jedamzik}}]{soj97primordial}%
  \BibitemOpen
  \bibfield  {author} {\bibinfo {author} {\bibfnamefont {G.}~\bibnamefont
  {Sigl}}, \bibinfo {author} {\bibfnamefont {A.~V.}\ \bibnamefont {Olinto}},\
  and\ \bibinfo {author} {\bibfnamefont {K.}~\bibnamefont {Jedamzik}},\ }\href
  {https://doi.org/10.1103/PhysRevD.55.4582} {\bibfield  {journal} {\bibinfo
  {journal} {Phys. Rev.}\ }\textbf {\bibinfo {volume} {D55}},\ \bibinfo {pages}
  {4582} (\bibinfo {year} {1997})},\ \Eprint
  {https://arxiv.org/abs/astro-ph/9610201} {arXiv:astro-ph/9610201 [astro-ph]}
  \BibitemShut {NoStop}%
\bibitem [{\citenamefont {Safran}(1994)}]{Safran94Statistical}%
  \BibitemOpen
  \bibfield  {author} {\bibinfo {author} {\bibfnamefont {S.~A.}\ \bibnamefont
  {Safran}},\ }\href@noop {} {\emph {\bibinfo {title} {{Statistical
  Thermodynamics of Surfaces, Interfaces, and Membranes}}}},\ Frontiers in
  physics\ (\bibinfo  {publisher} {Addison-Wesley Pub.},\ \bibinfo {year}
  {1994})\BibitemShut {NoStop}%
\bibitem [{\citenamefont {Jinno}\ and\ \citenamefont
  {Takimoto}(2017)}]{jt17gravitational}%
  \BibitemOpen
  \bibfield  {author} {\bibinfo {author} {\bibfnamefont {R.}~\bibnamefont
  {Jinno}}\ and\ \bibinfo {author} {\bibfnamefont {M.}~\bibnamefont
  {Takimoto}},\ }\href {https://doi.org/10.1103/PhysRevD.95.024009} {\bibfield
  {journal} {\bibinfo  {journal} {Phys. Rev.}\ }\textbf {\bibinfo {volume}
  {D95}},\ \bibinfo {pages} {024009} (\bibinfo {year} {2017})},\ \Eprint
  {https://arxiv.org/abs/1605.01403} {arXiv:1605.01403 [astro-ph.CO]}
  \BibitemShut {NoStop}%
\bibitem [{\citenamefont {M\'egevand}\ and\ \citenamefont
  {Membiela}(2020)}]{mm20bubble}%
  \BibitemOpen
  \bibfield  {author} {\bibinfo {author} {\bibfnamefont {A.}~\bibnamefont
  {M\'egevand}}\ and\ \bibinfo {author} {\bibfnamefont {F.~A.}\ \bibnamefont
  {Membiela}},\ }\href {https://doi.org/10.1103/PhysRevD.102.103514} {\bibfield
   {journal} {\bibinfo  {journal} {Phys. Rev. D}\ }\textbf {\bibinfo {volume}
  {102}},\ \bibinfo {pages} {103514} (\bibinfo {year} {2020})},\ \Eprint
  {https://arxiv.org/abs/2008.01873} {arXiv:2008.01873 [astro-ph.CO]}
  \BibitemShut {NoStop}%
\bibitem [{\citenamefont {Megevand}\ and\ \citenamefont
  {Membiela}(2021{\natexlab{b}})}]{mm21model}%
  \BibitemOpen
  \bibfield  {author} {\bibinfo {author} {\bibfnamefont {A.}~\bibnamefont
  {Megevand}}\ and\ \bibinfo {author} {\bibfnamefont {F.~A.}\ \bibnamefont
  {Membiela}},\ }\href {https://doi.org/10.1103/PhysRevD.104.123532} {\bibfield
   {journal} {\bibinfo  {journal} {Phys. Rev. D}\ }\textbf {\bibinfo {volume}
  {104}},\ \bibinfo {pages} {123532} (\bibinfo {year} {2021}{\natexlab{b}})},\
  \Eprint {https://arxiv.org/abs/2108.07034} {arXiv:2108.07034 [astro-ph.CO]}
  \BibitemShut {NoStop}%
\bibitem [{\citenamefont {Navas}\ \emph {et~al.}(2024)\citenamefont {Navas}
  \emph {et~al.}}]{n+24review}%
  \BibitemOpen
  \bibfield  {author} {\bibinfo {author} {\bibfnamefont {S.}~\bibnamefont
  {Navas}} \emph {et~al.} (\bibinfo {collaboration} {Particle Data Group}),\
  }\href {https://doi.org/10.1103/PhysRevD.110.030001} {\bibfield  {journal}
  {\bibinfo  {journal} {Phys. Rev. D}\ }\textbf {\bibinfo {volume} {110}},\
  \bibinfo {pages} {030001} (\bibinfo {year} {2024})}\BibitemShut {NoStop}%
\bibitem [{\citenamefont {Hobbs}\ \emph {et~al.}(2010)\citenamefont {Hobbs}
  \emph {et~al.}}]{haa+10international}%
  \BibitemOpen
  \bibfield  {author} {\bibinfo {author} {\bibfnamefont {G.}~\bibnamefont
  {Hobbs}} \emph {et~al.},\ }\href
  {https://doi.org/10.1088/0264-9381/27/8/084013} {\bibfield  {journal}
  {\bibinfo  {journal} {Classical and Quantum Gravity}\ }\textbf {\bibinfo
  {volume} {27}},\ \bibinfo {pages} {084013} (\bibinfo {year}
  {2010})}\BibitemShut {NoStop}%
\bibitem [{\citenamefont {Manchester}(2013)}]{manchester13international}%
  \BibitemOpen
  \bibfield  {author} {\bibinfo {author} {\bibfnamefont {R.~N.}\ \bibnamefont
  {Manchester}},\ }\href {https://doi.org/10.1088/0264-9381/30/22/224010}
  {\bibfield  {journal} {\bibinfo  {journal} {Classical and Quantum Gravity}\
  }\textbf {\bibinfo {volume} {30}},\ \bibinfo {pages} {224010} (\bibinfo
  {year} {2013})}\BibitemShut {NoStop}%
\bibitem [{\citenamefont {Verbiest}\ \emph {et~al.}(2016)\citenamefont
  {Verbiest} \emph {et~al.}}]{vlh+16international}%
  \BibitemOpen
  \bibfield  {author} {\bibinfo {author} {\bibfnamefont {J.~P.~W.}\
  \bibnamefont {Verbiest}} \emph {et~al.},\ }\href
  {https://doi.org/10.1093/mnras/stw347} {\bibfield  {journal} {\bibinfo
  {journal} {Monthly Notices of the Royal Astronomical Society}\ }\textbf
  {\bibinfo {volume} {458}},\ \bibinfo {pages} {1267} (\bibinfo {year}
  {2016})}\BibitemShut {NoStop}%
\bibitem [{\citenamefont {Carilli}\ and\ \citenamefont
  {Rawlings}(2004)}]{cr04motivation}%
  \BibitemOpen
  \bibfield  {author} {\bibinfo {author} {\bibfnamefont {C.~L.}\ \bibnamefont
  {Carilli}}\ and\ \bibinfo {author} {\bibfnamefont {S.}~\bibnamefont
  {Rawlings}},\ }\href {https://doi.org/10.1016/j.newar.2004.09.001} {\bibfield
   {journal} {\bibinfo  {journal} {New Astronomy Reviews}\ }\textbf {\bibinfo
  {volume} {48}},\ \bibinfo {pages} {979} (\bibinfo {year} {2004})}\BibitemShut
  {NoStop}%
\bibitem [{\citenamefont {Weltman}\ \emph {et~al.}(2020)\citenamefont {Weltman}
  \emph {et~al.}}]{wbc+20fundamental}%
  \BibitemOpen
  \bibfield  {author} {\bibinfo {author} {\bibfnamefont {A.}~\bibnamefont
  {Weltman}} \emph {et~al.},\ }\bibfield  {journal} {\bibinfo  {journal}
  {Publications of the Astronomical Society of Australia}\ }\textbf {\bibinfo
  {volume} {37}},\ \href {https://doi.org/10.1017/pasa.2019.42}
  {10.1017/pasa.2019.42} (\bibinfo {year} {2020})\BibitemShut {NoStop}%
\bibitem [{\citenamefont {Garcia-Bellido}\ \emph {et~al.}(2021)\citenamefont
  {Garcia-Bellido}, \citenamefont {Murayama},\ and\ \citenamefont
  {White}}]{gmw21exploring}%
  \BibitemOpen
  \bibfield  {author} {\bibinfo {author} {\bibfnamefont {J.}~\bibnamefont
  {Garcia-Bellido}}, \bibinfo {author} {\bibfnamefont {H.}~\bibnamefont
  {Murayama}},\ and\ \bibinfo {author} {\bibfnamefont {G.}~\bibnamefont
  {White}},\ }\href {https://doi.org/10.1088/1475-7516/2021/12/023} {\bibfield
  {journal} {\bibinfo  {journal} {Journal of Cosmology and Astroparticle
  Physics}\ }\textbf {\bibinfo {volume} {2021}}\bibinfo  {number} { (12)},\
  \bibinfo {pages} {023}}\BibitemShut {NoStop}%
\bibitem [{\citenamefont {Sesana}\ \emph {et~al.}(2021)\citenamefont {Sesana}
  \emph {et~al.}}]{s+21unveiling}%
  \BibitemOpen
\bibfield  {number} {  }\bibfield  {author} {\bibinfo {author} {\bibfnamefont
  {A.}~\bibnamefont {Sesana}} \emph {et~al.},\ }\href
  {https://doi.org/10.1007/s10686-021-09709-9} {\bibfield  {journal} {\bibinfo
  {journal} {Exper. Astron.}\ }\textbf {\bibinfo {volume} {51}},\ \bibinfo
  {pages} {1333} (\bibinfo {year} {2021})},\ \Eprint
  {https://arxiv.org/abs/1908.11391} {arXiv:1908.11391 [astro-ph.IM]}
  \BibitemShut {NoStop}%
\bibitem [{\citenamefont {Robson}\ \emph {et~al.}(2019)\citenamefont {Robson},
  \citenamefont {Cornish},\ and\ \citenamefont {Liu}}]{rcl19construction}%
  \BibitemOpen
  \bibfield  {author} {\bibinfo {author} {\bibfnamefont {T.}~\bibnamefont
  {Robson}}, \bibinfo {author} {\bibfnamefont {N.~J.}\ \bibnamefont
  {Cornish}},\ and\ \bibinfo {author} {\bibfnamefont {C.}~\bibnamefont {Liu}},\
  }\href {https://doi.org/10.1088/1361-6382/ab1101} {\bibfield  {journal}
  {\bibinfo  {journal} {Classical and Quantum Gravity}\ }\textbf {\bibinfo
  {volume} {36}},\ \bibinfo {pages} {105011} (\bibinfo {year}
  {2019})}\BibitemShut {NoStop}%
\bibitem [{\citenamefont {Caprini}\ \emph {et~al.}(2016)\citenamefont {Caprini}
  \emph {et~al.}}]{c+16science}%
  \BibitemOpen
  \bibfield  {author} {\bibinfo {author} {\bibfnamefont {C.}~\bibnamefont
  {Caprini}} \emph {et~al.},\ }\href
  {https://doi.org/10.1088/1475-7516/2016/04/001} {\bibfield  {journal}
  {\bibinfo  {journal} {JCAP}\ }\textbf {\bibinfo {volume} {2016}}\bibfield
  {number} {\bibinfo  {number} { (04)},\ \bibinfo {pages} {001}},\ }\Eprint
  {https://arxiv.org/abs/1512.06239} {arXiv:1512.06239 [astro-ph.CO]}
  \BibitemShut {NoStop}%
\bibitem [{\citenamefont {Caprini}\ \emph {et~al.}(2020)\citenamefont {Caprini}
  \emph {et~al.}}]{c+20detecting}%
  \BibitemOpen
  \bibfield  {author} {\bibinfo {author} {\bibfnamefont {C.}~\bibnamefont
  {Caprini}} \emph {et~al.},\ }\href
  {https://doi.org/10.1088/1475-7516/2020/03/024} {\bibfield  {journal}
  {\bibinfo  {journal} {JCAP}\ }\textbf {\bibinfo {volume} {2020}}\bibfield
  {number} {\bibinfo  {number} { (03)},\ \bibinfo {pages} {024}},\ }\Eprint
  {https://arxiv.org/abs/1910.13125} {arXiv:1910.13125 [astro-ph.CO]}
  \BibitemShut {NoStop}%
\bibitem [{\citenamefont {Crowder}\ and\ \citenamefont
  {Cornish}(2005)}]{cc05lisa}%
  \BibitemOpen
  \bibfield  {author} {\bibinfo {author} {\bibfnamefont {J.}~\bibnamefont
  {Crowder}}\ and\ \bibinfo {author} {\bibfnamefont {N.~J.}\ \bibnamefont
  {Cornish}},\ }\href {https://doi.org/10.1103/PhysRevD.72.083005} {\bibfield
  {journal} {\bibinfo  {journal} {Phys. Rev. D}\ }\textbf {\bibinfo {volume}
  {72}},\ \bibinfo {pages} {083005} (\bibinfo {year} {2005})}\BibitemShut
  {NoStop}%
\bibitem [{\citenamefont {Corbin}\ and\ \citenamefont
  {Cornish}(2006)}]{cc06detecting}%
  \BibitemOpen
  \bibfield  {author} {\bibinfo {author} {\bibfnamefont {V.}~\bibnamefont
  {Corbin}}\ and\ \bibinfo {author} {\bibfnamefont {N.~J.}\ \bibnamefont
  {Cornish}},\ }\href {https://doi.org/doi:10.1088/0264-9381/23/7/014}
  {\bibfield  {journal} {\bibinfo  {journal} {Classical and Quantum Gravity}\
  }\textbf {\bibinfo {volume} {23}},\ \bibinfo {pages} {2435} (\bibinfo {year}
  {2006})}\BibitemShut {NoStop}%
\bibitem [{\citenamefont {Harry}\ \emph {et~al.}(2006)\citenamefont {Harry},
  \citenamefont {Fritschel}, \citenamefont {Shaddock}, \citenamefont
  {Folkner},\ and\ \citenamefont {Phinney}}]{hfs+06laser}%
  \BibitemOpen
  \bibfield  {author} {\bibinfo {author} {\bibfnamefont {G.~M.}\ \bibnamefont
  {Harry}}, \bibinfo {author} {\bibfnamefont {P.}~\bibnamefont {Fritschel}},
  \bibinfo {author} {\bibfnamefont {D.~A.}\ \bibnamefont {Shaddock}}, \bibinfo
  {author} {\bibfnamefont {W.}~\bibnamefont {Folkner}},\ and\ \bibinfo {author}
  {\bibfnamefont {E.~S.}\ \bibnamefont {Phinney}},\ }\href
  {https://doi.org/10.1088/0264-9381/23/15/008} {\bibfield  {journal} {\bibinfo
   {journal} {Class. Quant. Grav.}\ }\textbf {\bibinfo {volume} {23}},\
  \bibinfo {pages} {4887} (\bibinfo {year} {2006})},\ \bibinfo {note}
  {[Erratum: Class.Quant.Grav. 23, 7361 (2006)]}\BibitemShut {NoStop}%
\bibitem [{\citenamefont {Kawamura}\ \emph {et~al.}(2011)\citenamefont
  {Kawamura} \emph {et~al.}}]{k+11japanese}%
  \BibitemOpen
  \bibfield  {author} {\bibinfo {author} {\bibfnamefont {S.}~\bibnamefont
  {Kawamura}} \emph {et~al.},\ }\href
  {https://doi.org/10.1088/0264-9381/28/9/094011} {\bibfield  {journal}
  {\bibinfo  {journal} {Class. Quant. Grav.}\ }\textbf {\bibinfo {volume}
  {28}},\ \bibinfo {pages} {094011} (\bibinfo {year} {2011})}\BibitemShut
  {NoStop}%
\bibitem [{\citenamefont {Yagi}\ and\ \citenamefont
  {Seto}(2011)}]{ys11detector}%
  \BibitemOpen
  \bibfield  {author} {\bibinfo {author} {\bibfnamefont {K.}~\bibnamefont
  {Yagi}}\ and\ \bibinfo {author} {\bibfnamefont {N.}~\bibnamefont {Seto}},\
  }\bibfield  {journal} {\bibinfo  {journal} {Physical Review D}\ }\textbf
  {\bibinfo {volume} {83}},\ \href {https://doi.org/10.1103/physrevd.83.044011}
  {10.1103/physrevd.83.044011} (\bibinfo {year} {2011})\BibitemShut {NoStop}%
\bibitem [{\citenamefont {Isoyama}\ \emph {et~al.}(2018)\citenamefont
  {Isoyama}, \citenamefont {Nakano},\ and\ \citenamefont
  {Nakamura}}]{inn18multiband}%
  \BibitemOpen
  \bibfield  {author} {\bibinfo {author} {\bibfnamefont {S.}~\bibnamefont
  {Isoyama}}, \bibinfo {author} {\bibfnamefont {H.}~\bibnamefont {Nakano}},\
  and\ \bibinfo {author} {\bibfnamefont {T.}~\bibnamefont {Nakamura}},\
  }\bibfield  {journal} {\bibinfo  {journal} {Progress of Theoretical and
  Experimental Physics}\ }\textbf {\bibinfo {volume} {2018}},\ \href
  {https://doi.org/10.1093/ptep/pty078} {10.1093/ptep/pty078} (\bibinfo {year}
  {2018})\BibitemShut {NoStop}%
\bibitem [{\citenamefont {Punturo}\ \emph {et~al.}(2010)\citenamefont {Punturo}
  \emph {et~al.}}]{paa+10einstein}%
  \BibitemOpen
  \bibfield  {author} {\bibinfo {author} {\bibfnamefont {M.}~\bibnamefont
  {Punturo}} \emph {et~al.},\ }\href
  {https://doi.org/10.1088/0264-9381/27/19/194002} {\bibfield  {journal}
  {\bibinfo  {journal} {Classical and Quantum Gravity}\ }\textbf {\bibinfo
  {volume} {27}},\ \bibinfo {pages} {194002} (\bibinfo {year}
  {2010})}\BibitemShut {NoStop}%
\bibitem [{\citenamefont {Maggiore}\ \emph {et~al.}(2020)\citenamefont
  {Maggiore} \emph {et~al.}}]{m+20science}%
  \BibitemOpen
  \bibfield  {author} {\bibinfo {author} {\bibfnamefont {M.}~\bibnamefont
  {Maggiore}} \emph {et~al.} (\bibinfo {collaboration} {ET}),\ }\href
  {https://doi.org/10.1088/1475-7516/2020/03/050} {\bibfield  {journal}
  {\bibinfo  {journal} {JCAP}\ }\textbf {\bibinfo {volume} {2020}}\bibfield
  {number} {\bibinfo  {number} { (03)},\ \bibinfo {pages} {050}},\ }\Eprint
  {https://arxiv.org/abs/1912.02622} {arXiv:1912.02622 [astro-ph.CO]}
  \BibitemShut {NoStop}%
\bibitem [{\citenamefont {Abac}\ \emph {et~al.}(2026)\citenamefont {Abac} \emph
  {et~al.}}]{aaa+26science}%
  \BibitemOpen
  \bibfield  {author} {\bibinfo {author} {\bibfnamefont {A.}~\bibnamefont
  {Abac}} \emph {et~al.},\ }\href
  {https://doi.org/10.1088/1475-7516/2026/03/081} {\bibfield  {journal}
  {\bibinfo  {journal} {Journal of Cosmology and Astroparticle Physics}\
  }\textbf {\bibinfo {volume} {2026}}\bibinfo  {number} { (03)},\ \bibinfo
  {pages} {081}}\BibitemShut {NoStop}%
\bibitem [{\citenamefont {Abbott}\ \emph {et~al.}(2017)\citenamefont {Abbott}
  \emph {et~al.}}]{aaa+17exploring}%
  \BibitemOpen
\bibfield  {number} {  }\bibfield  {author} {\bibinfo {author} {\bibfnamefont
  {B.~P.}\ \bibnamefont {Abbott}} \emph {et~al.},\ }\href
  {https://doi.org/10.1088/1361-6382/aa51f4} {\bibfield  {journal} {\bibinfo
  {journal} {Classical and Quantum Gravity}\ }\textbf {\bibinfo {volume}
  {34}},\ \bibinfo {pages} {044001} (\bibinfo {year} {2017})}\BibitemShut
  {NoStop}%
\bibitem [{\citenamefont {Reitze}\ \emph {et~al.}(2019)\citenamefont {Reitze}
  \emph {et~al.}}]{r+19cosmic}%
  \BibitemOpen
  \bibfield  {author} {\bibinfo {author} {\bibfnamefont {D.}~\bibnamefont
  {Reitze}} \emph {et~al.},\ }\href@noop {} {\bibfield  {journal} {\bibinfo
  {journal} {Bull. Am. Astron. Soc.}\ }\textbf {\bibinfo {volume} {51}},\
  \bibinfo {pages} {035} (\bibinfo {year} {2019})},\ \Eprint
  {https://arxiv.org/abs/1907.04833} {arXiv:1907.04833 [astro-ph.IM]}
  \BibitemShut {NoStop}%
\bibitem [{\citenamefont {Thrane}\ and\ \citenamefont
  {Romano}(2013)}]{tr13sensitivity}%
  \BibitemOpen
  \bibfield  {author} {\bibinfo {author} {\bibfnamefont {E.}~\bibnamefont
  {Thrane}}\ and\ \bibinfo {author} {\bibfnamefont {J.~D.}\ \bibnamefont
  {Romano}},\ }\bibfield  {journal} {\bibinfo  {journal} {Physical Review D}\
  }\textbf {\bibinfo {volume} {88}},\ \href
  {https://doi.org/10.1103/physrevd.88.124032} {10.1103/physrevd.88.124032}
  (\bibinfo {year} {2013})\BibitemShut {NoStop}%
\bibitem [{\citenamefont {Schmitz}(2021)}]{schmitz21new}%
  \BibitemOpen
  \bibfield  {author} {\bibinfo {author} {\bibfnamefont {K.}~\bibnamefont
  {Schmitz}},\ }\href {https://doi.org/10.1007/JHEP01(2021)097} {\bibfield
  {journal} {\bibinfo  {journal} {JHEP}\ }\textbf {\bibinfo {volume}
  {2021}}\bibfield  {number} {\bibinfo  {number} { (01)},\ \bibinfo {pages}
  {097}},\ }\Eprint {https://arxiv.org/abs/2002.04615} {arXiv:2002.04615
  [hep-ph]} \BibitemShut {NoStop}%
\bibitem [{\citenamefont {Coffey}\ \emph {et~al.}(2004)\citenamefont {Coffey},
  \citenamefont {Kalmykov},\ and\ \citenamefont
  {Waldron}}]{CoffeyKalmykovWaldron04Langevin}%
  \BibitemOpen
  \bibfield  {author} {\bibinfo {author} {\bibfnamefont {W.}~\bibnamefont
  {Coffey}}, \bibinfo {author} {\bibfnamefont {Y.~P.}\ \bibnamefont
  {Kalmykov}},\ and\ \bibinfo {author} {\bibfnamefont {J.~T.}\ \bibnamefont
  {Waldron}},\ }\href@noop {} {\emph {\bibinfo {title} {The Langevin Equation:
  With Applications in Physics, Chemistry and Electrical Engineering}}},\
  \bibinfo {edition} {2nd}\ ed.,\ World Scientific Series In Contemporary
  Chemical Physics\ (\bibinfo  {publisher} {World Scientific Publishing
  Company},\ \bibinfo {year} {2004})\BibitemShut {NoStop}%
\bibitem [{\citenamefont
  {Milton~Abramowitz}(1970)}]{MiltonAbramowitz70Handbook}%
  \BibitemOpen
  \bibfield  {author} {\bibinfo {author} {\bibfnamefont {I.~A.~S.}\
  \bibnamefont {Milton~Abramowitz}},\ }\href@noop {} {\emph {\bibinfo {title}
  {{Handbook of Mathematical Functions with Formulas, Graphs, and Mathematical
  Table}}}}\ (\bibinfo  {publisher} {Dover Publications},\ \bibinfo {year}
  {1970})\BibitemShut {NoStop}%
\end{thebibliography}%

\end{document}